\documentclass[acmsmall, screen]{acmart}    
\AtBeginDocument{%
  \providecommand\BibTeX{{%
    \normalfont B\kern-0.5em{\scshape i\kern-0.25em b}\kern-0.8em\TeX}}}

\makeatletter
\let\@authorsaddresses\@empty
\makeatother

\setcopyright{acmlicensed}
\copyrightyear{2025}
\acmYear{2025}
\acmDOI{XXXXXXX.XXXXXXX}
\acmJournal{TODAES}

\ccsdesc[500]{Hardware~Very large scale integration design}
\ccsdesc[500]{Computing methodologies~Machine learning}

\usepackage{hhline}
\usepackage{colortbl}
\usepackage{centernot}
\usepackage{pbox}
\usepackage{amsmath}
\usepackage{amsfonts}
\usepackage{url}
\usepackage{bm}
\usepackage{comment}
\usepackage{subfig}
\usepackage{graphicx}
\usepackage{cleveref}
\usepackage{epstopdf}
\usepackage{multirow}
\usepackage{color}
\usepackage{tabu}
\usepackage{mdframed}
\usepackage{syntax}
\usepackage{fancyvrb}

\usepackage[ruled,linesnumbered,algo2e]{algorithm2e}
\usepackage[noend]{algpseudocode}
\usepackage{algorithmicx}
\usepackage{algpseudocode}
\usepackage{paralist}
\usepackage{stmaryrd}
\usepackage{tikz}
\usepackage[referable]{threeparttablex}
\usepackage{tabularx}
\usepackage{booktabs}
\usepackage{array}
\usepackage{fancyhdr}
\usepackage{float}
\usepackage{xspace}
\usepackage{pifont}
\usepackage{balance}
\usepackage{xcolor}
\usepackage[figuresright]{rotating}
\usepackage{tablefootnote}
\usepackage[normalem]{ulem}

\usepackage[utf8]{inputenc} 
\usepackage[T1]{fontenc}    
\usepackage{hyperref}       
\usepackage{graphicx}
\usepackage{tikz}
\usepackage{psfrag}
\usepackage{forest}
\usepackage{makecell}
\usepackage{adjustbox}
\usepackage{booktabs}
\usepackage{multirow}
\usepackage{enumitem}

\usepackage{url}            
\usepackage{booktabs}       
\usepackage{amsfonts}       
\usepackage{nicefrac}       
\usepackage{microtype}      
\usepackage{xcolor}         
\usepackage{color}
\usepackage{amsmath}

\usepackage{tikz}
\usepackage{xcolor}

\usepackage{stfloats}

\usepackage{graphicx}
\usepackage{adjustbox}

\usepackage{float}

\newfloat{figtab}{htb}{fgtb}
\makeatletter
  \newcommand\figcaption{\def\@captype{figure}\caption}
  \newcommand\tabcaption{\def\@captype{table}\caption}
\makeatother

\definecolor{princetonorange}{RGB}{255,143,0}

\definecolor{lightgreen}{RGB}{198, 224, 183}

\definecolor{lightred}{RGB}{240, 205, 176}

\definecolor{normal-purple}{RGB}{132, 66, 245}

\newcommand{\gray}[1]{\textcolor{gray}{#1}}

\usepackage{pgfplots}
\pgfplotsset{compat=1.17} 
\usepgfplotslibrary{dateplot} 

\definecolor{USTgold}{RGB}{153,102,0}
\definecolor{USTyellow}{RGB}{204,153,0}
\definecolor{USTyellowlight}{RGB}{255,212,0}
\definecolor{USTorange}{RGB}{255,166,26}
\definecolor{USTpink}{RGB}{255,157,157}
\definecolor{USTblue}{RGB}{0,51,102}
\definecolor{USTmiddle}{RGB}{0,116,188}
\definecolor{USTlight}{RGB}{99,202,225}
\definecolor{USTgray}{RGB}{204,204,204}
\definecolor{USTred}{RGB}{237,27,47}
\definecolor{USTdarkred}{RGB}{124,35,72}

\usepackage{tcolorbox}
\tcbuselibrary{skins,breakable}
    {\endtcolorbox}

\newcommand{\new}[1]{{#1}}

\begin{document}




\title{A Survey of Circuit Foundation Model: \\Foundation AI Models for VLSI Circuit Design and EDA}

\author{Wenji Fang}
\authornotemark[2]
\affiliation{%
  \institution{Hong Kong University of Science and Technology (HKUST)}
  \country{Hong Kong}
}
\orcid{0000-0002-8380-9395}

\author{Jing Wang}
\authornotemark[2]
\affiliation{%
    \institution{Hong Kong University of Science and Technology (HKUST)}
  \country{Hong Kong}
}
\orcid{0009-0000-3117-3340}

\author{Yao Lu}
\affiliation{%
    \institution{Hong Kong University of Science and Technology (HKUST)}
  \country{Hong Kong}
}
\orcid{0009-0007-3230-7786}

\author{Shang Liu}
\affiliation{%
    \institution{Hong Kong University of Science and Technology (HKUST)}
  \country{Hong Kong}
}
\orcid{0009-0000-0057-7844}

\author{Yuchao Wu}
\affiliation{%
    \institution{Hong Kong University of Science and Technology (HKUST)}
  \country{Hong Kong}
}
\orcid{0009-0000-8040-6198}

\author{Yuzhe Ma}
\affiliation{%
    \institution{Hong Kong University of Science and Technology (Guangzhou) (HKUST(GZ))}
  \country{China}
}
\orcid{0000-0002-3612-4182}

\author{Zhiyao Xie}
\authornotemark[1]
\affiliation{%
    \institution{Hong Kong University of Science and Technology (HKUST)}
  \country{Hong Kong}
}
\email{eezhiyao@ust.hk}
\orcid{0000-0002-4442-592X}

\renewcommand{\thefootnote}{\fnsymbol{footnote}}

\footnotetext[2]{These authors contributed equally to this work.}
\footnotetext[1]{Corresponding author (eezhiyao@ust.hk).}

\renewcommand{\thefootnote}{\arabic{footnote}}


\begin{abstract}


Artificial intelligence (AI)-driven electronic design automation (EDA) techniques have been extensively explored for VLSI circuit design applications. Most recently, \textbf{foundation AI models for circuits} have emerged as a new technology trend. Unlike traditional task-specific AI solutions, these new AI models are developed through two stages: 1) \emph{self-supervised pre-training} on a large amount of unlabeled data to learn intrinsic circuit properties; and 2) \emph{efficient fine-tuning} for specific downstream applications, such as early-stage design quality evaluation, circuit-related context generation, and functional verification. 
This new paradigm brings many advantages: model generalization, less reliance on labeled circuit data, efficient adaptation to new tasks, and unprecedented generative capability. In this paper, we propose referring to AI models developed with this new paradigm as \textbf{\textit{circuit foundation models (CFMs)}}. 
This paper provides a comprehensive survey of the latest progress in circuit foundation models, unprecedentedly covering over 160 relevant works. Over 90\% of our introduced works were published in or after 2022, indicating that this emerging research trend has attracted wide attention in a short period. In this survey, we propose to categorize all existing circuit foundation models into two primary types: 1) \textbf{encoder-based methods} performing general \emph{circuit representation learning for predictive tasks}; and 2) \textbf{decoder-based methods} leveraging \emph{large language models (LLMs) for generative tasks}. For our introduced works, we cover their input modalities, model architecture, pre-training strategies, domain adaptation techniques, and downstream design applications. In addition, this paper discussed the unique properties of \emph{circuits} from the data perspective. These circuit properties have motivated many works in this domain and differentiated them from general AI techniques. Finally, we shared our observed challenges and potential future research directions about developing foundation AI models for EDA methodologies. 


\end{abstract}




\maketitle

\section{Introduction}\label{sec:intro}  


Integrated circuit (IC) is the foundation of our information society. Its complexity has been continuously growing, recently exceeding 100 billion transistors~\cite{tirumala2024nvidia}. 
Such increases in IC complexity have led to sky-rocketing IC design costs, which are estimated to surpass US\$500 million for 3nm technology~\cite{DesignCost}. These challenges result in a compelling need to improve IC design efficiency, possibly achieved by ground-breaking next-generation electronic design automation (EDA) techniques. Many EDA practitioners in both academia and industry have placed high hopes in new artificial intelligence (AI) or machine learning (ML) methods in IC design and EDA techniques, targeting more agile design for lower IC \emph{design costs}, less \emph{human efforts}, and shorter \emph{turnaround time}.

\textbf{AI for EDA/chip design.} In recent years, AI for chip design, also named AI/ML for EDA or AI-assisted EDA~\cite{huang2021machine, rapp2021mlcad}, has been viewed as a highly promising technique, owing to its ability to \emph{reuse knowledge} from prior circuit design data. Relevant AI-driven EDA techniques are also adopted in commercial EDA tools~\cite{dsoai, cerebrus}. Various ML models can be trained to provide early predictions or optimizations for circuits, bypassing time-consuming downstream design and simulation steps. 
Learning from prior design solutions, ML models can perform circuit quality evaluations at early design stages and thus guide early design optimizations. 
%
Existing AI for EDA techniques have been extensively explored for almost all standard VLSI design stages (e.g., architecture stage, high-level synthesis (HLS) code, register-transfer level (RTL) code, gate-level netlist, post-placement layout, clock tree, and post-routing layout) and all primary circuit design objectives (e.g., timing, power, area, congestion, IR drop, signal integrity, and functionality).

\begin{figure}[!t]
  \centering
  \includegraphics[width=0.9\linewidth]{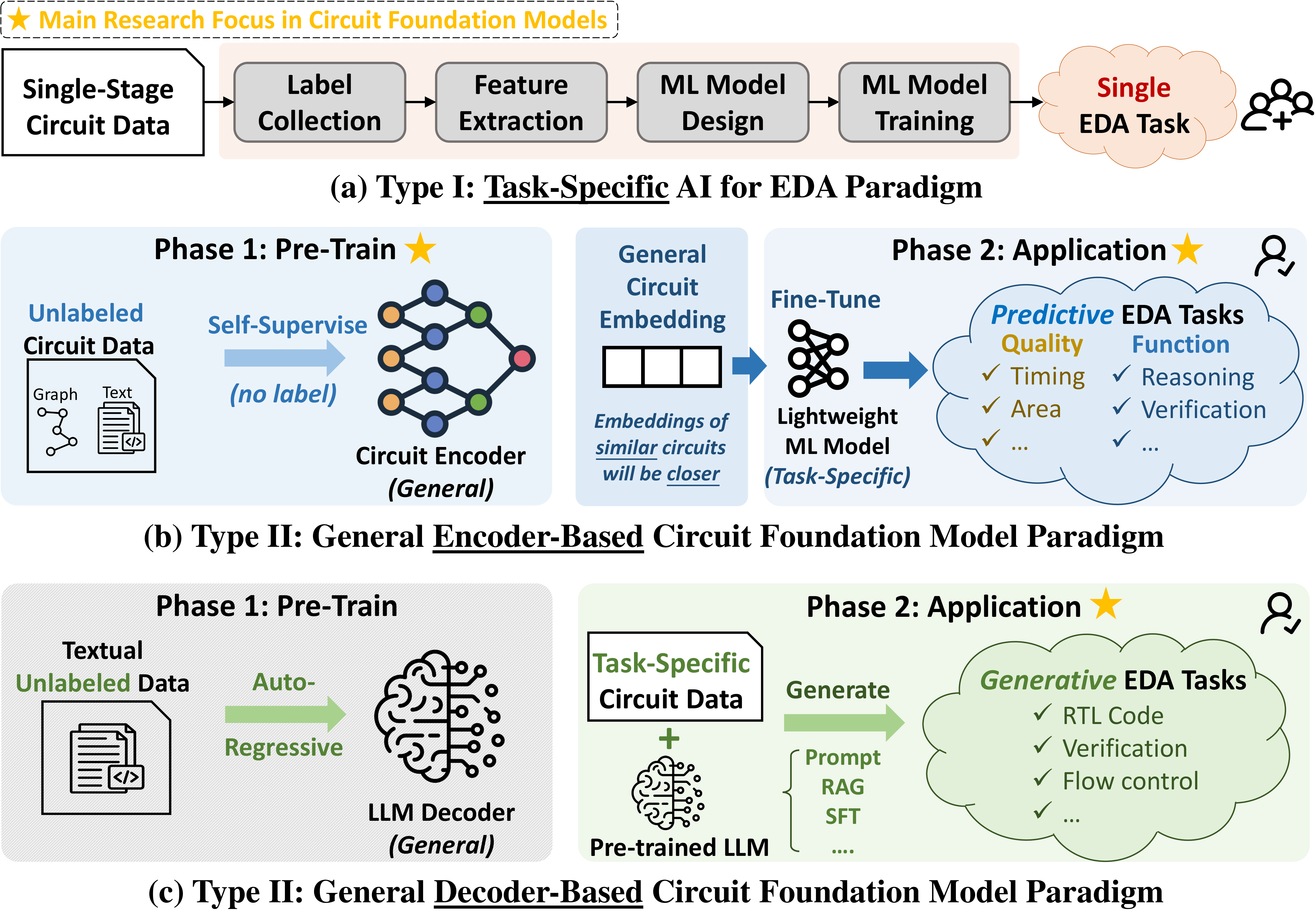}
    \vspace{-.07in}
  \caption{Different paradigms of AI for EDA techniques. (a) Type I: Supervised Predictive AI Techniques for EDA. This type of work has been extensively studied. \textbf{(b)\,\&\,(c) Type II: Foundation AI Techniques for EDA (i.e., Circuit Foundation Models)}. This type of work includes two paradigms, named encoder-based and decoder-based circuit models. Both paradigms develop the foundation AI model through two stages: self-supervised \emph{pre-training} and \emph{fine-tuning}. Our survey will focus on the emerging type II methods.} 
  \label{fig:paradigm}
      \vspace{-.2in}
\end{figure}


\begin{figure}[!t]
\captionsetup[subfigure]{labelformat=empty}
\vspace{-.15in}
	\centering
	\subfloat[(a) \emph{Encoder-based} circuit foundation model, covered in~\Cref{sec:encoder}.]{\includegraphics[width=0.82\linewidth]{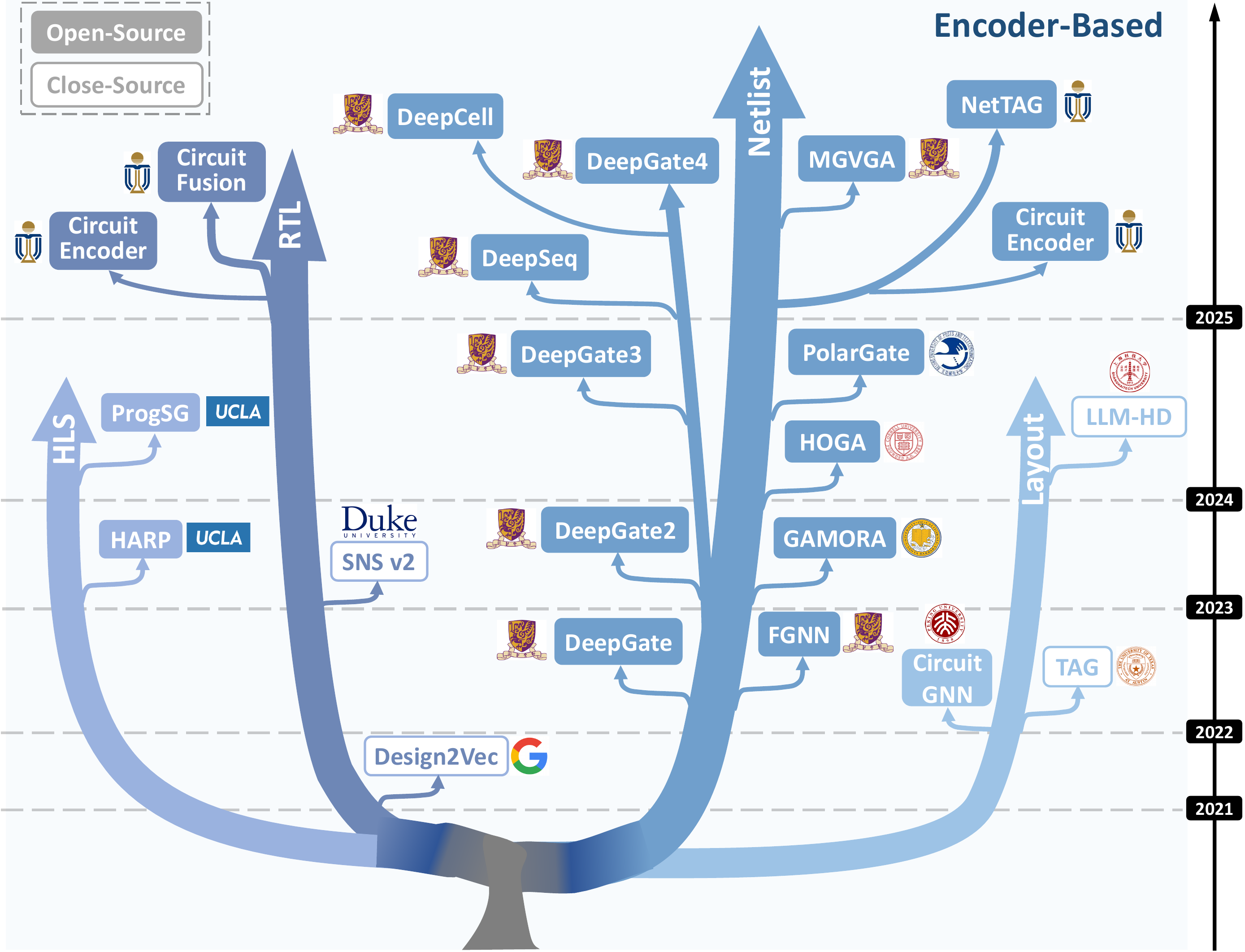}\label{fig:tree-enc}}\hspace{0pt}
	\subfloat[(b) \emph{Decoder-based} circuit foundation model, covered in~\Cref{sec:ai-decoder}.]{\includegraphics[width=0.85\linewidth]{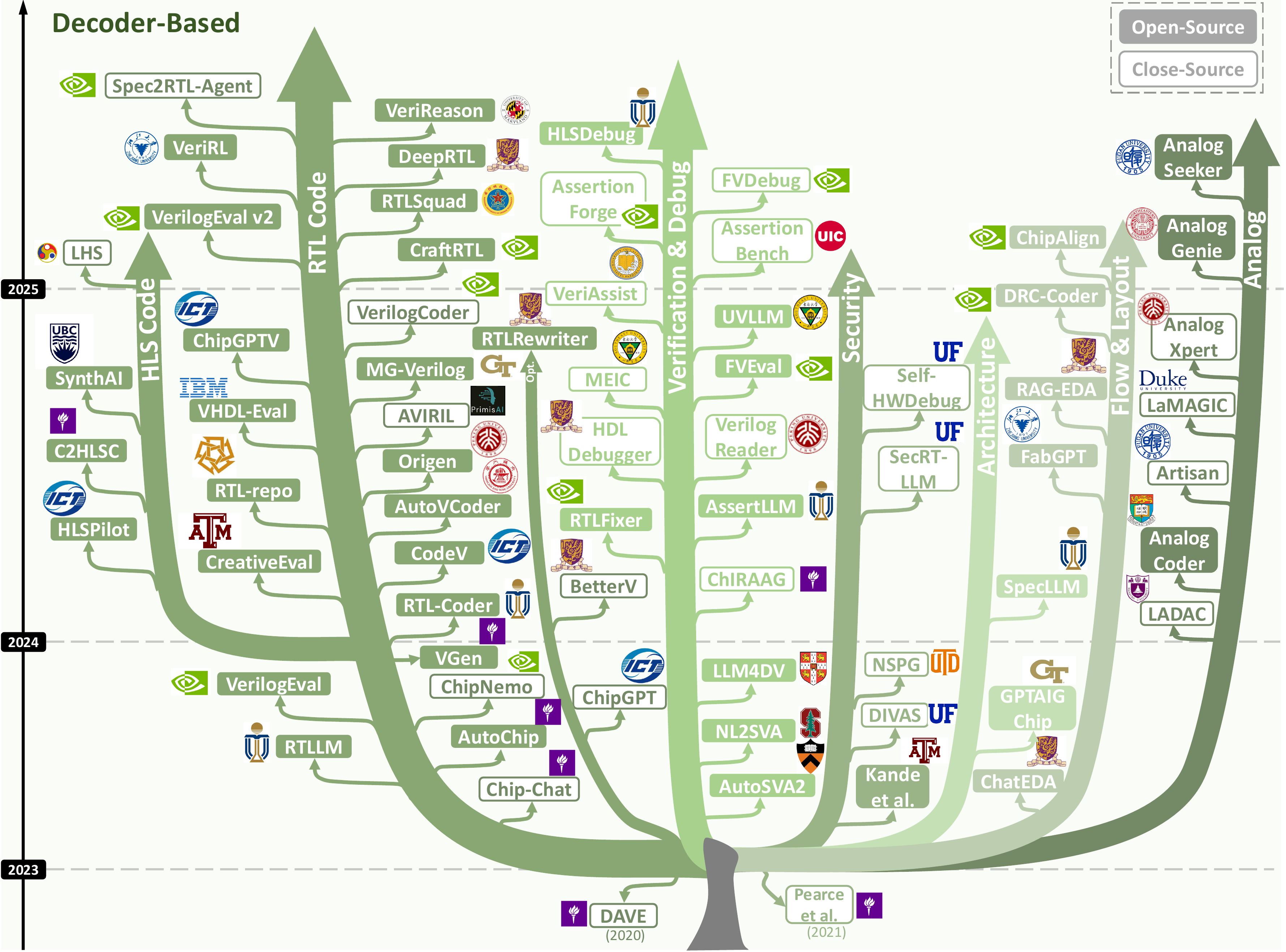}\label{fig:tree-dec.pdf}}\hspace{0pt}
    \vspace{-.1in}
    \caption{Evolutionary tree  of foundation AI models for VLSI design and EDA.}
    \label{fig:evo-tree}
         \vspace{-.25in}
\end{figure}

\begin{figure}[!t]
  \centering
  \includegraphics[width=0.99\linewidth]{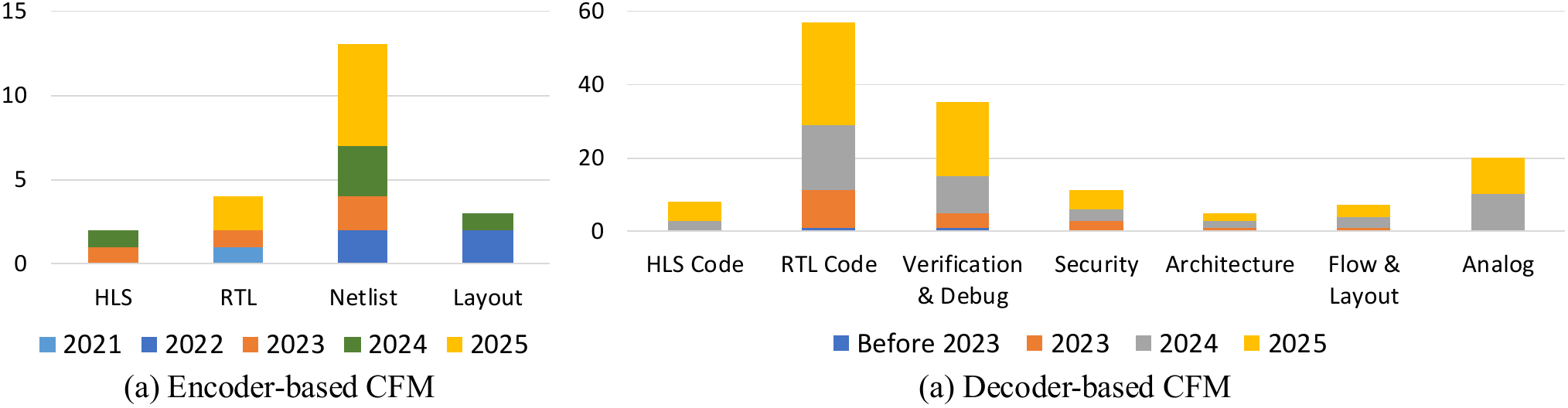}
    \vspace{-.13in}
  \caption{Quantitative coverage of surveyed circuit foundation models.
(a) Encoder-based CFMs: number of works by design stage and publication year.
(b) Decoder-based CFMs: number of works by primary application category and publication year.}
  \label{fig:hist}
      \vspace{-.2in}
\end{figure}

\textbf{Foundation AI for EDA/chip design: a new trend and our focus.} Recently, general foundation AI models in natural language processing (NLP) and computer vision (CV) (e.g., BERT~\cite{devlin2018bert}, CLIP~\cite{radford2021learning}, DALLE~\cite{ramesh2022hierarchical}, and ChatGPT~\cite{achiam2023gpt}) have emerged and represent a significant leap in AI techniques. These foundation models, characterized by their large model scale and application scopes, have demonstrated an incredible ability to understand, predict, and generate content~\cite{chen2024large}. In comparison with these foundation models in NLP and CV, previous AI applications in \emph{circuits} lag far behind well-explored general natural languages and images. This has motivated the latest trend of exploring foundation AI models for EDA techniques and circuit design applications.


The trending works on foundation AI for EDA have demonstrated unprecedented ability in model generalization, few-shot learning, and generation tasks. As shown in~\Cref{fig:paradigm}, these models typically leverage a two-stage paradigm of \emph{pre-training} on large-scale datasets followed by \emph{fine-tuning} for specific applications, significantly enhancing adaptability across various EDA tasks. Their great potential has attracted wide attention from the EDA community. Some representative works~\cite{lu2024rtllm, liu2023chipnemo, liu2023verilogeval, thakur2023benchmarking} are relatively highly cited since their publication, compared with average EDA publications. However, there is a lack of systematic definition, analysis, or survey on this series of latest works, leading to confusion when discussing many concepts in our communities (e.g., large circuit model vs. LLM-aided design vs. AI agents for EDA). 
In this survey paper, we will cover all representative works on foundation AI for EDA. We propose referring to this type of work as \textbf{circuit foundation models (CFMs)}. 
\Cref{fig:evo-tree} illustrates the evolutionary tree of existing circuit foundation models, including both \emph{encoder-based} and \emph{decoder-based} paradigms.
This paper also covers the potential and challenges of CFMs from our perspective.

\Cref{fig:hist} further visualizes the quantitative coverage of the surveyed CFMs, showing the number of works by publication year, design stage, and application category. As shown in \Cref{fig:hist}, decoder-based CFMs significantly outnumber encoder-based ones.
This imbalance largely reflects the higher barrier to developing encoder-based CFMs, which typically require training from scratch on large-scale circuit datasets with carefully designed pre-training objectives, whereas many decoder-based works can directly leverage existing pre-trained LLMs and focus on task adaptation.
Moreover, encoder-based CFMs must learn robust, reusable representations that generalize across stages and tasks, which remains technically challenging and is still under active exploration.

\textbf{Definition of circuit foundation model.}
We first define a \emph{foundation AI model} as a large-capacity model that (i) is \emph{pre-trained} on large-scale data, (ii) is designed to be \emph{adaptable} to a wide range of downstream tasks via finetuning, prompting, or other lightweight adaptation. The foundation AI model aims to provide \emph{general-purpose representations or generative capabilities} that can be reused across tasks and domains.
We then specialize this notion to \emph{circuit foundation models} by constraining both the data modalities and downstream tasks to the circuit/EDA domain: a CFM is a foundation model that is pre-trained on circuit-related data (or adapted from a general foundation model to circuit data) and intended to support multiple circuit/EDA tasks.
We use this operational definition throughout the survey to decide which works to include in our taxonomy.


\textbf{Structure of Section~\ref{sec:intro}.} In this Introduction, we will first propose our own taxonomy of existing AI for EDA techniques in Section~\ref{subsec:taxonomy}, categorizing all existing AI for EDA techniques into two major types. Then we will briefly introduce the already extensively studied Type I techniques (supervised AI for EDA) in Section~\ref{subsec:type1} and elaborate on the emerging Type II techniques (foundation AI for EDA, the focus of our survey) in Section~\ref{subsec:type2}. After that, in Section~\ref{subsec:survey}, we will summarize all existing surveys that cover similar topics and elaborate on the contributions of this survey. In \Cref{subsec:overview}, we will introduce the overall structure of this whole survey paper.





\subsection{Our Taxonomy of AI for EDA Techniques: Two Different Types} 
\label{subsec:taxonomy}

In this survey, we propose to categorize existing AI for EDA techniques into two main types, as listed below. Figure~\ref{fig:paradigm} summarizes and compares all three paradigms of these two types of works. 
\vspace{.06in}
\begin{itemize}
    \item \textbf{Type I: Supervised Predictive AI Techniques for EDA.} The mainstream paradigm of previous AI for EDA solutions adopts \emph{supervised predictive} AI models. These supervised predictive models have been developed for various applications, including early-stage design quality prediction, fast design quality simulation, design space exploration, etc. Relevant works have been extensively studied and covered in existing surveys~\cite{huang2021machine, rapp2021mlcad} and book~\cite{ren2022machine}. 
    \vspace{.06in}
    \item \textbf{Type II: Foundation AI Techniques for EDA (\emph{Circuit Foundation Model}).} This trending technique is the focus of this survey. The development of foundation AI solutions, according to our proposed definition, involves two phases: 1) Pre-training phase; 2) Fine-tuning phase. 
    The first \emph{pre-training} step, which is typically \emph{self-supervised} on a large amount of unlabeled data, enables the AI model to learn more general circuit intrinsic patterns. The subsequent \emph{fine-tuning} step can efficiently make the model adapt to specific EDA tasks. \\

    \vspace{-.08in}
    Figure ~\ref{fig:paradigm}(b)~\&~(c) summarize two different paradigms of foundation AI models for circuits. 
    We propose to incorporate both paradigms into the scope of \emph{circuit foundation models}:
    \vspace{.03in}
    \begin{itemize}
        \item \textbf{Encoder-based circuit foundation models.} One primary paradigm performs \emph{circuit representation learning} to support \emph{predictive tasks}. They typically encode a circuit design into a general embedding (i.e., a vector with rich circuit information). This embedding will be the input to lightweight downstream models for various EDA applications.\looseness=-1     
        \vspace{.03in}
        \item \textbf{Decoder-based circuit foundation models.} The other primary paradigm performs decoding tasks, thus supporting \emph{generative tasks}. 
        They typically adopt decoder-based large language models (LLMs) to help \emph{generate} circuits, including design HLS or RTL code, design functionality descriptions, verification assertions, EDA tool scripts, etc. 
        
    \end{itemize}
\end{itemize}


\subsection{Type I: Supervised Predictive AI Techniques for EDA (covered in prior surveys)}
\label{subsec:type1}

In recent years, supervised learning for predictive EDA tasks has been extensively studied, spanning almost all major design stages and key design tasks. Such models have proven highly effective in delivering early-stage estimates of design quality and in guiding subsequent optimization decisions.
Existing supervised learning methods are mostly tailored to specific tasks, such as early prediction of various design quality metrics (e.g, timing~\cite{fang2024annotating, wang2023restructure, guo2022timing, barboza2019machine, kahng2018using, xie2021timing}, area~\cite{fang2023masterrtl, fang2024transferable, xu2023fast, sengupta2022good}, power~\cite{zhang2025autopower, zhang2025firepower, du2024powpredict, zhang2023panda, xie2021apollo, zhou2019primal, kim2019simmani, xie2022deep, zhang2020grannite}, IR drop~\cite{xie2020powernet, ho2019incpird, xie2023ir, chhabria2021mavirec, fang2018machine}, routability~\cite{xie2018routenet, liu2021global, zheng2023lay, chen2020pros, chang2021auto, pan2022towards, huang2019routability},  crosstalk~\cite{liang2020routing, kuhlmann2001exact, kahng2015si}, and manufacturability~\cite{yang2017layout, geng2020hotspot, yang2017lithography}) or the reasoning of circuit functionalities~\cite{wu2023gamora, alrahis2021gnn, chowdhury2021reignn, he2021graph, ma2019high} for verification applications. Additionally, tasks for circuit optimization (e.g., flow tuning~\cite{xie2020fist, neto2022flowtune, liang2021flowtuner}, design space exploration~\cite{bai2021boom, liu2013learning, schafer2019high}, design quality optimization~\cite{lu2021rl, lu2023rl}) also largely rely on the prediction of circuit quality to provide feedback. 

As~\Cref{fig:paradigm} (a) shows, these methods are typically developed by supervised training, which requires extensive label collection, model customization, and model development for every single task. Despite obvious effectiveness, this mainstream supervised paradigm has several inter-related general limitations:  
\begin{enumerate}
    \item \textbf{Difficulty to get sufficient labeled data.} 
    It is typically difficult to accumulate sufficient labeled training data: 1) Many coarse-grained prediction tasks do not support many labels. For example, to predict the layout area of a netlist, each circuit layout only provides one label (i.e., its layout area). 2) The label generation process is inherently highly time-consuming. 
    A dilemma is, most predictive AI models are trained to bypass the the slowest design/simulation steps. However, these slowest steps are exactly required to collect labels. 
    \vspace{.03in}
    \item \textbf{Time-consuming AI model development process.} The development process of supervised task-specific solutions is tedious and time-consuming. The development steps include circuit collection, label generation, feature engineering, model architecture design, model training, and model testing. This whole process easily takes months of engineering efforts. 
    \vspace{.03in}
    \item \textbf{Lack of generalization across tasks.} Since supervised task-specific models cannot be directly generalized to other tasks, it leads to an inefficient repetitive development of ML solutions. Moreover, from the methodology perspective, it implies that these supervised ML solutions only \emph{learned} task-specific patterns, instead of \emph{understanding} more general knowledge of target circuit designs.
\end{enumerate}

Due to the page limit and the large number of extensively explored type I works, we will not exhaustively cover all prior type I works. For a more comprehensive list of type I supervised predictive works, we refer our readers to prior surveys~\cite{huang2021machine, rapp2021mlcad} and a book~\cite{ren2022machine} co-authored by many researchers in this domain. 



\subsection{Type II: Foundation AI Techniques for EDA (the focus of this paper)}  
\label{subsec:type2}
This survey focuses on the emerging paradigms of foundation AI techniques for EDA. As illustrated in~\Cref{fig:paradigm} (b) and (c), this type of technique leverages pre-trained foundation AI models for circuits (referred to as \emph{circuit foundation models}), which can be efficiently fine-tuned using a small amount of task-specific labeled circuit data. Compared to traditional task-specific supervised AI for EDA solutions, this type II techniques offer significant advantages:
\begin{enumerate}
    \item \textbf{Learning unlabeled circuit intrinsics.} Circuit foundation models are typically pre-trained on a large amount of unlabeled data, enabling them to capture the underlying intrinsic information about circuits, without requiring expensive labeled datasets.
    \vspace{.03in}
    \item \textbf{Efficient fine-tuning for solving EDA task.} Well-pre-trained models require only a small amount of labeled data for fine-tuning. It significantly reduces the time and resources needed to solve each specific EDA task compared to training models from scratch. 
    \vspace{.03in}
    \item \textbf{Generalization across various tasks.} 
    General circuit intrinsics learned by foundation models can be adapted to multiple tasks, making the models versatile and reducing the need for repetitive task-specific model development. 
    \vspace{.03in}
    \item \textbf{Unprecedented generative capability for EDA tasks.} Some circuit foundation models exhibit remarkable generative capabilities, unprecedentedly automating tasks such as circuit code generation, assertion generation for verification, and design flow script generation. These models go beyond existing predictive tasks, enabling innovative AI-driven solutions that enhance design productivity and streamline circuit development flow.
\end{enumerate}

\textbf{Encoder-based circuit foundation model.} 
\Cref{fig:paradigm} (b) demonstrates the paradigm of circuit encoders.
Circuit encoders transform various circuit modalities (e.g., graphs or text formats) into generalized embeddings that contain rich intrinsic circuit properties. These encoders are typically \emph{pre-trained on circuit data}. 
Due to the uniqueness of the circuit data compared with well-studied images or natural language, encoder models have to be specifically customized to handle circuit data. 
Research works primarily focus on two aspects: (1) in phase 1, developing specialized ML architectures and pre-training techniques to effectively capture circuit semantics, structural information, and physical attributes, and (2) in phase 2, leveraging pre-trained circuit encoders to support various predictive EDA tasks, including design quality evaluation and functional reasoning. 
In this survey, we systematically categorize existing circuit encoders according to their respective design stages and provide a comprehensive analysis of their supported downstream tasks.

\textbf{Decoder-based circuit foundation model.}
\Cref{fig:paradigm} (c) illustrates the paradigm of circuit decoders.
Circuit decoders typically leverage LLMs as their backbone, which are typically \emph{extensively pre-trained on vast text datasets} spanning multiple domains. Leveraging the powerful pre-trained LLMs, circuit decoders mainly focus on domain adaptation to circuit-related generative tasks, such as prompt engineering, fine-tuning, retrieval-augmented generation, etc. In this survey, we categorize existing decoder-based methods based on their application domains, covering key areas such as circuit code generation, verification, design flow automation, etc. For each category, we analyze representative benchmarks, model development techniques, and the latest advancements.



\textbf{Key differences} between encoder- and decoder-based models are summarized below:
\begin{enumerate}[topsep=1pt]
    \item \textbf{Circuit modality as input}: Encoders primarily process \emph{graph-based} circuit structures, such as netlists and control-data flow graphs, often leveraging graph learning models. Some recent works integrate multimodal learning, combining structural graphs with textual descriptions. In contrast, decoders focus on \emph{text-based} formats like HDL code and natural language specifications, utilizing LLMs for interpretation and generation.
    \item \textbf{Circuit learning techniques}: Encoders require customized pre-training and fine-tuning on circuit data. They are typically built from scratch using graph AI models. There is no standard architecture for circuit encoding, leading to diverse model designs and self-supervised learning techniques. In contrast, decoders typically rely on LLMs already extensively pre-trained on vast text datasets. Relevant works rely on existing pre-trained LLMs in the public domain, including both open-sourced (e.g., Llama, Mistral, DeepSeek) and commercial (e.g., GPT-3.5, GPT-4o) LLMs. These works focus on adaptation to the circuit domain through prompt engineering, fine-tuning, and retrieval augmented generation (RAG).
    \item \textbf{Target downstream tasks}: Encoders typically support predictive tasks such as design quality evaluation and functional reasoning, leveraging encoded circuit embeddings. Decoders are typically tailored for generative tasks, such as circuit code generation, verification automation, design flow generation, etc.\looseness=-1
\end{enumerate}

Despite these advancements, research on CFMs is still in an early stage: existing models cover only a subset of design stages and tasks, and many design choices (e.g., architectures, objectives, datasets, and evaluation protocols) are far from settled. We further discuss the current limitations and open challenges of circuit foundation models in~\Cref{sec:challenges}.




\subsection{Comparison of Existing Relevant Surveys and This Paper.}\label{subsec:survey}

\textbf{Surveys on AI for EDA.}
There are various surveys that summarize recent trends in AI for EDA, including early general surveys such as~\cite{huang2021machine, rapp2021mlcad, ren2022machine}.
Other surveys focus on specific aspects of AI for EDA, such as graph learning for EDA~\cite{lopera2021survey, ren2022graph, ma2020understanding, sanchez2023comprehensive}, reinforcement learning (RL) for EDA~\cite{zhu2024survey}, and AI techniques for logic synthesis~\cite{wu2022ai}.
Supervised learning methods (e.g., GNNs for logic synthesis) and reinforcement learning methods (e.g., RL for floorplanning) have already demonstrated strong performance on specific EDA tasks.
Compared with these surveys and techniques, our work focuses on \emph{circuit foundation models} for EDA and positions them as a \emph{new paradigm} within AI for EDA.
Building on the successes of task-specific supervised and RL-based approaches, circuit foundation models introduce a more general pretrain–finetune paradigm in which a single pre-trained circuit model can be adapted to a wide range of predictive and generative EDA tasks.
In this way, CFMs extend the current AI for EDA landscape by shifting from widely adopted task-specific solutions to general, reusable models that can leverage cross-task and cross-stage knowledge, while complementing existing supervised and graph-based methods.

\textbf{Surveys on foundation AI models for EDA.}
Table~\ref{tab:tab-survey} compares all existing survey papers~\cite{zhong2023llm4eda, pan2025survey, abdollahi2025hardware, saha2024llm, kande2024llms, wang2024llms, paria2024navigating, chen2024large} about foundation AI models for circuit applications. Notably, almost all surveys~\cite{zhong2023llm4eda, pan2025survey, abdollahi2025hardware, saha2024llm, kande2024llms, wang2024llms, paria2024navigating} focus only on decoder-based models (i.e., LLM for EDA). 
This trend reflects the rapid evolution of LLMs and their significant potential for generative EDA tasks, such as HDL code generation, verification, debugging, etc. Among these surveys on decoder-based LLMs for EDA, some surveys~\cite{zhong2023llm4eda, abdollahi2025hardware, pan2025survey} try to provide comprehensive reviews on multiple relevant tasks, while some others~\cite{saha2024llm, paria2024navigating, kande2024llms, wang2024llms} focus on one specific topic, mostly about \emph{circuit security}. 
The only exception is a special perspective paper~\cite{chen2024large} co-authored by many EDA researchers. It advocates for an ambitious framework of multiple \emph{encoder}-based foundation models aligned across design stages. This envisioned concept is named large circuit model (LCM)~\cite{chen2024large}. Different from existing surveys, our survey paper incorporates both encoder-based and decoder-based circuit foundation models, analyzing their similarities and differences.\looseness=-1

Table~\ref{tab:tab-survey} also reports the number of CFM-related works covered in each survey paper. We only count works in the scope of the circuit foundation model (i.e., pre-train and fine-tuned AI models targeting circuit design tasks). Partially due to the fast development in this emerging direction, most existing surveys only covered less than 40 related works. 
In comparison, our comprehensive survey unprecedentedly introduces the largest number of (i.e., over 140) relevant works, covering all key circuit design tasks listed in Table~\ref{tab:tab-survey}. We briefly introduce each existing survey and highlight the unique contributions of our study below.


\begin{table}[!t]
    \centering
    \resizebox{1.0 \textwidth}{!}
    {\begin{tabular}{c||c|c|c|c|c|c||c|c||c||c}
    \toprule
        \multirow{2}{*}{Surveys} & Design & Design & Design & Design & Design & Design & Encoder & Decoder & Time & No. of CFM \\
         & Generation & Verification  & Debugging & Security & Optim. & Flow & -based & -based & Published &  Works Covered \\
    \hline
    \hline
        \cite{saha2024llm} & & & & \checkmark & & & & \checkmark & 2023-10 & 15  \\
        \hline
        \cite{zhong2023llm4eda} & \checkmark & \checkmark & \checkmark & & \checkmark & & & \checkmark & 2023-12 & 22 \\
        \hline
        \multirow{2}{*}{\cite{chen2024large}} & \multirow{2}{*}{\checkmark} & & & & & \multirow{2}{*}{\checkmark} & \multirow{2}{*}{\checkmark} & \multirow{2}{*}{\checkmark} & \multirow{2}{*}{2024-03} & 6 (Encoder) \\
      & & & & & & & & & & + 21 (Decoder) \\
        \hline
        \cite{kande2024llms} & \checkmark & & \checkmark & \checkmark &  &   &  & \checkmark & 2024-04 &  14 \\
        \hline
        \cite{wang2024llms} & \checkmark & & & \checkmark & \checkmark & \checkmark  & & \checkmark & 2024-05 &  32 \\
        \hline
        \cite{paria2024navigating} & \checkmark & \checkmark & \checkmark & \checkmark & & & & \checkmark & 2024-06  & 24  \\
        \hline
        \cite{xu2024llm} & \checkmark & \checkmark & \checkmark &  &  & \checkmark & & \checkmark & 2024-10  & 29 \\
        \hline
         \cite{abdollahi2025hardware} & \checkmark & \checkmark & \checkmark & &&  & & \checkmark & 2024-12 & 71 \\
         \hline
        \cite{pan2025survey} & \checkmark & & & & & \checkmark & & \checkmark & 2025-01 & 39 \\
        \hline
        \hline
        \multirow{2}{*}{\textbf{Ours}} & \multirow{2}{*}{\checkmark} & \multirow{2}{*}{\checkmark} & \multirow{2}{*}{\checkmark} & \multirow{2}{*}{\checkmark} & \multirow{2}{*}{\checkmark} & \multirow{2}{*}{\checkmark} & \multirow{2}{*}{\checkmark} &  \multirow{2}{*}{\checkmark} &  \multirow{2}{*}{2025-10 } & 21 (Encoder) \\
        & & & & & & &  & & & + 140 (Decoder)\\
        \bottomrule
    \end{tabular}}
    \vspace{.02in}
    \caption{Comparison of \emph{existing surveys} on foundation models for chip design, covered in Section~\ref{subsec:survey}.} 
    \label{tab:tab-survey}
    \vspace{-.35in}
\end{table}

\textbf{Surveys on decoder models covering broad tasks.} \emph{LLM4EDA}~\cite{zhong2023llm4eda} is an early \emph{comprehensive review}, covering various EDA tasks such as chatbot-based methods, circuit code and script generation, and circuit verification. However, since it was published in 2023 and this direction and developed very fast, it only covered 22 works. 
\emph{Xu et al.}~\cite{xu2024llm} summarize 29 early-stage studies on circuit code generation, debugging, verification, and physical implementation. While it provides insights into these areas, it lacks coverage of the latest developments and broader topics such as security, design optimization, and architecture design.
\emph{Abdollahi et al.}~\cite{abdollahi2025hardware} provide a more \emph{extensive} survey, analyzing 71 studies on LLM-assisted circuit design, including applications in circuit generation, verification, and debugging. However, possibly due to their automated literature screening process, we observed several incorrect descriptions in this survey. For example, the survey~\cite{abdollahi2025hardware} incorrectly categorizes works of~\cite{mudigere2022software, lai2024lcm, heo2024neupims} as LLM-aided design methodologies, while these works actually primarily focus on the acceleration of LLMs (i.e., designing hardware accelerators). 
A recent survey by \emph{Pan et al.}~\cite{pan2025survey} reviews LLM applications in EDA. Despite its recency, it still only covers a limited number of works (39 works), primarily focusing on design generation and design flow automation. It lacks a broader discussion on design verification, security, architecture design, and analog tasks.


\textbf{Surveys on decoder models covering a specific task.} In addition to surveys targeting broad EDA applications, the other series of surveys~\cite{saha2024llm, paria2024navigating, kande2024llms, wang2024llms} focus specifically on \emph{circuit security topics} with LLM-assisted techniques.
\emph{Saha et al.}~\cite{saha2024llm} pioneered the discussion of integrating LLMs into the SoC security verification paradigm in 2023. At the time, not many specialized LLM-based solutions had been customized for SoC security. Therefore this work~\cite{saha2024llm} primarily summarizes applying \emph{general LLM techniques} in hardware security tasks. 
In 2024, three short surveys (all less than 7 pages) ~\cite{kande2024llms, wang2024llms, paria2024navigating} cover LLM methods for hardware security, each covering 10 to 30 works.

\textbf{A perspective paper on LCM.} In 2024, a special perspective paper~\cite{chen2024large} co-authored by many EDA researchers proposes and advocates an interesting and ambitious concept of \emph{large circuit model (LCM)}. This LCM can be viewed as an envisioned framework of multiple aligned encoder-based circuit foundation models, each devoted to one design stage. 
This paper also reviews both supervised task-specific AI solutions and foundation AI models for EDA (i.e., 6 encoders and 21 decoders). It identifies key challenges in developing large-scale circuit encoders and sets the stage for future advancements in encoder-based circuit foundation models.

\textbf{The contributions of this survey}, compared with prior surveys, can be summarized below:   
\begin{enumerate}[topsep=0pt]
    \item This survey proposes the concept of \emph{circuit foundation model}. It incorporates both encoder-based circuit representation learning techniques and decoder-based LLM for EDA methods into a unified framework, enabling comparison between these two paradigms. 
    \item This comprehensive survey systematically introduces over 160 works. All existing circuit foundation models cited in prior surveys~\cite{zhong2023llm4eda, pan2025survey, abdollahi2025hardware, saha2024llm, kande2024llms, wang2024llms, paria2024navigating, chen2024large} have been covered in this work. 
    \item The 21 encoder-based models span all standard design stages, including HLS, RTL, netlist, and layout stages, along with their supporting predictive EDA tasks.
    \item The 111 decoders-based models cover all mainstream EDA applications, including VLSI circuit code processing (generation, optimization, verification, and debugging), hardware security, design flow automation, physical design, architecture design, and analog design.  
    \item  Besides the in-depth analysis of these approaches, we highlight key advancements, challenges, and future research directions to further enhance circuit foundation models and their impact on modern VLSI design automation.
\end{enumerate}

\begin{figure}[!t]
  \centering
  \includegraphics[width=0.99\linewidth]{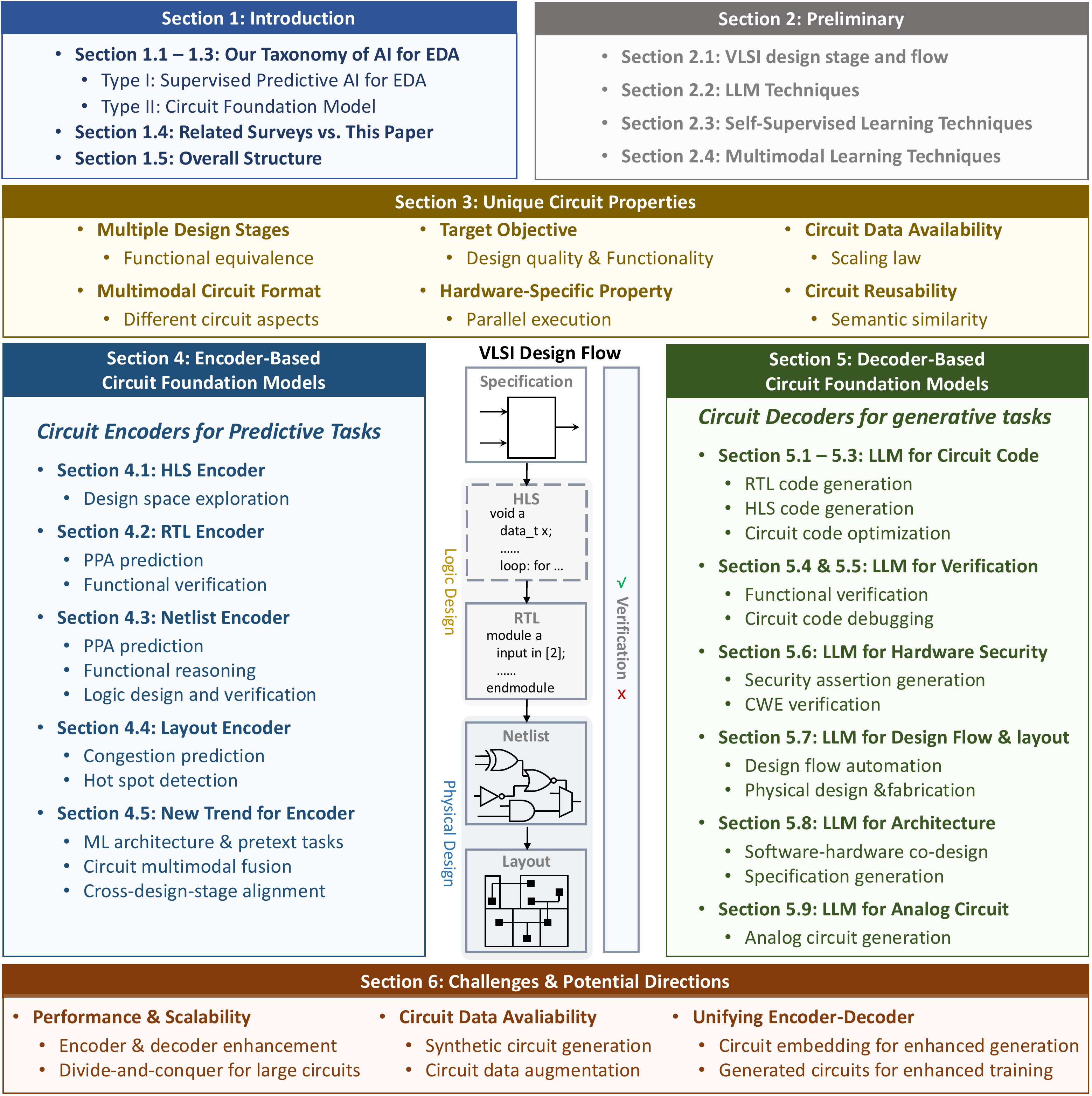}
    \vspace{-.13in}
  \caption{Overview of this survey paper. \Cref{sec:preliminary} provides the background of VLSI circuit design and foundation AI model techniques. \Cref{sec:unique-circuit} discusses the unique properties of circuit data that motivate AI-driven solutions. \Cref{sec:encoder} and \Cref{sec:ai-decoder} comprehensively review existing circuit encoders and decoders, respectively. Finally, \Cref{sec:challenges} explores key challenges and future directions in circuit foundation models.} 
  \label{fig:overview}
\end{figure}

This survey tries to cover all publications within the scope of the circuit foundation model, including journals, transactions, conference and workshop proceedings, thesis, and pre-prints. However, very short articles (e.g., late-breaking results, experiment reports) that are equal to or less than 3 pages may not be covered. For the same work with multiple versions and possibly different titles, we will avoid duplicated citations and tend to \emph{cite the latest version}. When counting the publication date, we use the \emph{date when the \textbf{earliest} version} gets released to the public.


\subsection{Overall Structure of This Survey Paper}
\label{subsec:overview}

Figure~\ref{fig:overview} provides the overall structure of this paper. 
\begin{itemize}
    \item In Section~\ref{sec:preliminary}, we will summarize related \textbf{preliminary knowledge}, covering both standard VLSI circuit design flow (Section~\ref{sec:prelim-vlsi}) and basic techniques of general foundation AI models, including LLM techniques (\Cref{sec:prelim-LLM}), self-supervised learning techniques  (\Cref{sec:prelim-self}), multimodal learning techniques  (\Cref{sec:prelim-multimodal}). 
    \item In Section~\ref{sec:unique-circuit}, we will introduce all our observed \textbf{unique properties of circuit data}. These properties have largely motivated many CFM works in this survey, and differentiate these works from general AI solutions in other domains (e.g., CV, NLP). 
    \item In Section~\ref{sec:encoder}, we will cover all existing \textbf{encoder-based circuit foundation models}, covering the HLS stage (Section~\ref{sec:enc-hls}), RTL stage (Section~\ref{sec:enc-rtl}), netlist stage (Section~\ref{sec:enc-netlist}) and layout stage (Section~\ref{sec:enc-layout}). The emerging and more advanced circuit encoder techniques will be covered in~\Cref{sec:encoder-advanced}. 
    \item In Section \ref{sec:ai-decoder}, we will cover all existing \textbf{decoder-based circuit foundation models}, covering all application domains: RTL code generation (Section~\ref{sec:llm-rtl}), HLS code generation (Section~\ref{sec:llm-hls-gen}), design optimization (Section~\ref{sec:llm-opt}), hardware code verification (Section~\ref{sec:llm-verification}), hardware code debugging (Section~\ref{sec:llm-debug}), hardware design security (Section~\ref{sec:llm-hardware-security}), design flow automation and layout design (Section~\ref{sec:llm-flow}), hardware architecture design (Section~\ref{sec:llm-hardware-architecture}), and analog circuit design (Section~\ref{sec:llm-analog}). 
    \item  In Section~\ref{sec:challenges}, we will analyze the \textbf{challenges and opportunities} of the circuit foundation models, based on our own research experience. 
\end{itemize}



\section{Preliminary}
\label{sec:preliminary}


Before covering specific CFM works, in this Section, we first summarize \textbf{preliminary knowledge} related to circuit foundation model, covering both standard VLSI circuit design flow in Section \ref{sec:prelim-vlsi} and the basic techniques of foundation AI models, including LLM techniques in \Cref{sec:prelim-LLM}, self-supervised learning techniques in \Cref{sec:prelim-self}, multimodal learning techniques in \Cref{sec:prelim-multimodal}. 
In addition, we provide a glossary table in \Cref{tab:tab-glossary} that collects frequently used terms and acronyms along with their definitions. It includes both EDA-related terms and AI-related terms, helping readers quickly look up notations used throughout the survey.

\begin{table}[!t]
    \centering

    \resizebox{1.0 \textwidth}{!}{

    \begin{tabular}{c|c|c||c|c|c} \toprule 
\multirow{19}{*}{\textbf{EDA}} & CFM   & circuit foundation model           & \multirow{17}{*}{\textbf{AI}} & NLP  & natural language processing                \\
                               & LCM   & large circuit model                &                               & CV   & computer vision                            \\
                               & RTL   & register-transfer level            &                               & LLM  & large language model                       \\
                               & HLS   & high-level synthesis               &                               & SFT  & supervised fine-tuning                     \\
                               & ECO   & Engineering Change Order           &                               & RAG  & retrieval augmented generation             \\
                               & HDL   & hardware description language      &                               & RL   & reinforcement learning                     \\
                               & CWE   & Common Weakness Enumeration        &                               & RLHF & reinforcement learning from human feedback \\
                               & CDFG  & control-data flow graph            &                               & GNN  & graph neural network                       \\
                               & PPA   & performance power and area         &                               & RNN  & recurrent neural network                   \\
                               & IP    & intellectual property              &                               & LSTM & long short-term memory network             \\
                               & AST   & abstract syntax tree               &                               & MAE  & mean absolute error                        \\
                               & AIG   & and-inverter graph                 &                               & MAPE & mean absolute percentage error             \\
                               & GDSII & graphic design system              &                               & UFT  & unsupervised fine-tuning                   \\
                               & DUT   & design under test                  &                               & SSL  & self-supervised learning                   \\
                               & SVA   & SystemVerilog assertion            &                               & OCR  & optical character   recognition            \\
                               & UVM   & universal verification methodology &                               & CoT  & chain-of-thought                           \\
                               & DRC   & design rule check                  &                               & ToT  & tree-of-thought                            \\
                               & SCA   & side-channel attack                &                               &      &                                            \\
                               & OPC   & optical   proximity correction     &                               &      &         \\ \bottomrule                                  
\end{tabular}
    
    }
    \caption{Glossary of acronyms frequently used in this survey, grouped by EDA terms and AI terms.}
    \label{tab:tab-glossary}
    \vspace{-.35in}
\end{table}

\subsection{Standard VLSI Design Stage and Flow}
\label{sec:prelim-vlsi}

A standard VLSI circuit design flow comprises several stages: specification definition, RTL design, logic synthesis, and physical design, as shown in the center of~\Cref{fig:overview}. At each stage, the design is represented in the corresponding format: specification, RTL code, netlist, and layout. In addition to these standard stages, high-level synthesis (HLS) is sometimes employed for more agile design or FPGA prototyping, based on HLS code in C/C++/SystemC. Verification and design quality analysis are carried out at various stages to ensure functional correctness and meet design quality constraints, respectively. Together, these stages transform the initial design specifications into a manufacturable and verified digital circuit layout. We introduce each design stage below.

\textbf{Specification definition.} The design process begins with a clear natural language specification that outlines the expected functionality, as well as performance, power, and area (PPA) requirements for a target digital circuit. This specification serves as the blueprint for subsequent design steps.

\textbf{HLS code design.} The specification can be translated into an abstract design using high-level programming languages or description languages like C/C++ or SystemC. Designers develop \emph{algorithms} that meet the functional requirements. The algorithms are described at a high level, focusing on functionality rather than hardware specifics.

\textbf{RTL design.} RTL design is the process of translating the high-level specification into a more detailed and implementable representation using hardware description languages (HDLs) such as Verilog or VHDL. These HDLs describe the behavior of digital circuits at the register-transfer level. 
The HDL code captures how data moves between registers (i.e., sequential registers) and how logic gates operate on that data within each clock cycle (i.e., combinational logic). 
%
Viewing each design as a finite-state machine, RTL defines the state transitions across clock cycles, ensuring that the circuit responds correctly to changes in input signals and synchronizes with the clock.


\textbf{Logic synthesis.}
Logic synthesis converts high-level RTL designs into low-level, optimized gate-level netlists. This process consists of three key steps: translation (i.e., elaboration), optimization, and technology mapping.
First, the synthesis process begins by translating the RTL code into an intermediate representation, such as the AND-Inverter Graph (AIG) in synthesis tools like ABC~\cite{brayton2010abc}. The synthesis tool then optimizes the logic based on constraints like delay and logic depth. Finally, technology mapping is performed, where the optimized logic is mapped to specific gates from a technology library provided by semiconductor foundries. This library contains various gate types, each with unique characteristics. The final output is a gate-level netlist, which represents the circuit in terms of logic gates and their interconnections.

\textbf{Physical design.}
Physical design translates the gate-level netlist into a manufacturable physical layout. This process includes several key steps: floor planning, placement, clock tree synthesis (CTS), and routing.
The first step, floor planning, involves arranging the major functional blocks of the chip in a way that optimizes performance while minimizing area. Designers determine the approximate locations of various components. Following floor planning, placement positions individual gates and components within the predefined floorplan, aiming to minimize wire length and ensure efficient placement.
CTS follows placement, where the clock distribution network is designed to ensure the proper synchronization of all clock signals across the chip.
Finally, routing connects the placed components using metal layers to form the required electrical connections. During routing, considerations such as signal integrity, minimization of crosstalk, and adherence to design rules are essential to ensure the layout is functionally correct and manufacturable.

\textbf{Verification.} 
Verification ensures the design meets specifications~\cite{hachtel2005logic} and includes functional and physical verification.
Functional verification checks if the design meets its specifications, using testbench simulations to model real-world conditions. Formal verification applies mathematical techniques, with equivalence checking to ensure consistency between design representations (e.g., RTL and gate-level).
Physical verification ensures the layout complies with manufacturing constraints using design rule checking (DRC) and layout versus schematic (LVS) to detect and correct violations for manufacturability.


\textbf{Analysis.} Analysis evaluates the design against performance metrics to ensure it meets specifications.
Static timing analysis (STA) verifies that timing constraints are met, ensuring signals propagate within required time limits. Power analysis estimates both dynamic (switching activity) and static (leakage currents) power consumption. Signal integrity analysis checks for issues like crosstalk, noise, and electromagnetic interference. Additionally, thermal analysis assesses heat generation and dissipation to ensure proper thermal management and reliable operation.

\subsection{LLM Techniques in AI Foundation Models} 
\label{sec:prelim-LLM}

The evolution of LLMs marks a pivotal advancement in artificial intelligence, particularly in NLP. Before delving into their applications in circuit design, it's essential to understand their techniques. 
Below, we introduce the brief evolution history of LLM and the key techniques employed in the morden advanced LLM models. This foundational understanding highlights the transformative potential of LLMs across various domains, including circuit design.

\textbf{A brief history of LLM.}
LLMs have evolved from early rule-based approaches to modern deep learning-driven foundation models, significantly advancing natural language processing (NLP). These advancements have enabled models like BERT and GPT to capture complex semantic and contextual nuances, leading to breakthroughs in various language-related tasks. Below, we summarize the key evolutionary stages of LLM development.
\begin{enumerate}
    \item \textbf{Rule-based method:} The earliest NLP systems relied on manually crafted linguistic rules and statistical models~\cite{NLP-book}. These approaches defined explicit syntactic and semantic rules for processing text but were limited in scalability and adaptability. While rule-based methods could handle predefined patterns effectively, they struggled with the complexity and variability of natural language, making them unsuitable for large-scale applications.
    \item \textbf{ML solution by manual feature engineering: } The introduction of statistical machine learning improved NLP by enabling data-driven language modeling. Early machine learning solutions required extensive manual feature engineering, where domain experts designed handcrafted features such as n-grams, part-of-speech tags, and dependency structures. Traditional models, including Hidden Markov Models and Support Vector Machines, demonstrated better adaptability than rule-based methods but still relied on human-designed representations, limiting their generalization capabilities.
    \item \textbf{Task-specific deep learning:} The emergence of deep learning revolutionized NLP by replacing manual feature engineering with automatic representation learning. Models like Word2Vec~\cite{mikolov2013efficientWord2Vec} and GloVe~\cite{pennington-etal-2014-glove} introduced word embeddings, representing words in continuous vector spaces to capture semantic relationships. Recurrent Neural Networks (RNNs)~\cite{Mikolov2010RecurrentNN} and Long Short-Term Memory Networks (LSTMs)~\cite{sak2014long} further improved sequence modeling by capturing contextual dependencies in text. However, these models faced challenges with long-range dependencies and computational efficiency due to their sequential nature.
    \item \textbf{General transformer-based foundation model:} The introduction of the Transformer architecture marked a paradigm shift in NLP. Transformers utilize self-attention mechanisms to process entire sequences in parallel, capturing global dependencies efficiently. Encoder-based models like BERT excel in understanding context through bidirectional masked language modeling, while decoder-based models like GPT specialize in generative tasks using autoregressive token prediction. These foundation models are pre-trained on massive datasets and fine-tuned for various downstream applications, eliminating the need for task-specific model development. Their success has extended beyond NLP, inspiring new research directions in domains such as circuit design, where they are increasingly being used for tasks such as RTL code generation, verification, and design optimization.
\end{enumerate}

\textbf{Key techniques in decoder-only LLMs.} 
Modern decoder-only LLMs, such as GPT, leverage a range of advanced techniques to enhance their performance and adaptability across various tasks. Below, we summarize five key techniques used in state-of-the-art decoder-based language models. \textbf{(1) Auto-regressive generation:} Decoder-only LLMs follow an auto-regressive approach, where they generate text sequentially, predicting one token at a time based on previously generated tokens. This autoregressive process allows models to produce coherent and contextually relevant text, making them highly effective for generative tasks such as text completion, summarization, and code generation. \textbf{(2) Prompt engineering:} Prompt engineering involves carefully crafting input text (prompts) to guide LLMs toward producing desired outputs. Since decoder-based models lack inherent task-specific fine-tuning for every possible use case, effective prompting helps steer model behavior without requiring additional training. Techniques such as zero-shot prompting (providing a task description), few-shot prompting (including examples), and chain-of-thought prompting (explicit reasoning steps) have been widely explored to enhance model performance across different applications. \textbf{(3) Supervised fine-tuning (SFT):} SFT refines pre-trained LLMs on specific datasets with labeled examples, enabling better adaptation to specialized tasks. By providing high-quality training examples, SFT improves accuracy and reliability in domain-specific applications, such as HDL code generation, circuit verification, and design optimization. Many domain-adapted LLMs, including those for EDA tasks, leverage SFT to improve performance on structured data and technical domains. \textbf{(4) Retrieval-augmented generation (RAG):} RAG enhances LLMs by incorporating external knowledge sources during inference. Instead of relying solely on pre-trained knowledge, the model retrieves relevant documents or contextual information from databases, augmenting its response with up-to-date and factual content. This technique is particularly useful for knowledge-intensive applications, such as circuit debugging and design flow optimization, where dynamic information retrieval improves response accuracy and relevance. \textbf{(5) Reinforcement learning from human feedback (RLHF):} RLHF refines LLM behavior using human preferences to optimize response quality. In this approach, human annotators rank model outputs, and reinforcement learning algorithms adjust the model’s reward function to align responses with human expectations. RLHF has been instrumental in making LLMs more aligned with human intent, improving coherence, factual correctness, and ethical considerations in generated outputs.

\subsection{Self-Supervised Learning Techniques in AI Foundation Model}
\label{sec:prelim-self}

In the development of AI foundation models, a variety of machine learning techniques are employed to enable these models to generalize across a wide range of tasks. These techniques are categorized into two primary phases: self-supervised learning for pre-training and supervised fine-tuning for downstream tasks.

Self-supervised learning is a powerful technique that allows models to learn from unlabeled data by generating labels from the data itself, often using auxiliary tasks. This phase helps the model understand general representations that can later be fine-tuned for specific tasks. We demonstrate the representative self-supervised learning techniques below.
\begin{itemize}
    \item \textbf{Contrastive learning}. This method learns representations by comparing similar (positive) and dissimilar (negative) pairs. For example, in image processing, positive pairs may be different augmentations of the same image, while negative pairs come from different classes. This method helps models generate useful embeddings for downstream tasks like retrieval or classification. It has been successful in computer vision and natural language processing (e.g., SimCLR~\cite{chen2020simple}, CLIP~\cite{radford2021learning}, MoCo~\cite{he2020momentum}).
    \item \textbf{Mask-reconstruction.} This method involves randomly masking parts of the input and training the model to predict the missing information. This forces the model to learn context from the surrounding data, improving its ability to understand structure and relationships. In NLP, BERT~\cite{devlin2018bert} predicts masked words in sentences, while in vision tasks, Masked Autoencoders (MAE)~\cite{he2022masked} show how masking portions of an image can lead to effective representation learning, aiding tasks like classification or segmentation.
    \item \textbf{Auto-regressive.} The auto-regressive method involves predicting the next element in a sequence, given the previous elements. In natural language processing, models like GPT~\cite{chatgpt} generate coherent text by predicting the next word based on the preceding context. In vision tasks, pixel-based auto-regressive models predict pixel values given prior pixels, such as in PixelCNN~\cite{van2016conditional}. The strength of auto-regressive models lies in their ability to learn complex dependencies within sequential or spatial data, allowing them to generate high-quality outputs for tasks like text generation, image synthesis, and beyond.
\end{itemize}

After pre-training with self-supervised methods, foundation AI models are fine-tuned on labeled data to adapt to specific tasks. This fine-tuning process enhances their performance across various applications, such as in the domain of NLP and CV.

\subsection{Multimodal Learning Techniques in AI Foundation Model}
\label{sec:prelim-multimodal}

Multimodal learning techniques are essential for AI foundation models, as they enable the integration and processing of multiple data modalities such as text, images, and video. We summarize the key multimodal learning techniques into two categories: multimodal encoders for representation learning and multimodal decoders for generation.\looseness=-1

\textbf{Multimodal encoders for representation learning} focus on learning joint representations across multiple modalities, allowing the model to extract and relate information effectively. Notable examples include CLIP~\cite{radford2021learning}, which aligns visual and textual representations to enable zero-shot learning for tasks like image classification and retrieval. ALBEF~\cite{li2021align} builds upon CLIP by aligning text and image representations and then performing multimodal fusion, improving performance in multimodal reasoning tasks such as visual question answering. These techniques lay the foundation for multimodal circuit representation learning, where textual descriptions (e.g., HDL code), structural graphs (e.g., netlists), and layout images can be effectively integrated for comprehensive circuit analysis and optimization.

\textbf{Multimodal decoders for generation} utilize one modality as input to generate content in another modality, such as describing images with text or synthesizing images and videos from textual descriptions. The BLIP family~\cite{li2022blip, li2023blip} bridges image understanding and text generation by introducing a connector that adapts image embeddings for frozen LLMs, enabling accurate textual descriptions of images. LLaVA~\cite{liu2024visual} enhances image understanding by fine-tuning LLMs with visual-text instruction pairs, improving the model’s ability to process and describe images. Extending this approach to video, Video-LLaVA~\cite{lin2023video} generates video content based on textual or image-based inputs.
Beyond generating text from visual inputs, some models focus on the reverse task—creating visual content from textual descriptions. DALL·E~\cite{ramesh2021zero} pioneers this field by generating diverse and high-quality images from textual prompts, facilitating creative content synthesis. Parti~\cite{yu2022scaling} further refines this capability, enabling the generation of high-resolution, contextually accurate images from detailed prompts. 
These advancements in multimodal generation highlight the potential of circuit foundation models, where similar approaches could be employed to detect and summarize layout defects~\cite{jiang2024fabgpt}, analyze EDA documents for question answering~\cite{pu2025mm}, or facilitate design debugging by linking textual analysis with waveform visualization~\cite{fang2024assertllm}.

\vspace{.25in}

\section{Unique Circuit Data Properties}
\label{sec:unique-circuit}

In this section, we will summarize the unique properties of circuits,  especially from the data perspective. We will compare the circuit data with other common data formats, such as general images or natural languages. Understanding these unique properties of circuits is important, since they largely motivate many circuit foundation models and thus differentiate these CFM from general AI solutions in other domains like CV or NLP. 




\textbf{Equivalence across design stages.} In the standard digital IC design flow, which includes specification, HLS code, RTL, netlist, and layout, ensuring equivalence across these stages is crucial for maintaining the integrity of the design. Each design stage refines the design from an abstract specification into a more detailed representation, but the underlying functionality and performance must remain consistent. This concept of equivalence has led to the use of circuit equivalent transformations as a data augmentation technique, allowing for the generation of multiple, functionally equivalent representations of a circuit. Furthermore, it has inspired cross-design-stage alignment in circuit foundation models, enabling these models to capture and align information across different stages of the design flow. This alignment enhances the model’s ability to transfer knowledge between stages and improves cross-stage consistency. 

\begin{figure}[!t]
    \centering
    \includegraphics[width=0.8\linewidth]{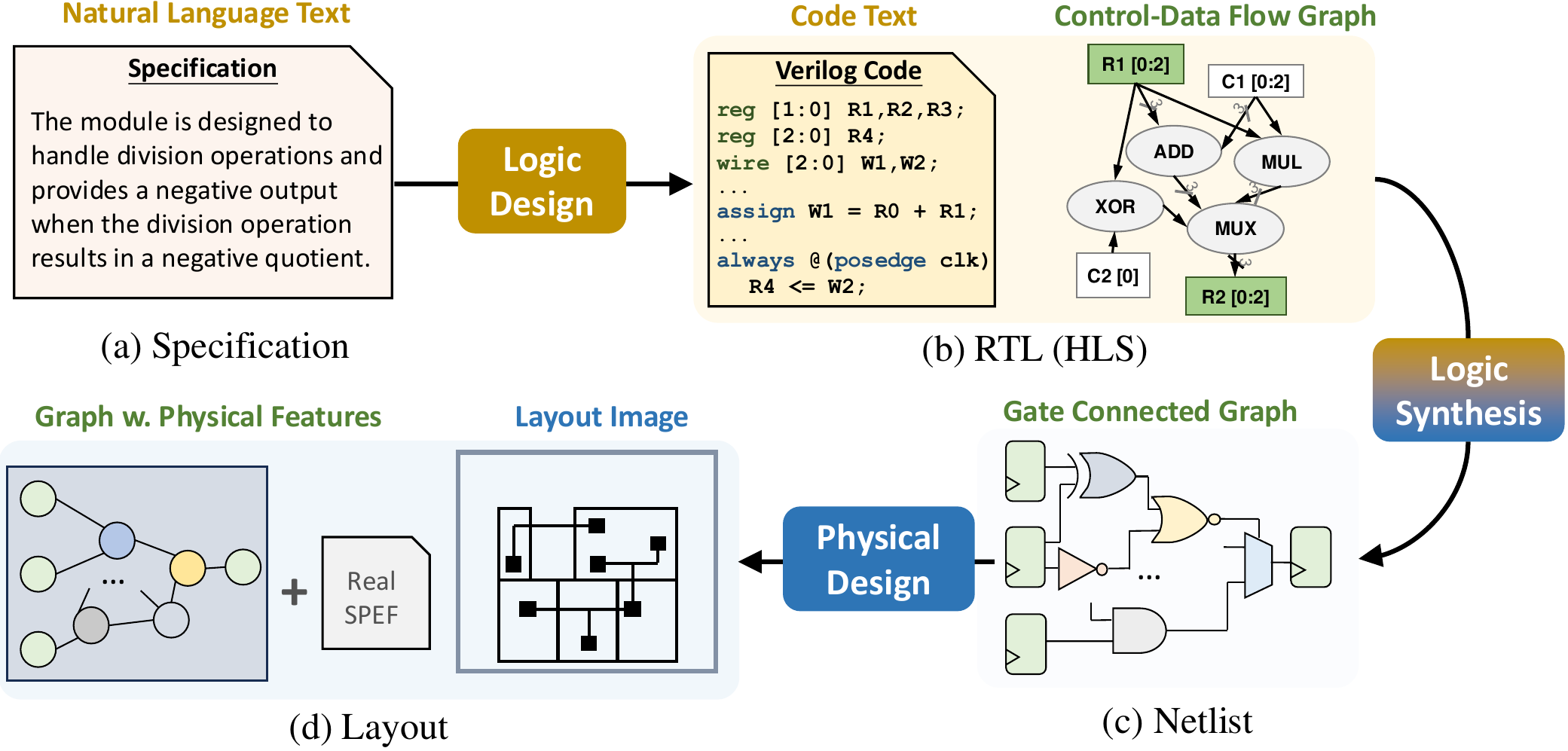}
   \vspace{-.1in}
    \caption{VLSI design stages and corresponding modalities.}
    \label{fig:modality}
    \vspace{-.15in}
\end{figure}

\textbf{Multimodal circuit format.} As shown in~\Cref{fig:modality}, circuit data inherently can be represented in multiple formats and modalities, each capturing different aspects of the circuit, including:
    \begin{itemize}[topsep=0pt]
        \item \textbf{Text.} This modality includes hardware description languages such as Verilog and VHDL, along with high-level specifications in natural language. Text-based representations define circuit functionality, behavioral constraints, and design requirements, emphasizing semantic information of circuits.
        \item \textbf{Graph.} Circuit structures are naturally represented as graphs, where nodes correspond to components (e.g., logic gates, registers, functional blocks) and edges capture connectivity (e.g., data flow, control dependencies). Graph-based formats, including control-data flow graphs and gate-connected graphs, preserve the topological relationships, which are crucial for structural reasoning in EDA tasks.
        \item \textbf{Image.} The physical layout of circuits, particularly at the post-synthesis stage, can be represented in two-dimensional visual formats, similar to the format of images. These `images' capture geometric features, including component placement and interconnect routing, which are critical for the physical design process and manufacturability.
    \end{itemize}
Each of these modalities provides a unique perspective of the circuit, and fusing them enables a comprehensive understanding of the design, facilitating advancing foundation AI techniques for circuits.\looseness=-1

\textbf{Multiple objectives.}
The ultimate target objectives in circuit design are \textit{PPA} (i.e., power, performance, and area) and \textit{functionality}. PPA metrics are crucial for optimizing the overall design, ensuring that the circuit meets the required performance standards while minimizing power consumption and chip area. 
Functionality metric targets fulfilling the intended specifications, guaranteeing that the circuit behaves correctly under various conditions. Achieving both PPA optimization and functional correctness is essential for delivering robust and efficient hardware.

\textbf{Parallel execution of hardware.} Hardware circuits inherently operate with parallelism, clearly distinguishing them from the sequential execution of software code. In combinational logic, multiple logic operations are computed simultaneously, enabling high-speed parallel data processing. 
Meanwhile, sequential elements, such as registers and flip-flops, update synchronously at each clock cycle, ensuring efficient and coordinated circuit state transition. This fundamental parallelism plays a crucial role in defining circuit behavior, making it essential for accurately capturing circuit intrinsic properties in AI-driven design automation.

\textbf{Circuit data availability.}
AI-driven EDA solutions depend on access to high-quality, diverse, and representative circuit data for both model development and evaluation. However, the scarcity of open circuit datasets remains a significant technical bottleneck. This challenge primarily arises from the semiconductor industry’s reluctance to share proprietary circuit designs, which are considered valuable commercial IP. The absence of publicly available datasets hampers AI-driven EDA advancements, as collecting labeled data is both time-consuming and resource-intensive. Moreover, the limited diversity of open-source circuit designs restricts model generalization and performance.\looseness=-1

As circuit foundation models gain traction in agile IC design, data availability becomes even more critical, particularly in the context of scaling laws for circuit foundation models, as observed in several existing works~\cite{fang2025circuitfusion, shi2024deepgate3}. These laws suggest that model performance improves with larger datasets, making the shortage of diverse and extensive circuit data a fundamental limitation in training highly capable models. Addressing this challenge is crucial for unlocking the full potential of AI-driven EDA solutions.

\textbf{Circuit reusability.}
Reusability is a key factor in practical circuit development, as companies often rely on pre-designed IP blocks rather than building circuits from scratch. This inherent reusability presents an opportunity for circuit foundation models to exploit semantic similarities across designs, enhancing performance on downstream tasks. By leveraging patterns and shared features within circuit datasets, these models can improve efficiency and adaptability in various EDA applications. 

\section{Foundation Model as a Circuit Encoder} 
\label{sec:encoder}

\begin{figure}[!b]
    \centering
    \includegraphics[width=0.97\linewidth]{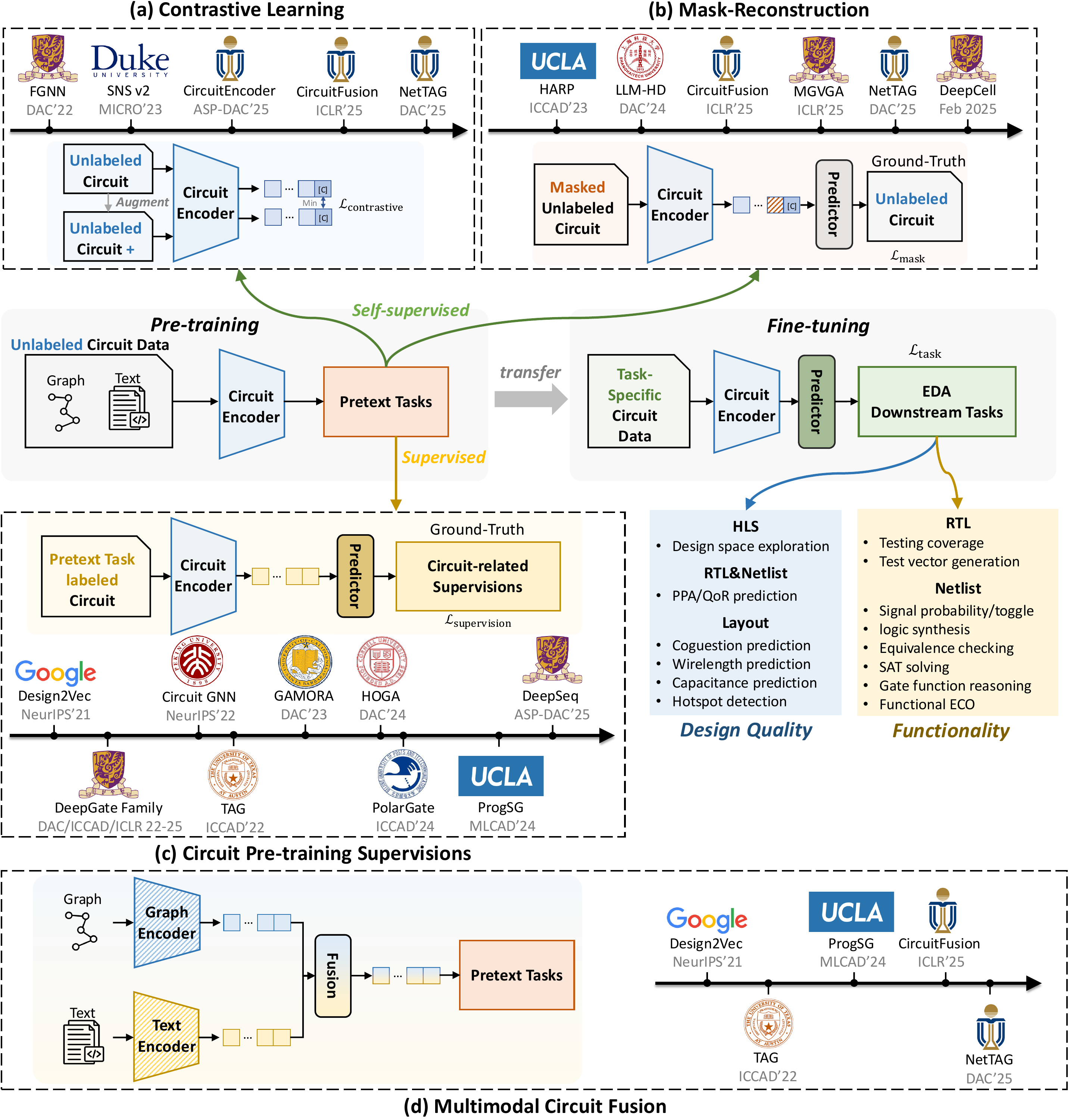}
    \vspace{-.1in}
    \caption{Summary of pre-training techniques used in circuit encoders, covered in~\Cref{sec:encoder}. Representative pre-training techniques include (a) self-supervised contrastive learning, (b) self-supervised mask-reconstruction, (c) circuit-related supervisions, and (d) multimodal circuit fusion.}
    \label{fig:encoder}
\end{figure}

In this section, we will cover all encoder-based circuit foundation models. 
The circuit encoder paradigm consists of two major stages: (1) It first \textit{pre-trains} AI models to encode circuits into generalized embedding vectors that capture rich intrinsic properties of circuits. These embeddings provide a flexible representation that can further be \textit{fine-tuned} with task-specific supervision. (2) The \textit{fine-tuning} process enables the embeddings to support various \textit{predictive} downstream tasks, such as early-stage design quality prediction and functional reasoning, thus supporting design space exploration. The goal is to predict specific outcomes based on given circuit data. Figure~\ref{fig:encoder} summarizes our covered encoder works based on their proposed pre-training techniques.

Existing works mainly focus on exploring the pre-training techniques that capture the structural, semantic, and physical aspects of circuits.
As summarized in~\Cref{fig:encoder}, we categorize the existing circuit encoder pre-training techniques into four types, each focusing on different aspects of learning circuit representations:
\begin{itemize}
    \item \new{\textbf{Self-supervised contrastive learning}~\cite{wang2022functionality, xu2023fast, fang2025circuitencoder, fang2025circuitfusion, fang2025nettag} in \Cref{fig:encoder} (a).} This technique minimizes the distance between embeddings of similar circuits while maximizing the distance between embeddings of dissimilar circuits based on the circuits’ functionality. This process pre-trains the model to differentiate between functionally equivalent and non-equivalent circuits. In this way, the pre-trained model learns meaningful representations that reflect the intrinsic functional properties of the circuit designs.
    \item \new{\textbf{Self-supervised mask-reconstruction}~\cite{sohrabizadeh2023robust, chen2024llm, wu2025circuit, shi2025deepcell, fang2025circuitfusion, fang2025nettag} in \Cref{fig:encoder} (b).} In this technique, parts of the circuit representation are masked, and the model is pre-trained to reconstruct the masked missing parts. This process pre-trains the model to learn robust, complete representations of circuits, capturing both the structural and functional aspects. The pre-training task typically involves masking graph nodes or textual tokens of a circuit and using the remaining circuit information to predict the missing parts.
    \item \new{\textbf{Supervised circuit pre-training tasks}~\cite{vasudevan2021learning, li2022deepgate, yang2022versatile, zhu2022tag, wu2023gamora, deng2024less, PolarGate, qin2024cross, khan2024deepseq} in \Cref{fig:encoder} (c).}
    In addition to self-supervised pre-training techniques, certain approaches incorporate task-related supervision to pre-train circuit encoders. Unlike direct target-task supervision in supervised methods, these pre-training tasks provide generalizable guidance to help the model learn circuit properties from labeled data. For example, predicting the truth-table distance between circuit pairs pre-trains the model to capture functional properties, which can then be leveraged for functional tasks such as SAT solving and logic synthesis.
    \item \new{\textbf{Multimodal circuit fusion}~\cite{vasudevan2021learning, zhu2022tag, qin2024cross, fang2025circuitfusion, fang2025nettag} in \Cref{fig:encoder} (d).} This technique integrates multiple modalities of circuit data, such as textual, structural, and physical information, to create richer, more comprehensive representations. The model is pre-trained to fuse these different modalities, enabling it to capture a broader range of circuit characteristics. 
    In this way, the model supports complex tasks that require information from different modalities.
\end{itemize}

In the following subsections, we summarize existing circuit encoders based on their target circuit design stages, including HLS stage (Section~\ref{sec:enc-hls}), RTL stage (Section~\ref{sec:enc-rtl}), netlist stage (Section~\ref{sec:enc-netlist}) and layout stage (Section~\ref{sec:enc-layout}). A detailed comparison and summary of these circuit encoders are provided in ~\Cref{tbl:encoder} and ~\Cref{tbl:encoder2}, respectively.
Section~\ref{sec:encoder-advanced} will cover the emerging and more advanced circuit encoder techniques. For each stage, we first summarize the employed circuit dataset, including detailed statistics and data collection process, then detail the proposed encoding techniques, including circuit preprocessing, ML model architecture, and pre-training techniques, and finally discuss the supported downstream tasks with evaluation metrics.


\subsection{Circuit Encoder for HLS}
\label{sec:enc-hls}

\begin{table}[!t]
\vspace{-.1in}
\center
\vspace{.05in}
\resizebox{0.95\textwidth}{!}{

\begin{tabular}{c||c|cc|cc|cc} \toprule
\multirow{2}{*}{\textbf{\begin{tabular}[c]{@{}c@{}}Target   \\      Stage\end{tabular}}} & \multirow{2}{*}{\textbf{Method}}                          & \multicolumn{2}{c|}{\textbf{Modality}} & \multicolumn{2}{c|}{\textbf{Pre-Training}}      & \multicolumn{2}{c}{\textbf{Downstream Task}}     \\ \cline{3-8}
                                                                                         &                                                           & \textbf{Graph}     & \textbf{Text}    & \textbf{Self-Supervised} & \textbf{Supervised} & \textbf{Design Quality} & \textbf{Functionality} \\ \hline \hline
\multirow{2}{*}{\textbf{HLS}}                                                            & HARP~\cite{sohrabizadeh2023robust}                        & \checkmark         &                  & \checkmark               &                     & \checkmark              &                        \\
                                                                                         & ProgSG~\cite{qin2024cross}                                & \checkmark         & \checkmark       & \checkmark               &                     & \checkmark              &                        \\ \hline
\multirow{4}{*}{\textbf{RTL}}                                                            & Design2Vec~\cite{vasudevan2021learning}                   & \checkmark         & \checkmark       &                          & \checkmark          &                         & \checkmark             \\
                                                                                         & SNS v2~\cite{xu2023fast}                                  & \checkmark         &                  &                          &                     & \checkmark              &                        \\
                                                                                         & CircuitEncoder~\cite{fang2025circuitencoder}              & \checkmark         &                  & \checkmark               &                     & \checkmark              &                        \\
                                                                                         & CircuitFusion~\cite{fang2025circuitfusion}                & \checkmark         & \checkmark       & \checkmark               &                     & \checkmark              &                        \\ \hline
\multirow{12}{*}{\textbf{Netlist}}                                                       & DeepGate~\cite{li2022deepgate}                            & \checkmark         &                  &                          & \checkmark          &                         & \checkmark             \\
                                                                                         & DeepGate2~\cite{shi2023deepgate2}                         & \checkmark         &                  &                          & \checkmark          &                         & \checkmark             \\
                                                                                         & DeepGate3/4~\cite{shi2024deepgate3,   zheng2025deepgate4} & \checkmark         &                  &                          & \checkmark          &                         & \checkmark             \\
                                                                                         & GAMORA~\cite{wu2023gamora}                                  & \checkmark         &                  &                          & \checkmark          &               & \checkmark             \\
                                                                                         & HOGA~\cite{deng2024less}                                  & \checkmark         &                  &                          & \checkmark          & \checkmark              & \checkmark             \\
                                                                                         & PolarGate~\cite{PolarGate}                                & \checkmark         &                  &                          & \checkmark          &                         & \checkmark             \\
                                                                                         & DeepSeq~\cite{khan2024deepseq,   khan2025deepseq2}        & \checkmark         &                  &                          & \checkmark          & \checkmark              &                        \\
                                                                                         & FGNN~\cite{wang2022functionality,   wang2024fgnn2}        & \checkmark         &                  & \checkmark               &                     &                         & \checkmark             \\
                                                                                         & CircuitEncoder~\cite{fang2025circuitencoder}              & \checkmark         &                  & \checkmark               &                     &                         & \checkmark             \\
                                                                                         & MGVGA~\cite{wu2025circuit}                                & \checkmark         & \checkmark       & \checkmark               &                     & \checkmark              & \checkmark             \\
                                                                                         & NetTAG~\cite{fang2025nettag}                              & \checkmark         & \checkmark       & \checkmark               &                     & \checkmark              & \checkmark             \\
                                                                                         & DeepCell~\cite{shi2025deepcell}                           & \checkmark         &                  & \checkmark               &                     &                         & \checkmark             \\ \hline
\multirow{3}{*}{\textbf{Layout}}                                                         & Circuit GNN~\cite{yang2022versatile}                      & \checkmark         &                  &                          & \checkmark          & \checkmark              &                        \\
                                                                                         & TAG~\cite{zhu2022tag}                                     & \checkmark         & \checkmark       &                          &                     & \checkmark              &                        \\
                                                                                         & LLM-HD~\cite{chen2024llm}                                 &                    & \checkmark       & \checkmark               &                     & \checkmark              &                \\ \bottomrule       
\end{tabular}

}
\vspace{.05in}
\caption{Comparison of modality, pre-training techniques, and supported downstream tasks for existing encoder-based circuit foundation models, as covered in~\Cref{sec:encoder}.}
\vspace{-.35in}
\label{tbl:encoder}
\end{table}

In the context of HLS, the circuit encoder plays a pivotal role in representing and optimizing the design space for HLS circuits. HLS involves the transformation of high-level programming languages (e.g., C/C++) into hardware description languages (e.g., Verilog), with the goal of improving the design, performance, and power efficiency of hardware systems. Efficient exploration of this design space is critical, and HLS encoders are explored to learn meaningful representations of the circuit designs, enabling better optimization and decision-making. As shown in~\Cref{fig:encoder-tl1} (a), two notable methods in this domain are HARP~\cite{sohrabizadeh2023robust} and ProgSG~\cite{qin2024cross}, both pre-train HLS encoders with self-supervised learning, improving the exploration of the HLS design space.

\subsubsection{Dataset for HLS circuits} 

\ 

The HLS dataset~\cite{bai2023towards} used in these works consists of 42 unique kernels, each with multiple optimization pragmas generated by the AMD/Xilinx HLS tool, resulting in over 10,000 design configurations. The HLS designs serve both text and graph modalities. In the text modality, the data consists of C/C++ code, averaging 1,286 tokens per program. In the graph modality, the programs are converted into the control-data flow graphs (CDFG), with an average of 354 nodes and 1,246 edges.\looseness=-1

\subsubsection{Encoding techniques for HLS circuits}

\ 

Both HARP~\cite{sohrabizadeh2023robust} and ProgSG~\cite{qin2024cross} employ self-supervised learning techniques to pre-train HLS encoders. HARP~\cite{sohrabizadeh2023robust} focuses on encoding the graph format of HLS CDFG, while ProgSG~\cite{qin2024cross} extends this by adding textual input for richer multimodal circuit representation learning. We detail the HLS encoding techniques below.

\textbf{Self-supervised HLS graph encoder with masked pragma reconstruction.}
HARP~\cite{sohrabizadeh2023robust} focuses on HLS control-data flow hierarchical graphs for representing circuit designs. Specifically, HARP~\cite{sohrabizadeh2023robust} utilizes a hierarchical graph representation of HLS designs, incorporating both high-level and low-level views, where the high-level view combines C/C++ code and LLVM intermediate representation (IR) to capture the program’s structure and semantics, and the low-level view focuses on LLVM IR to capture detailed implementation details. This dual-level representation helps mitigate long-range dependencies within the program. The model employs a GNN to encode this hierarchical graph into circuit embeddings. During pre-training, it applies a self-supervised learning technique called masked pragma reconstruction, with paradigm demonstrated in~\Cref{fig:encoder} (b). In this approach, certain pragmas (compiler directives) are masked, and the GNN model is trained to predict these masked pragmas based on the surrounding node embeddings in the graph. This enables the model to learn the specific effects of each pragma, enhancing its performance and improving its ability to transfer knowledge across tasks.

\textbf{Self-supervised HLS encoder enhanced via HLS graph-text mulimodal fusion.}
ProgSG~\cite{qin2024cross} builds upon HARP~\cite{sohrabizadeh2023robust} by integrating multimodal learning to improve HLS encoding.
It combines two modalities: CDFG hierarchical graph used in HARP~\cite{sohrabizadeh2023robust} and HLS C/C++ source code text, allowing the model to capture both structural and semantic aspects of the design. ProgSG~\cite{qin2024cross} uses a GNN for graph encoding and an LLM for text encoding. It introduces a node-token message passing mechanism for multimodal fusion, where information is exchanged through block nodes and tokens from the high-level view before being propagated to normal nodes and tokens via GNN and transformer layers.
To address the scarcity of labeled designs, ProgSG~\cite{qin2024cross} employs a self-supervised pre-training technique based on compiler-generated data flow analysis tasks. This static analysis task predicts the relationship between two nodes in a CDFG, such as reachability and data dependencies, enabling the model to learn how data moves through the program. This pre-training improves the model’s ability to generalize, boosting its performance in downstream tasks such as design space exploration and design optimization.

\subsubsection{Downstream tasks for HLS encoders} 

\ 

The two methods, HARP~\cite{sohrabizadeh2023robust} and ProgSG~\cite{qin2024cross}, support downstream tasks that predict various HLS design quality metrics, including latency (in cycle counts), block RAM utilization, digital signal processor utilization, flip-flop utilization, and lookup-table utilization. These metrics are critical for evaluating the performance and efficiency of HLS designs. The models are assessed using the regression metric root mean square error (RMSE), which measures the accuracy of the design performance predictions.

In addition to performance prediction, these HLS encoders are further used for design space exploration, a task aimed at finding the optimal design for a given kernel. This process involves exploring various design configurations to identify the best-performing design in terms of resource utilization and latency.

\begin{figure}[!t]
    \centering
    \includegraphics[width=1.0\linewidth]{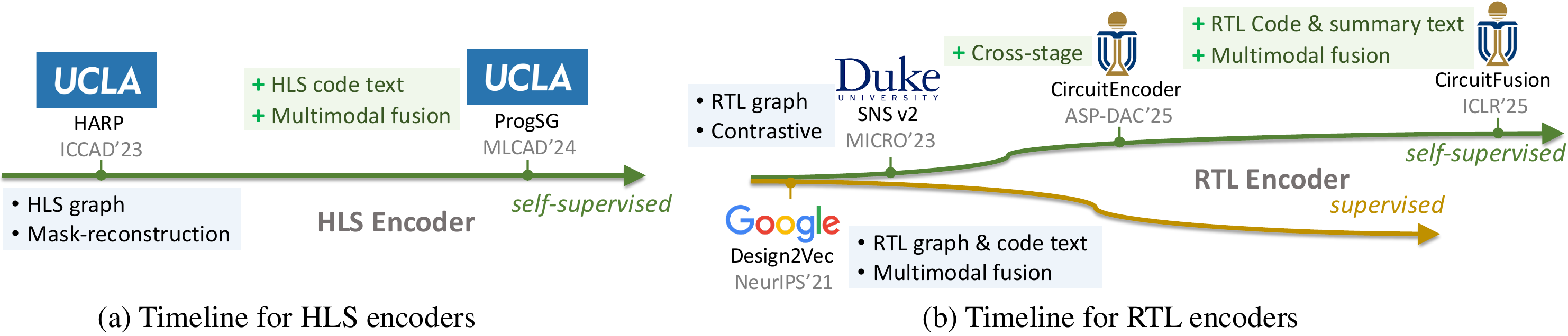}
    \vspace{-.25in}
    \caption{Timeline for HLS (\Cref{sec:enc-hls}) and RTL (\Cref{sec:enc-rtl}) encoders.}
    \label{fig:encoder-tl1}
    \vspace{-.25in}
\end{figure}

\subsection{Circuit Encoder for RTL Stage}\label{sec:enc-rtl}

In the RTL stage of VLSI design, the RTL encoder can capture both the semantics and structure of RTL circuits. As illustrated in \Cref{fig:encoder-tl1} (b), which shows the timeline of existing RTL encoders, four notable methods have emerged in this domain: Design2Vec~\cite{vasudevan2021learning} utilizes supervised pre-training tasks for functional verification tasks. In contrast, SNS v2~\cite{xu2023fast}, CircuitEncoder~\cite{fang2025circuitencoder}, and CircuitFusion~\cite{fang2025circuitfusion} employ self-supervised learning techniques for design quality prediction tasks.

\subsubsection{Dataset for RTL circuits} 

\ 

The RTL designs are used in both text and graph modalities: the text modality directly adopts the HDL code (e.g., Verilog), while the graph modality converts the RTL code into a CDFG based on the abstract syntax tree.
For functional verification tasks, Design2Vec~\cite{vasudevan2021learning} employs three designs, including two RISC-V CPUs and one TPU. For each design, the authors generated random tests and sampled each test parameter uniformly. They used a testbench to randomly sample input test stimuli and a Verilog RTL simulator to obtain ground-truth labels of whether a cover point was covered by that test, resulting 4118 cover points in total. 
As for design quality evaluation tasks, SNS v2~\cite{xu2023fast} and CircuitFusion~\cite{fang2025circuitfusion} collect various types of RTL designs from various open-sourced benchmarks, including ITC'99~\cite{corno2000rt}, OpenCores~\cite{URL:opencore}, Chipyard~\cite{ amid2020chipyard}, VexRiscv~\cite{vexriscv}, XiangShan~\cite{xu2022towards}, and other open-sourced designs. The RTL designs are synthesized using logic synthesis tools such as Synopsys Design Compiler, and the design quality metrics (i.e., PPA values) are obtained from the post-synthesis netlists.
In the latest work CircuitFusion~\cite{fang2025circuitfusion}, the dataset scale includes up to 500K nodes for the circuit graph and up to 20M tokens for Verilog code text.

\subsubsection{Encoding techniques for RTL} 

\ 

As shown in~\Cref{fig:encoder-tl1} (b), in the supervised encoding branch, Design2Vec~\cite{vasudevan2021learning} pioneers RTL encoding by learning functional semantics through pre-training supervisions for verification tasks.
In the self-supervised learning branch, SNS v2~\cite{xu2023fast} proposes to leverage functional contrastive learning on RTL graphs to capture RTL circuit representations, while CircuitEncoder~\cite{fang2025circuitencoder} enhances this by introducing cross-stage alignment with netlist stage, incorporating implementation details from netlists.
CircuitFusion~\cite{fang2025circuitfusion} further improves by integrating code text and functional summaries with the RTL graph for multimodal fusion, and adds additional self-supervised techniques to learn RTL circuits at multiple modalities and granularities. We detail the key techniques for RTL encoders below.\looseness=-1

\textbf{Supervised RTL semantic encoder with functional supervisions.}
Design2Vec~\cite{vasudevan2021learning} learns semantic representations of RTL circuits for functional verification. The input to Design2Vec~\cite{vasudevan2021learning} includes the hardware design represented as a CDFG derived from the RTL Verilog code, along with the corresponding source code text. The CDFG captures both the control and data flow aspects of the design, providing a comprehensive view of the hardware’s functionality.
Design2Vec~\cite{vasudevan2021learning} employs a GNN  to process the RTL CDFG, with each node augmented by RTL code text embeddings obtained from an LSTM for multimodal fusion. To capture the sequence dependency of circuit functionality, an additional LSTM is used to generate final node embeddings.
During pre-training, Design2Vec~\cite{vasudevan2021learning} uses a supervised learning pre-training task that predicts the coverage of specific points in the design when simulated on test inputs. This task requires the model to integrate both the structural and functional aspects of the hardware design, effectively learning the interactions between control and data flow.

\textbf{Self-supervised RTL encoder with contrastive learning.}
Although the supervised pre-training task is designed to learn the circuit functionality, it cannot be generalized to other function-unrelated tasks. The other three RTL encoders (i.e., SNS v2~\cite{xu2023fast}, CircuitEncoder~\cite{fang2025circuitencoder}, and CircuitFusion~\cite{fang2025circuitfusion}) employ self-supervised learning techniques that learn a generalized circuit embedding. 
The pioneering work SNS v2~\cite{xu2023fast} first introduces self-supervised contrastive learning to learn generalized circuit embeddings. 
The input to the SNS v2~\cite{xu2023fast} model is the graph format of HDL code based on the abstract syntax tree. It proposes a hierarchical graph format for RTL designs, where the low-level graph consists of subgraphs sampled from registers, and the high-level graph represents register dependency.
The model uses a two-level hierarchical GNN architecture. The low level processes small subgraphs, capturing local structural and functional features, while the high level aggregates these embeddings to predict quality metrics such as power, area, and timing for the entire design.
For pre-training, SNS v2~\cite{xu2023fast} employs a contrastive learning approach to pre-train the subgraph GNN on unlabeled hardware designs. The model learns to create functionally equivalent circuit representations, where similar embeddings are assigned to functionally equivalent circuits, despite differences in their representation. This self-supervised learning task enables the model to understand circuit equivalence. After pre-training, the model is fine-tuned using labeled datasets and adapted to new domains, allowing it to predict design quality metrics for various RTL circuits.

\textbf{Self-supervised RTL encoder enhanced with cross-stage alignment.}
Following SNS v2~\cite{xu2023fast}, CircuitEncoder~\cite{fang2025circuitencoder} also converts circuit RTL code into a graph-based representation using the abstract syntax tree. The model processes the graph using a graph transformer, which allows it to learn from the structural relationships of RTL designs.
For pre-training, CircuitEncoder~\cite{fang2025circuitencoder} employs graph contrastive learning on RTL designs, similar to SNS v2~\cite{xu2023fast}. In addition, CircuitEncoder~\cite{fang2025circuitencoder} introduces multi-stage contrastive learning, which involves learning embeddings both within the same design stage (intra-stage) and across different stages (inter-stage) between RTL and netlist designs. This technique helps align the embeddings from different design stages into a shared latent space, improving the model’s ability to transfer learning between different stages and enhancing its generalization across the hardware design process.

\textbf{Self-supervised RTL encoder enhanced with multimodal fusion.}
Another recent work CircuitFusion~\cite{fang2025circuitfusion} proposes self-supervised learning and advances the RTL encoder by fusing multiple modalities of RTL designs to enhance chip design workflows. Specifically, it processes three input modalities: HDL code, representing circuit functionality in textual form (e.g., Verilog); graph format, capturing the circuit’s structure through an abstract syntax tree; and functionality summary, a high-level textual abstraction of the design’s function. The model uses unimodal encoders for each modality, including graph, code, and summary encoders, followed by a multimodal fusion encoder with a cross-attention mechanism to combine the outputs into a unified latent space. During pre-training, CircuitFusion~\cite{fang2025circuitfusion} utilizes several self-supervised tasks: (1) Intra-modal learning, including contrastive learning and masked graph modeling to capture the internal structure of each modality, (2) Cross-modal alignment, where contrastive learning aligns the different modalities in a shared space, (3) Multimodal fusion, which involves tasks like masked summary modeling and mixup-embedding matching to combine structural and semantic information from all modalities, and (4) Implementation-aware alignment, which aligns RTL and netlist representations to ensure the design’s functionality maps accurately to its physical implementation.

\subsubsection{Downstream tasks for RTL encoders}

\ 

The RTL stage is crucial for implementing the functionality of the specification and serves as the foundation for design quality optimization, such as PPA. The primary downstream tasks for RTL encoders focus on functional verification and early-stage PPA prediction.
For \textbf{functional verification}, the Design2Vec~\cite{vasudevan2021learning} model uses the semantic representations learned through pre-training tasks to predict whether a given test covers specific portions of the design and to generate test vectors. The model is evaluated based on its ability to predict coverage and detect bugs in hardware designs. By improving test generation and bug detection efficiency, the model enhances the overall verification process.
For \textbf{design quality prediction}, SNS v2~\cite{xu2023fast}, CircuitEncoder~\cite{fang2025circuitencoder}, and CircuitFusion~\cite{fang2025circuitfusion} all focus on predicting key synthesis results, including area, power consumption, and timing for hardware designs. These models are evaluated using performance metrics such as the correlation coefficient (R) and Mean Absolute Percentage Error (MAPE), providing insights into the accuracy of design quality predictions and contributing to early-stage optimization for PPA.

\subsection{Circuit Encoder for Netlist Stage}\label{sec:enc-netlist}
\begin{figure}
    \centering
    \includegraphics[width=1\linewidth]{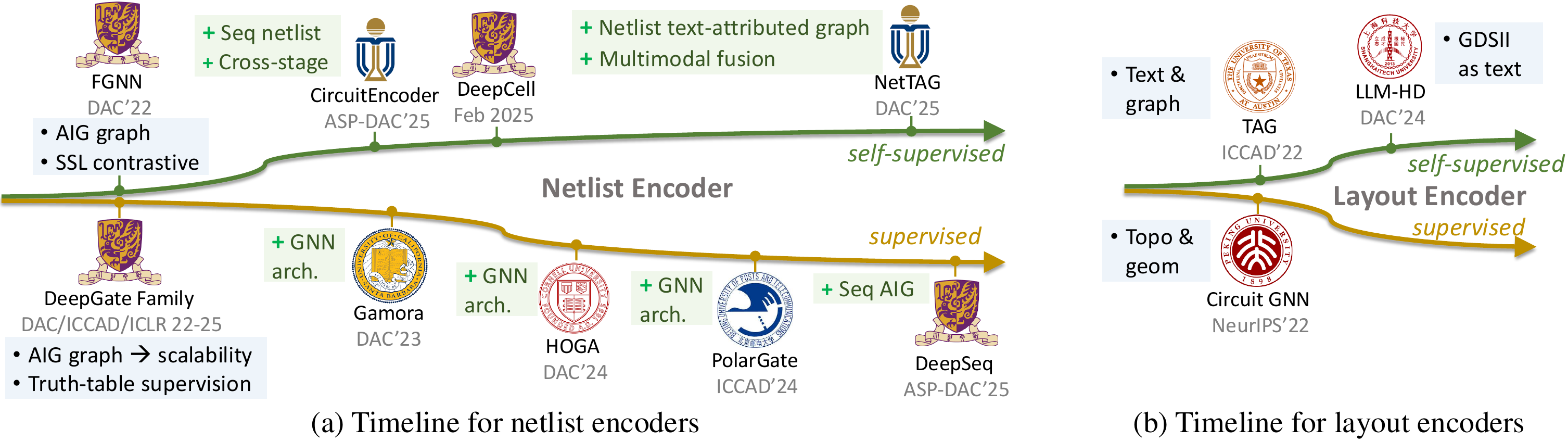}
    \vspace{-.25in}
    \caption{Timeline for netlist (\Cref{sec:enc-netlist}) and layout (\Cref{sec:enc-layout}) encoders.}
    \label{fig:encoder-tl2}
    \vspace{-.25in}
\end{figure}

The netlist stage is one of the most actively explored stages in circuit encoders, with circuit encoding playing a critical role in extracting meaningful representations from the structural and functional properties of logic circuits. In recent years, several methods have been developed that apply graph learning techniques (e.g., GNNs and Graph Transformers), to improve netlist analysis.
As shown in \Cref{fig:encoder-tl2} (a), the timeline for netlist encoders includes both supervised methods, such as the DeepGate Family~\cite{li2022deepgate, shi2023deepgate2, shi2024deepgate3, zheng2025deepgate4}, HOGA~\cite{deng2024less}, PolarGate~\cite{PolarGate}, and DeepSeq~\cite{khan2024deepseq, khan2025deepseq2}, as well as self-supervised methods like FGNN~\cite{wang2022functionality, wang2024fgnn2}, CircuitEncoder~\cite{fang2025circuitencoder}, NetTAG~\cite{fang2025nettag}, and DeepCell~\cite{shi2025deepcell}. These encoders have evolved from encoding simple AND-Inverter Graphs (AIGs) of netlists to more complex post-synthesis netlists involving various types of gates.
For downstream tasks, these netlist encoders support a wide range of applications. These include functional reasoning and verification tasks, such as arithmetic block identification, SAT solving, and logic synthesis, as well as netlist-stage design quality evaluation tasks like timing, power, and area estimation.

\subsubsection{Dataset for circuit netlists} 

\ 

Most of the netlist encoders~\cite{li2022deepgate, shi2023deepgate2, shi2024deepgate3, zheng2025deepgate4, wang2022functionality, wang2024fgnn2, deng2024less, PolarGate, khan2024deepseq, khan2025deepseq2} target the And-Inverter Graph (AIG) format of the netlist, which is an intermediate representation commonly used in logic synthesis and verification. Recently, works~\cite{fang2025nettag, shi2025deepcell} have expanded their scope beyond basic AIG gates to handle more complex post-synthesis netlists, which include various standard cells. 
For AIG datasets, the encoders gather data from various benchmarks like OpenABC-D~\cite{chowdhury2021openabc}, ITC'99~\cite{corno2000rt}, IWLS~\cite{albrecht2005iwls}, OpenCores~\cite{URL:opencore}, EPFL~\cite{amaru2015epfl}, GAMORA~\cite{wu2023gamora}, arithmetic modules~\cite{wang2022functionality}, Chipyard~\cite{amid2020chipyard}, and LGSynth-93~\cite{mcelvain1993lgsynth93}. RTL designs from these benchmarks are typically converted into AIG formats using the ABC open-source logic synthesis tool. These encoders primarily focus on combinational logic within AIGs, with DeepSeq~\cite{khan2025deepseq2} also considering sequential registers in its encoding process.
As for the post-synthesis netlist datasets, they are obtained from RTL benchmarks like ITC'99~\cite{corno2000rt}, IWLS~\cite{albrecht2005iwls}, OpenCores~\cite{URL:opencore}, EPFL~\cite{amaru2015epfl}, Chipyard~\cite{amid2020chipyard} and VexRiscv~\cite{vexriscv}.  Logic synthesis is conducted using technology libraries to generate the post-synthesis netlists. NetTAG~\cite{fang2025nettag} processes both combinational and sequential netlist gates, while DeepCell~\cite{shi2025deepcell} focuses on the combinational aspects. This expansion enables the models to handle a wider range of netlist formats and to predict design quality more accurately at post-synthesis stages.

\subsubsection{Encoding techniques for netlist}

\ 

As shown in \Cref{fig:encoder-tl2} (a), we categorize the encoding methods into supervised pre-training tasks and self-supervised learning methods.
In the supervised encoding branch, DeepGate family~\cite{li2022deepgate, shi2023deepgate2, shi2024deepgate3, zheng2025deepgate4} pioneers AIG encoding for netlists, learning functional semantics for logic synthesis and verification tasks. They have improved scalability with advanced supervision, better graph learning models, and optimized memory consumption. Other methods, like HOGA~\cite{deng2024less} and PolarGate~\cite{PolarGate}, refine AIG encoding with customized GNN architectures and message-passing mechanisms to capture both structural and functional properties. 
DeepSeq~\cite{khan2024deepseq, khan2025deepseq2} extends the DeepGate Family to handle sequential circuits, improving the model’s ability to process more complex circuit behaviors.
In the self-supervised learning branch, FGNN~\cite{wang2022functionality, wang2024fgnn2} first introduces functional contrastive learning to solve the arithmetic block identification problem. CircuitEncoder~\cite{fang2025circuitencoder} enhances this with cross-stage alignment, incorporating RTL-stage information to improve netlist encoding. NetTAG~\cite{fang2025nettag} and DeepCell~\cite{shi2025deepcell} push the boundaries of AIG encoding by advancing it to handle more complex post-synthesis netlists, allowing for the processing of designs with various standard cells and more intricate gate structures, thus improving the prediction and optimization of hardware designs in post-synthesis stages.

\begin{table}[!t]
\centering
\resizebox{1\textwidth}{!}{

\begin{tabular}{c||c|c|c} \toprule
\multirow{2}{*}{\textbf{\begin{tabular}[c]{@{}c@{}}Target   \\      Stage\end{tabular}}} & \multirow{2}{*}{\textbf{Method}}                          & \textbf{Technique}                                                                                                                                                                      & \textbf{Downstream Task}                                                                                    \\
                                                                                         &                                                           & \textbf{Pre-train objective}                                                                                                                                                            & \textbf{}                                                                                                   \\ \hline \hline
\multirow{2}{*}{\textbf{HLS}}                                                            & HARP~\cite{sohrabizadeh2023robust}                        & Masked pragma reconstruction                                                                                                                                                            & HLS design space exploration                                                                                \\ 
                                                                                         & ProgSG~\cite{qin2024cross}                                & Data flow analysis tasks for graph and node                                                                                                                                             & HLS design space exploration                                                                                \\
& Design2Vec~\cite{vasudevan2021learning}                   & Testing cover point prediction                                                                                                                                                          & Verification coverage prediction and test generation                                                        \\ \hline
                                                                                                                                                     & SNS v2~\cite{xu2023fast}                                  & Functional contrastive learning                                                                                                                                                         & Post-synthesis PPA prediction                                                                               \\                                                             
 \cline{2-4} \multirow{4}{*}{\textbf{RTL}} & CircuitEncoder~\cite{fang2025circuitencoder}              & \begin{tabular}[c]{@{}c@{}}Intra-stage functional   contrastive learning\\      Cross-stage functional contrastive alignment\end{tabular}                                               & Post-synthesis PPA prediction                                                                               \\ \cline{2-4}
                                                                                         & CircuitFusion~\cite{fang2025circuitfusion}                & \begin{tabular}[c]{@{}c@{}}Masked gate  reconstruction\\      Functional contrastive  for graph/   summary\\      Modality fusion\\      Cross-design-stage alignment\end{tabular}      & Post-synthesis PPA prediction                                                                               \\ \hline
                                                        & DeepGate~\cite{li2022deepgate}                            & Signal probability prediction                                                                                                                                                           & Signal probability prediction on large AIGs                                                                 \\
                                                                                         & DeepGate2~\cite{shi2023deepgate2}                         & Truth-table supervisions on node                                                                                                                                                        & Logic synthesis and SAT solving                                                                             \\
                                                                                         & DeepGate3/4~\cite{shi2024deepgate3, zheng2025deepgate4} & Truth-table supervisions on node and graph                                                                                                                                              & SAT solving                                                                                                 \\
                                                                                         & GAMORA~\cite{wu2023gamora}                                  & Task-specific supervisions                                                                                                                                                              & Logic functional reasoning                                                        \\
                                                                                         & HOGA~\cite{deng2024less}                                  & Task-specific supervisions                                                                                                                                                              & Logic synthesis QoR prediction, functional reasoning                                                        \\
                                \multirow{7}{*}{\textbf{Netlist}}                                                         & PolarGate~\cite{PolarGate}                                & Truth-table supervisions                                                                                                                                                                & Signal probability and truth-table distance prediction                                                      \\
                                                                                         & DeepSeq~\cite{khan2024deepseq,   khan2025deepseq2}        & Truth-table supervisions on node                                                                                                                                                        & Toggle rate prediction for power analysis                                                                   \\
                                                                                         & FGNN~\cite{wang2022functionality,   wang2024fgnn2}        & Functional contrastive learning                                                                                                                                                         & Gate function reasoning                                                                                     \\ \cline{2-4}
\textbf{}                                                                                & CircuitEncoder~\cite{fang2025circuitencoder}              & \begin{tabular}[c]{@{}c@{}}Intra-stage functional   contrastive learning\\      Cross-stage functional contrastive alignment\end{tabular}                                               & Register function  reasoning                                                                                \\ \cline{2-4}
\textbf{}                                                                                & MGVGA~\cite{wu2025circuit}                                & Masked gate    reconstruction                                                                                                                                                           & QoR prediction, logic equivalence identification                                                            \\ \cline{2-4}
\textbf{}                                                                                & NetTAG~\cite{fang2025nettag}                              & \begin{tabular}[c]{@{}c@{}}Logic expression   contrastive\\      Masked gate reconstruction\\      Netlist graph contrastive learning\\      Netlist graph size prediction\end{tabular} & \begin{tabular}[c]{@{}c@{}}Post-layout PPA prediction\\      Gate/Register function prediction\end{tabular} \\ \cline{2-4}
\textbf{}                                                                                & DeepCell~\cite{shi2025deepcell}                           & Masked circuit modeling                                                                                                                                                                 & Functional ECO                                                                                              \\ \hline
\multirow{3}{*}{\textbf{Layout}}                                                         & Circuit GNN~\cite{yang2022versatile}                      & Task-specific supervisions                                                                                                                                                              & Congestion and wirelength prediction                                                                        \\
                                                                                         & TAG~\cite{zhu2022tag}                                     & Layout instance distance prediction                                                                                                                                                     & Wirelength, and net parasitic capacitance  prediction                                                       \\
                                                                                         & LLM-HD~\cite{chen2024llm}                                 & Masked language modeling                                                                                                                                                                & Hotspot detection                                                                                         \\ \bottomrule 
\end{tabular}

}

\caption{Summary of the pre-training techniques and supported downstream tasks of circuit encoders, covered in~\Cref{sec:encoder}.}
\vspace{-.35in}
\label{tbl:encoder2}
\end{table}

\textbf{Supervised AIG encoder with functional supervision.}
The DeepGate family~\cite{li2022deepgate, shi2023deepgate2, shi2024deepgate3, zheng2025deepgate4} is one of the pioneers in netlist encoders. They handle circuit AIGs using customized graph learning models, which are pre-trained using supervised pre-training tasks. These works primarily focus on functional-related tasks, such as training on pairwise truth table differences between sampled logic gates. The DeepGate family continuously improves model performance and scalability, with DeepGate3~\cite{shi2024deepgate3} introducing a graph transformer to capture global circuit relationships, and DeepGate4~\cite{zheng2025deepgate4} optimizing the model by eliminating redundant computations.

Specifically, DeepGate~\cite{li2022deepgate} employs a GNN architecture specifically tailored for AIG graphs, incorporating an attention mechanism and recurrent layers to aggregate information across the graph. Each node’s embedding is computed based on its gate type and its relationships with neighboring nodes. The recurrent GNN is designed to capture the functional behavior of the circuit by using both forward and reversed propagation layers, simulating the logic behavior. Signal probability (the probability of a node being in logic ‘1’) is used as the supervision task, with signal probabilities derived from random logic simulations. These simulations are run on the circuits to obtain accurate probability values, allowing the model to learn functional behavior more effectively.

DeepGate2~\cite{shi2023deepgate2} enhances the functionality-awareness encoding by introducing the Hamming distance between the truth tables of logic gates as supervision. The model uses a one-round GNN architecture, which processes both functional and structural embeddings for each gate. Unlike the multi-round GNN used in the original DeepGate, this one-round architecture efficiently propagates embeddings in a single pass. The functional embeddings represent the logic behavior of gates, incorporating the pairwise truth table difference as a supervisory signal, while structural embeddings capture the topology of the circuit. A self-attention mechanism is used to aggregate information across different gates, enabling the model to focus more on controlling fan-in gates. 
During pre-training, a functionality-aware loss is proposed, which aligns gate embeddings with their functional equivalence. This loss minimizes the distance between embeddings corresponding to gates that perform similar logical operations, thereby improving the model’s ability to recognize functionally equivalent gates.

DeepGate3~\cite{shi2024deepgate3} improves upon DeepGate2~\cite{shi2023deepgate2} by enhancing both performance and scalability with a graph transformer model. It uses DeepGate2~\cite{shi2023deepgate2} as the AIG node tokenizer and refines the node embeddings with a graph transformer to capture long-range dependencies within the graph. For generating graph-level embeddings, another graph transformer is used for pooling. During pre-training, in addition to the gate-level supervisions used in DeepGate2, graph-level tasks are introduced. These tasks involve using fan-in cones to segment circuits into smaller subgraphs and predict intrinsic features such as the size and depth of these subgraphs, further improving the model’s ability to understand circuit structures at a broader level.

DeepGate4~\cite{zheng2025deepgate4} further improves the scalability and efficiency challenges of large-scale circuit AIG representation learning by integrating a GNN-based sparse transformer. By leveraging graph sparsity, the model reduces the time and memory complexity of the transformer, making it suitable for processing large circuits. The architecture also incorporates structural encodings for gates, such as level and out-degree, to enhance the learning of circuit properties. The circuit graph is partitioned into smaller cones based on logic levels, which are then processed by the sparse transformer. This approach significantly improves both accuracy and computational efficiency, particularly for large-scale circuit designs, outperforming previous methods in terms of scalability and overall performance.

\textbf{Supervised AIG encoder enhanced with GNN architecture.} In addition to DeepGate family, other works (i.e., GAMORA~\cite{wu2023gamora}, HOGA~\cite{deng2024less} and PolarGate~\cite{PolarGate}) explore to customize the GNN architecture and message-passing mechanism to enhance the scalability and performance of AIG encoding, combining with supervised pre-training tasks.
In GAMORA~\cite{wu2023gamora}, netlist AIGs are transformed into a graph representation and processed using a GNN. During pre-training, the GNN is trained with multiple functionally driven tasks, which jointly reason about Boolean function aggregation and structural topology. This enables efficient symbolic reasoning for large-scale Boolean networks. Specifically, GAMORA~\cite{wu2023gamora}’s model is designed to recognize fundamental functional components within circuits, including identifying adder root and leaf nodes and detecting XOR and MAJ functions. The multi-task learning framework enhances the model’s ability to generalize across various functional tasks, leveraging shared representations to improve both accuracy and scalability in processing large AIG-based netlists.

In HOGA~\cite{deng2024less}, hop-wise features are precomputed for each design to capture interactions over multiple hops before training. This step is done independently of the graph structure, enabling scalability for distributed training. The AIG format of circuits is processed using a customized GNN with a hop-wise aggregation scheme, which precomputes features based on multiple hops. It also employs gated self-attention to adaptively learn high-order circuit structures. This approach avoids recursive aggregation, which can be computationally expensive for large circuits. The model is then trained using task-specific labels, allowing it to be adapted for downstream tasks.

In PolarGate~\cite{PolarGate}, each node in the netlist AIGs represents two logical states: low level (0) and high level (1), which are fundamental for Boolean logic tasks. The model employs a GNN with a novel functionality-aware message passing mechanism that aggregates information from neighboring nodes while distinguishing between AND and NOT gates through specialized operators. To achieve this, PolarGate~\cite{PolarGate} introduces an ambipolar embedding space, where each node is mapped to both a positive and a negative embedding to represent the two logical states. It also uses differentiable logical operators, such as OPAND and OPNOT, that are designed to be differentiable and compatible with embedding propagation in the AIG structure. Additionally, the message passing strategy is modified to adhere to Boolean logical behavior, ensuring more accurate functional representation of the circuit. These innovations enable PolarGate~\cite{PolarGate} to effectively capture the logical operations of circuits and improve the model’s ability to process and learn from netlist-based designs.

\textbf{Supervised AIG encoder enhanced for sequential circuits.}
Beyond focusing on the combinational logics of AIGs, DeepSeq~\cite{khan2025deepseq2} explores capturing the sequential behavior of AIGs. DeepSeq~\cite{khan2025deepseq2} further advances this by using a directed acyclic GNN, which is optimized for sequential netlists. It incorporates a customized propagation scheme that avoids recursive propagation and handles cyclic sequential netlists in a single forward pass. The architecture separates learning into three distinct embedding spaces: structure embedding for circuit connectivity, function embedding for logic computations, and sequential embedding for capturing the temporal behavior between consecutive clock cycles. During pre-training, DeepSeq~\cite{khan2025deepseq2} uses multiple functional pre-training supervisions, including transition probability prediction to model sequential behavior, logic probability prediction to capture logic functionality, and pairwise truth-table difference to identify functional similarities among logic gates. These techniques enable DeepSeq~\cite{khan2025deepseq2} to effectively learn both the functional and sequential aspects of sequential AIG circuits.\looseness=-1

\textbf{Self-supervised AIG encoder with contrastive learning.}
In addition to customized supervised pre-training tasks based on circuit properties, another key approach for netlist encoders is leveraging self-supervised learning techniques to learn from unlabeled circuit data and capture the intrinsic information of the circuit.
FGNN~\cite{wang2022functionality, wang2024fgnn2} is a pioneer in adopting self-supervised contrastive learning for AIG netlist encoding. It uses a customized GNN to encode AIGs into embeddings, integrating a contrastive learning framework to enhance circuit functionality learning. The GNN architecture incorporates two types of learnable message aggregators: an ANG aggregator for AND gates and an INV aggregator for inverters. Asynchronous message passing is employed to efficiently propagate information through the graph while preserving functionality semantics. During pre-training, the model uses a contrastive learning scheme to learn circuit embeddings that reflect the Boolean functionality of the circuits. This scheme ensures that the embeddings of functionally equivalent circuits are close in the embedding space. Additionally, a new loss function is introduced to effectively capture the relative functional distance between circuits, taking into account input order invariance and circuits with different input widths, further improving the model’s ability to represent the circuits' functionality.

\textbf{Self-supervised netlist encoders with cross-stage alignment.} Recent advancements in self-supervised learning for netlist encoders have introduced cross-design-stage alignment to enhance model awareness of different abstraction levels in circuit design. CircuitEncoder~\cite{fang2025circuitencoder} and MGVGA~\cite{wu2025circuit} both propose novel alignment strategies to bridge the high-level abstract semantics of RTL with the low-level implementation details of netlists, enabling more robust circuit representation learning.
CircuitEncoder~\cite{fang2025circuitencoder} represents both RTL and netlist circuits as graph-based structures and processes them using a GNN. To learn circuit intrinsics, the model employs graph contrastive learning within each design stage, differentiating functionally similar and dissimilar circuit graphs. Additionally, it introduces intra-stage contrastive learning between RTL and netlist stages, effectively aligning representations across design stages. This cross-stage awareness significantly improves the encoder’s adaptability for downstream tasks following fine-tuning.
Similarly, MGVGA~\cite{wu2025circuit} proposes the concept of RTL-netlist alignment by integrating LLM-based processing for RTL descriptions and GNN-based encoding for AIG netlists. During pre-training, it introduces masked gate modeling, a technique that masks gates in the latent space while preserving logical equivalence, ensuring functional consistency throughout the representation learning process. Furthermore, a cross-modal learning strategy is implemented, where Verilog-based functional constraints guide AIG-based representation learning, reinforcing the structural-functionality alignment. These cross-stage alignment techniques enhance the capability of netlist encoders, improving their generalization across multiple circuit design stages and boosting performance in downstream EDA applications.

\begin{figure}[!t]
  \centering
  \includegraphics[width=\linewidth]{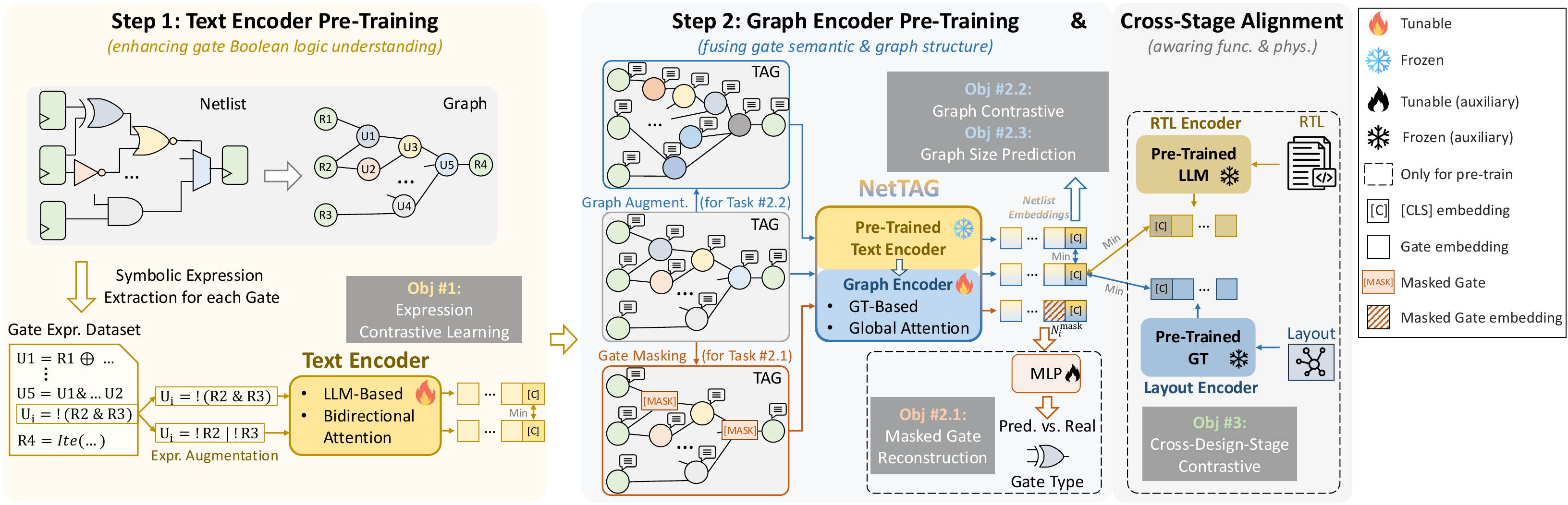}
    \vspace{-.3in}
  \caption{Multimodal pre-training techniques used in NetTAG~\cite{fang2025nettag}, including representative self-supervised learning methods such as contrastive learning, mask-reconstruction, and cross-design-stage alignment.}
  \label{fig:nettag}
   \vspace{-.2in}
\end{figure}

\textbf{Self-supervised post-synthesis netlist encoder with multimodal fusion.}
While many existing netlist encoders focus on simpler AND-Inverter gates, they struggle with more complex post-synthesis netlists that involve various gates from standard liberty cells. To address this challenge, two recent works have advanced netlist encoding by incorporating multimodal fusion (i.e., NetTAG~\cite{fang2025nettag}) or AIG-netlist alignment (i.e., DeepCell~\cite{shi2025deepcell}) to unprecedentedly handle the complexities of post-synthesis netlists.

As shown in~\Cref{fig:nettag}, in NetTAG~\cite{fang2025nettag}, post-synthesis netlists are represented as text-attributed graphs, where each node corresponds to a gate and is associated with attributes that include both functional symbolic logic expressions and physical characteristics (such as area, power, and delay). The model employs a two-stage multimodal hybrid architecture: first, an LLM-based text encoder processes the textual attributes of the gates to generate semantic-rich embeddings. Then, a graph transformer refines these embeddings by capturing the global circuit structure through graph-based attention mechanisms. During pre-training, the model utilizes four key self-supervised objectives. Expression contrastive learning enhances the LLM’s understanding of Boolean logic by contrasting symbolic expressions. Masked gate reconstruction is a graph-based task where certain gates are masked, and the model predicts the gate type, encouraging it to capture structural roles. Netlist graph contrastive learning aims to group similar netlists together while separating dissimilar ones, improving the model’s ability to recognize functional equivalence in different netlist structures. Finally, cross-stage contrastive alignment aligns netlist embeddings with RTL and layout embeddings, combining functional and physical information to improve performance across various design stages. These self-supervised tasks enable NetTAG~\cite{fang2025nettag} to learn both the functional and structural aspects of post-synthesis netlists, significantly enhancing its ability to predict design qualities and optimize circuits across different stages of the design process.

\textbf{Self-supervised post-synthesis netlist encoder with AIG-netlist alignment.}
DeepCell~\cite{shi2025deepcell} proposes multiview representation learning to simultaneously capture structural and functional information from both post-synthesis netlists and AIGs. The model uses two separate encoders: the PM Encoder, which is a GNN designed to capture the features of standard cells from the post-synthesis netlists, integrating both structural and functional embeddings through specialized aggregators, and the AIG Encoder, which is a pre-trained AIG encoder based on DeepGate2~\cite{shi2023deepgate2} that generates gate-level embeddings to provide additional structural information. During pre-training, DeepCell~\cite{shi2025deepcell} employs a self-supervised mask circuit modeling task, where a subset of the cell embeddings is masked, and the model reconstructs these embeddings using the information from the AIG encoder. This approach refines the post-synthesis netlist representations by integrating insights from both the local circuit view (i.e., netlist) and the global gate-level view (i.e., AIG), enhancing the overall quality of the circuit representation and improving downstream tasks such as design quality prediction and functional verification.

\subsubsection{Downstream tasks for netlist encoders}

\ 

Netlist encoders support a wide range of downstream tasks, including functional reasoning and verification tasks, as well as netlist-stage design quality evaluation tasks such as timing, power, and area estimation.
In \textbf{functional reasoning and verification}, key tasks ensure circuit functional correctness. Logic probability prediction estimates the likelihood that a gate outputs a logic ‘1’, evaluated by Mean Absolute Error (MAE). Equivalence class identification groups functionally equivalent gates, with performance assessed by classification accuracy. SAT solving checks Boolean satisfiability, evaluated by solving time and satisfiability accuracy. Arithmetic function block identification identifies components like adders, assessed using classification metrics. Finally, functional ECO identifies mismatches post-synthesis, with evaluation based on error reduction and change cost.
In \textbf{design quality prediction}, tasks focus on estimating key metrics like power, area, and delay. Logic synthesis QoR prediction uses MAPE to predict power, area, and delay after synthesis. Power evaluation estimates power based on toggle rate, evaluated by MAE and accuracy. Post-layout PPA prediction estimates power, performance, and area after layout, using MAPE to compare predicted versus actual results. These tasks enhance circuit optimization and validation.

\subsection{Circuit Encoder for Layout Stage}\label{sec:enc-layout}
In the layout stage of hardware design, circuit encoders process either the netlist or the GDSII format of circuit layouts. As shown in~\Cref{fig:encoder-tl2} (b), the timeline for layout encoders includes both supervised methods such as Circuit GNN~\cite{yang2022versatile} which handles layout topology and geometry, and self-supervised methods like TAG~\cite{zhu2022tag}, which employs text-graph multimodal encoding, and LLM-HD~\cite{chen2024llm} which treats layout GDSII data as text. These methods focus on effectively capturing the physical and structural properties of layout designs to improve design quality prediction.

\subsubsection{Dataset for layout circuits.}

\ 

The datasets used in Circuit GNN~\cite{yang2022versatile} come from the ISPD2011 benchmark~\cite{viswanathan2011ispd} for congestion prediction (12 designs) and the DAC2012 dataset~\cite{viswanathan2012dac} for net wirelength prediction (7 designs). These are preprocessed into a Circuit Graph that combines topological and geometrical information.
TAG~\cite{zhu2022tag} uses 447 industrial AMS circuits in sub-10nm technology. The data is processed with StarRC extraction tools to obtain placement coordinates and create spatial and text embeddings by annotating the netlists with instance names and device types.
For LLM-HD~\cite{chen2024llm}, the ICCAD 2012~\cite{torres2012iccad} and ICCAD 2020~\cite{hu2020iccad} benchmarks are used for layout hotspot detection, with GDSII layouts. The ICCAD 2012 dataset~\cite{torres2012iccad} focuses on metal layer hotspots, and ICCAD 2020~\cite{hu2020iccad} on via-layer patterns. The data is processed directly from GDSII to preserve spatial and geometric features, using semantic and hierarchical encoding.

\subsubsection{Encoding techniques for layout}

\ 

\Cref{fig:encoder-tl2} (b) illustrates the categories of layout encoders. In the supervised branch, Circuit GNN~\cite{yang2022versatile} customizes the GNN architecture to capture both the topology and geometry of the circuit layout. In the self-supervised branch, TAG~\cite{zhu2022tag} proposes text-graph multimodal learning, while LLM-HD~\cite{chen2024llm} focuses on leveraging LLMs for textual layout encoding. These approaches enable the models to effectively learn and represent the topology, geometry, and physical property of circuit layouts for downstream tasks.

\textbf{Supervised layout encoder with customized GNN architecture.}
In Circuit GNN~\cite{yang2022versatile}, the input modalities consist of topological data (from the netlist) and geometrical data (from the layout). These modalities are represented in a circuit graph, where cells and nets are the vertices, and topo-edges and geom-edges connect them. The model is built upon a GNN, which processes the heterogeneous circuit graph. The GNN utilizes message-passing to propagate information across both topological and geometrical edges. Topological information is passed through topo-edges, while geometrical information is transmitted via geom-edges. These messages are then fused to update the representations of cells and nets. During pre-training, the integration of topological and geometrical information is achieved through the message-passing paradigm, with topo-geom message-passing ensuring both types of data contribute to the final learned representation. Task-specific supervisions are employed to train the model.

\textbf{Self-supervised layout encoder with text-graph multimodal fusion.}
TAG~\cite{zhu2022tag} framework employs three primary modalities: (1) Text embedding, where the instance names and device/sub-circuit types from the circuit netlists are used as text inputs, processed using fastText to generate word embeddings, (2) Graph format, where the circuit is represented as a heterogeneous hierarchical graph encoding devices (nodes) and their connections (edges), including device types (e.g., NMOS, PMOS, capacitors) and hierarchical relationships between sub-circuits, and (3) Self-attention, where a multi-head self-attention layer is applied to the embeddings to capture global dependencies between instances within sub-circuits. The model architecture combines a GNN with self-attention to process both graph and text embeddings. GNN layers aggregate node information, while the self-attention mechanism ensures a global view of the circuit by considering the entire sub-circuit during training. During pre-training, the model is trained in an unsupervised manner with a focus on predicting the relative layout distance between instances within a sub-circuit. This distance prediction task is framed as a regression problem, where the embeddings are trained to predict the normalized relative distance between instances in manual layouts.

\textbf{Self-supervised layout encoder with text semantic encoding.}
In LLM-HD~\cite{chen2024llm}, the input modalities of this layout encoder include GDSII layout data and its semantic encoding. The GDSII data is transformed into a sequential format to make it suitable for processing by a language model. The key components include polygon shapes and spatial relationships between them, encoded as sequential tokens. The model architecture employs a BERT-based transformer specifically designed for layout patterns, utilizing multi-head self-attention to capture relationships between layout features both locally and globally. The architecture consists of an embedding layer, followed by LLM-HD~\cite{chen2024llm} layers, and concludes with a classification layer. During pre-training, the model uses masked language modeling, an unsupervised task where portions of the input data are randomly masked, and the model predicts the masked portions. This pre-training enables the model to learn representations of layout patterns, before fine-tuning for specific tasks such as hotspot detection.

\subsubsection{Downstream tasks for layout encoders}

\ 

Circuit GNN~\cite{yang2022versatile} supports both congestion prediction and net wirelength prediction tasks. For congestion, it predicts routing congestion during both the logic synthesis and placement stages, evaluated using correlations and classification metrics like precision, recall, and F1-score. 
TAG~\cite{zhu2022tag} handles three layout-stage tasks: layout matching prediction (binary classification of layout constraints, evaluated by accuracy, TPR, FPR, PPV, and F1-score), wirelength estimation (HPWL evaluated with R², MAE, and sMAPE), and net parasitic capacitance prediction (evaluated using R² and MAE).
LLM-HD~\cite{chen2024llm} focuses on hotspot detection, a binary classification task identifying layout areas prone to manufacturing defects.

\subsection{Summary of Trending Techniques for Advancing Circuit Encoders}
\label{sec:encoder-advanced}

\begin{figure}
    \centering
    \includegraphics[width=1\linewidth]{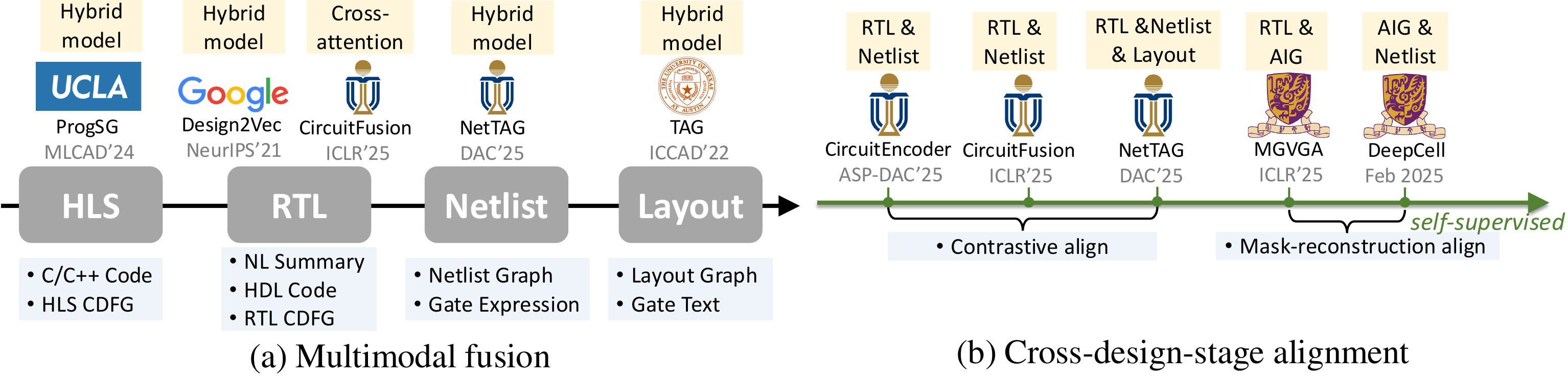}
    \vspace{-.3in}
    \caption{Timeline for multimodal fused and cross-stage aligned encoders.}
    \label{fig:encoder-advanced}
    \vspace{-.2in}
\end{figure}

\subsubsection{Trend 1: Customized ML model architecture and pre-training tasks for circuits}

\ 

\textbf{ML model architecture customized for circuits.}
To effectively capture the unique structural and functional properties of circuit data, various customized architectures have been developed, particularly in graph-based learning models. These architectures integrate specialized message-passing mechanisms to enhance the representation of circuit structures. For instance, GAMORA~\cite{wu2023gamora} and PolarGate~\cite{PolarGate} introduce customized GNN-based message passing tailored for AIGs, enabling the model to efficiently capture both Boolean functionality and structural connectivity.


\textbf{Pre-training tasks customized for circuits.}
Pre-training tasks for circuit encoders can be divided into supervised and self-supervised methods, with each method specifically designed to capture the unique properties of circuit data. In supervised learning, for example, DeepGate2~\cite{shi2023deepgate2} uses truth table supervision to train encoders by comparing the pairwise differences between truth tables of logic gates, thereby capturing the functional behavior of the circuit. This approach helps the encoder learn how different gates function in the context of their logic operations. On the other hand, for example, SNS v2~\cite{xu2023fast} introduces self-supervised learning for RTL encoders with functional contrastive learning. The model learns to cluster functionally similar circuits and separate dissimilar ones in the latent space. Another notable self-supervised pre-training task is masked circuit reconstruction, such as used in NetTAG~\cite{fang2025nettag}, where specific gates in a circuit’s netlist are masked, and the model learns to predict the missing gates based on the surrounding context. These pre-training tasks are vital for learning generalized representations that can be fine-tuned for various downstream tasks, such as design space exploration and functional verification.\looseness=-1


\subsubsection{Trend 2: circuit multimodal fusion}

\

Circuit design involves multiple modalities, including hardware description languages, graphical representations, and functional summaries, each capturing different aspects of the circuit. Recent works in circuit foundation models have focused on integrating these modalities through multimodal fusion techniques, enabling models to leverage both structural and semantic information for more robust representation learning. Two primary approaches have emerged in this area: hybrid ML model architecture that combines different encoders for various modalities and cross-attention-based fusion with self-supervised learning. The timeline of multimodal fused encoders is demonstrated in~\Cref{fig:encoder-advanced} (a).

\textbf{Multimodal fusion by hybrid models.} Hybrid models integrate distinct encoding architectures tailored to specific circuit modalities. For example, ProgSG~\cite{qin2024cross} and Design2Vec~\cite{vasudevan2021learning} focus on HLS and RTL-stage circuits, respectively, where the source code contains rich semantic information. These models employ LLMs to encode textual descriptions while using GNNs to capture structural information from control-data flow graphs. This dual-modality approach ensures that both functional intent and circuit topology are preserved in the learned representations. At the netlist and layout stages, where the netlist code provides limited functional information, NetTAG~\cite{fang2025nettag} and TAG~\cite{zhu2022tag} adopt a similar hybrid approach but with modifications suited for lower-level representations. NetTAG~\cite{fang2025nettag} extracts detailed symbolic logic expressions for each gate, encoding them using an LLM, while a GNN captures the circuit’s global structural dependencies. TAG~\cite{zhu2022tag} follows a similar strategy, leveraging textual attributes alongside graph-based structural encodings to improve netlist representation.

\textbf{Multimodal fusion by cross attention.} 
In addition to hybrid models, cross-attention-based fusion has been proposed as an alternative strategy for multimodal integration. CircuitFusion~\cite{fang2025circuitfusion}, designed for RTL-stage encoding, processes three primary modalities—HDL code, functional summaries, and structural graphs—in parallel. Each modality is first encoded independently, after which an additional multimodal fusion encoder with cross-attention mechanisms aligns and integrates the learned representations. The cross-attention mechanism ensures that the fused representation retains complementary information from all modalities while mitigating redundancy. CircuitFusion~\cite{fang2025circuitfusion} further enhances representation learning through self-supervised tasks such as masked summary modeling and embedding mixup, reinforcing the alignment between modalities.


\subsubsection{Trend 3: cross-design-stage alignment}

\ 

Cross-design-stage alignment has become a promising direction in circuit foundation models, enabling representations learned at earlier design stages (e.g., RTL) to be aligned with their corresponding lower-level implementations (e.g., netlist, layout). This alignment enhances generalizability, allowing models to better capture the functional and physical characteristics of circuits throughout the design process. Two primary approaches have been explored for achieving cross-stage alignment: contrastive learning-based alignment and mask-reconstruction-based alignment. The timeline of cross-stage aligned encoders is demonstrated in~\Cref{fig:encoder-advanced} (b).

\textbf{Cross-stage alignment via contrastive learning.}
Contrastive learning-based alignment has been effectively used for bridging different design stages by learning stage-invariant circuit representations. CircuitEncoder~\cite{fang2025circuitencoder} and CircuitFusion~\cite{fang2025circuitfusion} focus on RTL-to-netlist alignment by integrating structural and functional representations through self-supervised contrastive learning. These models enforce similarity constraints between functionally equivalent circuits across design stages, ensuring that embeddings capture both high-level design intent and low-level implementation details. NetTAG~\cite{fang2025nettag} extends this approach beyond RTL and netlist, incorporating layout information to enable RTL-netlist-layout alignment. By leveraging cross-modal contrastive learning, NetTAG~\cite{fang2025nettag} aligns representations across all three design stages, facilitating more accurate early-stage predictions of post-layout circuit characteristics.

\textbf{Cross-stage alignment via mask-reconstruction.}
Mask-reconstruction-based alignment, on the other hand, focuses on recovering masked portions of a circuit while maintaining logical and structural consistency across design stages. MGVGA~\cite{wu2025circuit} is designed for RTL-to-AIG alignment, where it employs Masked Gate Modeling to selectively mask gate-level details in AIG representations while preserving functional equivalence. This ensures that the learned embeddings retain both RTL-level semantics and AIG-level logic properties. DeepCell~\cite{shi2025deepcell} also employs this technique for AIG-to-netlist alignment, incorporating a self-supervised masking strategy to reconstruct standard cell representations from their lower-level gate descriptions. This approach enhances the model’s ability to understand structural variations while preserving functional equivalence across abstraction levels.

\clearpage
\newpage

\section{Foundation Model as a Circuit Decoder} 
\label{sec:ai-decoder}

\begin{figure}[!bh]
    \centering
{
     \small
     \resizebox{.93 \textwidth}{!}
    {\begin{forest}
for tree={
    grow'=east,
    draw,
    rounded corners,
    node options={font=\sffamily, fill=blue!20},
    edge={thick, color=black!600},
    l sep+=0.8cm,
    s sep+=0.1cm,
    parent anchor=east,
    child anchor=west,
    anchor=west,
    edge path={
        \noexpand\path [draw, \forestoption{edge}] (!u.parent anchor) -- +(5pt,0) |- (.child anchor)\forestoption{edge label};
    },
}
[\makecell{Foundation\\
model\\
as\\
circuit\\
decoder}, fill=black!20
    [\makecell{RTL code\\ generation\\
    (Section~\ref{sec:llm-rtl})}, for tree={fill=red!10}
            [\makecell{
            DAVE~\cite{pearce2020dave}, ChipGPT~\citep{chang2023chipgpt},
            VerilogEval~\citep{liu2023verilogeval},
            GPT4AIGChip~\citep{fu2023gpt4aigchip},
            \\
            Chip-Chat~\citep{blocklove2023chip}, AutoChip~\citep{thakur2023autochip},  
            RTLLM~\citep{lu2024rtllm}, VeriGen~\citep{thakur2024verigen},
            RapidGPT~\citep{RapidGPT-eda-aware}, CodeV~\cite{zhao2024codev}, 
            \\
        AutoVCoder~\cite{auto-v-coder},
         BetterV~\cite{pei2024betterv}, ChipNemo~\cite{liu2023chipnemo}, 
             Chang et al.~\cite{chang2024data}, OriGen~\cite{cui2024origen},   
             \\
             VerilogCoder~\cite{ho2024verilogcoder}, 
             Thakur et al.~\cite{thakur2023benchmarking}, 
             RTLCoder~\cite{liu2024rtlcoder}, MG-Verilog~\cite{zhang2024mg} \\
             CreativeEval~\cite{delorenzo2024creativeval}, VHDL-Eval~\cite{vijayaraghavan2024vhdl}, Chang et al.~\cite{chang2024natural}, 
             OPL4GPT~\cite{Tasnia2024opl4gpt}, 
             \\
             RTLSquad~\cite{wang2025rtlsquad}, MAGE~\cite{zhao2024mage}, 
             RTL-repo~\cite{allam2024rtl}, 
             OpenLLM-RTL~\cite{liu2024openllm}, 
             Sun et al.~\cite{sun2024classification}, 
             \\
             DeepRTL~\cite{liu2025deeprtl}, CraftRTL~\cite{liu2024craftrtl}, 
            Spec2RTL-Agent~\cite{yu2025spec2rtl}, RTLSeek~\cite{anonymous2025rtlseek}, VeriRL~\cite{teng2025verirl}, \\
            AutoSilicon~\cite{AutoSilicon}, FLAG~\cite{FLAG}, HyperPlace~\cite{HyperPlace}, 
            \cite{ 
            nair2023generating, nakkab2024rome, chain-of-descriptions,  zero-shot-rtl-code-generation-attention-sink-aug-llms,    make-every-move-count-llms-mcts, wang2024large, goh2024english, 
            pinckney2024revisiting, fang2025lintllm}, \cite{akyash2025rtl++, wang2025large, li2025deepcircuitx, li2025bridges, li2025bridges, ding2025iceberg, allam2025asic, ping2025hdlcore, openrtlset, TODAES-RTL-Blocklove, TODAES-RTL-Lopes, TODAES-Chang-Verilog-Gen}\\
            },
            for tree={fill=red!10}]
    ]
    [
    \makecell{HLS code generation\\
    (Section~\ref{sec:llm-hls-gen})}, for tree={fill=red!10}
    [
    \makecell{HLSPilot~\cite{xiong2024hlspilot}, 
         C2HLSC~\cite{collini2024c2hlsc},  
         SynthAI~\cite{sheikholeslam2024synthai}, \\
         Liao et al.~\cite{are-llms-any-good-for-hls},
         Gai et al.~\cite{gai2025exploring}
         , \new{LHS~\cite{LHS}, Qayyum et. al~\cite{todaes2024qayyum}}
         },
         for tree={fill=red!10}
    ]
    ]
     [
    \makecell{Design optimization\\
    (Section~\ref{sec:llm-opt})}, for tree={fill=pink!10}
    [
    \makecell{
    BetterV~\cite{ pei2024betterv},
    ChipGPT~\cite{chang2023chipgpt},
    RTLRewriter~\cite{yao2024rtlrewriter}, \\
    Martine et al.~\cite{martinez2024code},
    Xu et al.~\cite{xu2024optimizing}, Thorat et al.~\cite{thorat2023advanced},  \\
    Sandal et al.~\cite{zero-shot-rtl-code-generation-attention-sink-aug-llms},
    DeLorenzo et al.~\cite{make-every-move-count-llms-mcts} \\
    \new{HAPE~\cite{TODAES-HAPE}, HLSRewriter~\cite{HLSRewriter}, Yao et al.~\cite{TODAES-HLS-Optim-Yao}, ChatDSE~\cite{ChatDSE}}
    }
    ]
    ]
    [\makecell{Hardware \\ verification\\
    (Section~\ref{sec:llm-verification})}, for tree={fill=blue!15}
        [\makecell{ChipNeMo~\citep{liu2023chipnemo},
        RTLFixer~\citep{tsai2023rtlfixer},
        AutoSVA~\citep{orenes2023using},  \\
        NSPG~\citep{meng2023unlocking},
        DIVAS~\citep{paria2023divas}, SimEval~\cite{akyash2025simeval},  \\
        AssertLLM~\cite{fang2024assertllm},
        ChIRAAG~\cite{mali2024chiraag}, UVLLM~\cite{UVLLM},   \\
        LLM4DV~\cite{zhang2023llm4dv}, VerilogReader~\cite{ma2024verilogreader},  
        FVEval~\cite{kang2024fveval}, \\ OpenLLM-RTL~\cite{liu2024openllm}, AssertionBench~\cite{pulavarthi2024assertionbench},  \\
       \new{AssertionForge}~\cite{bai2025assertionforge}, \new{FVDebug~\cite{bai2025fvdebug}}, \new{ConfiBench~\cite{ConfiBench}},\\
       \new{Dick et al.~\cite{TODAES-SVA-Dataset}}
       \\
       \cite{kande2024security, 
        sun2023towards, liu2024domain, huang2024towards, kande2023llm, ahmad2023fixing, xiao2024llm, bhandari2024llm, evaluating-llms-for-hardware-design-test, zhou2025insights, pulavarthi2025llms}  \\
        }, for tree={fill=blue!12}]
    ]
    [
    \makecell{Circuit code debugging\\
    (Section~\ref{sec:llm-debug})}, for tree={fill=blue!12}
    [
    \makecell{
    MEIC~\cite{xu2024meic},  RTLFixer\cite{tsai2023rtlfixer}, VeriAssist~\cite{huang2024towards},  \\
    HDLdebugger~\cite{yao2024hdldebugger}, 
    Chrysalis~\cite{deming-chen-hls-debug},  \\ 
    Llm4sechw~\cite{fu2023llm4sechw}, \new{HLSDebugger~\cite{wang2025hlsdebugger}},  \cite{on-hardware-security-bug-code-fixes-by-prompting-llms, paria2023divas, qayyum2024bugs} \\}
    ]]
    [ 
    \makecell{Hardware security \\
    (Section~\ref{sec:llm-hardware-security})}, for tree={fill=blue!5}
    [\makecell{
    DIVAS~\cite{paria2023divas}, Saha et al.~\cite{saha2024empowering}, Self-HWDebug~\cite{akyash2024self}, \\
    Ahmad~\cite{ahmad2024hardware}, AutoSVA2~\cite{from-rtl-to-sva}, ChIAAG~\cite{mali2024chiraag},  \\
   Latibari et al.~\cite{latibari2024automated}, Netlist Whisperer~\cite{whisperer}, SCAR~\cite{srivastava2024scar}, \\
    Kande et al.~\cite{kande2024security},  Pearce et al.~\cite{pearce2023examining},
    \\~\cite{ahmad2023fixing, kokolakis2024harnessing, saha2024llm, wang2024llms, meng2023unlocking,
    paria2024navigating, tarek2024socurellm, kande2024llms, lin2023hw}, \\
    \new{HWREx~\cite{Todaes-Security-HWREx}}}]
    ]
        [
    \makecell{Design flow \& Layout\\
    (Section~\ref{sec:llm-flow})}, for tree={fill=orange!10}
    [
    \makecell{
    ChatEDA~\cite{he2023chateda}, SmartonAI~\cite{han2023new}, LLSM~\cite{huang2025llsm}, \\
    MetRex~\cite{abdelatty2025metrex}, DRC-Coder~\cite{chang2024drc}, \\ 
    ChipAlign~\cite{deng2024chipalign}, FabGPT~\cite{jiang2024fabgpt}, \\
    Chen et al.~\cite{chen2024intelligent}, Ho et al.~\cite{ho2024large},\\
    \new{Circuit Transformer~\cite{li2024circuit-transformer}, SEM-CLIP~\cite{jin2024semclip}, DCAFuse~\cite{DCAFuse-QiSun}}\\ 
    }
    ]]
    [
    \makecell{Architecture design
     \\
    (Section~\ref{sec:llm-hardware-architecture})}, for tree={fill=green!10}
    [
    \makecell{ AIGChip~\cite{fu2023gpt4aigchip}, ChatEDA~\cite{he2023chateda}, SpecLLM~\cite{li2024specllm}, \\
   \cite{ yan2023viability, liang2023unleashing, han2023new, pu2024customized}, \\
   \new{ChatArch~\cite{TODAES-ChatArch}, }
    }
    ]]
    [
    \makecell{Analog design
     \\
    (Section~\ref{sec:llm-analog})}, for tree={fill=yellow!10}
    [
    \makecell{ LADAC~\cite{Liu2024ladac}, AnalogCoder~\cite{lai2024analogcoder}, FLAG~\cite{mao2024flag}, \\ ADO-LLM~\cite{yin2024ado}, LaMAGIC~\cite{chang2024lamagic}, Artisan~\cite{chen2024artisan}, \\ LEDRO~\cite{kochar2024ledro}, AnalogXpert~\cite{zhang2024analogxpert}, AnalogGenie~\cite{topologiesanaloggenie}\\
    \new{Atelier~\cite{atelier}, Artisan~\cite{Artisan}}, \new{AnalogSeeker}~\cite{chen2025analogseeker}, \\
    \new{Golzan at al.~\cite{TODAES-Analog-Mix-Sig-Golzan}, LATENT~\cite{chaudhuri2025latent}} 
    }
    ]
    ]
]
\end{forest}}
}
\caption{Research tree of foundation models as circuit decoder, covered in Section~\ref{sec:ai-decoder}.}
\label{fig:tree}
\vspace{-.1in}
\end{figure}
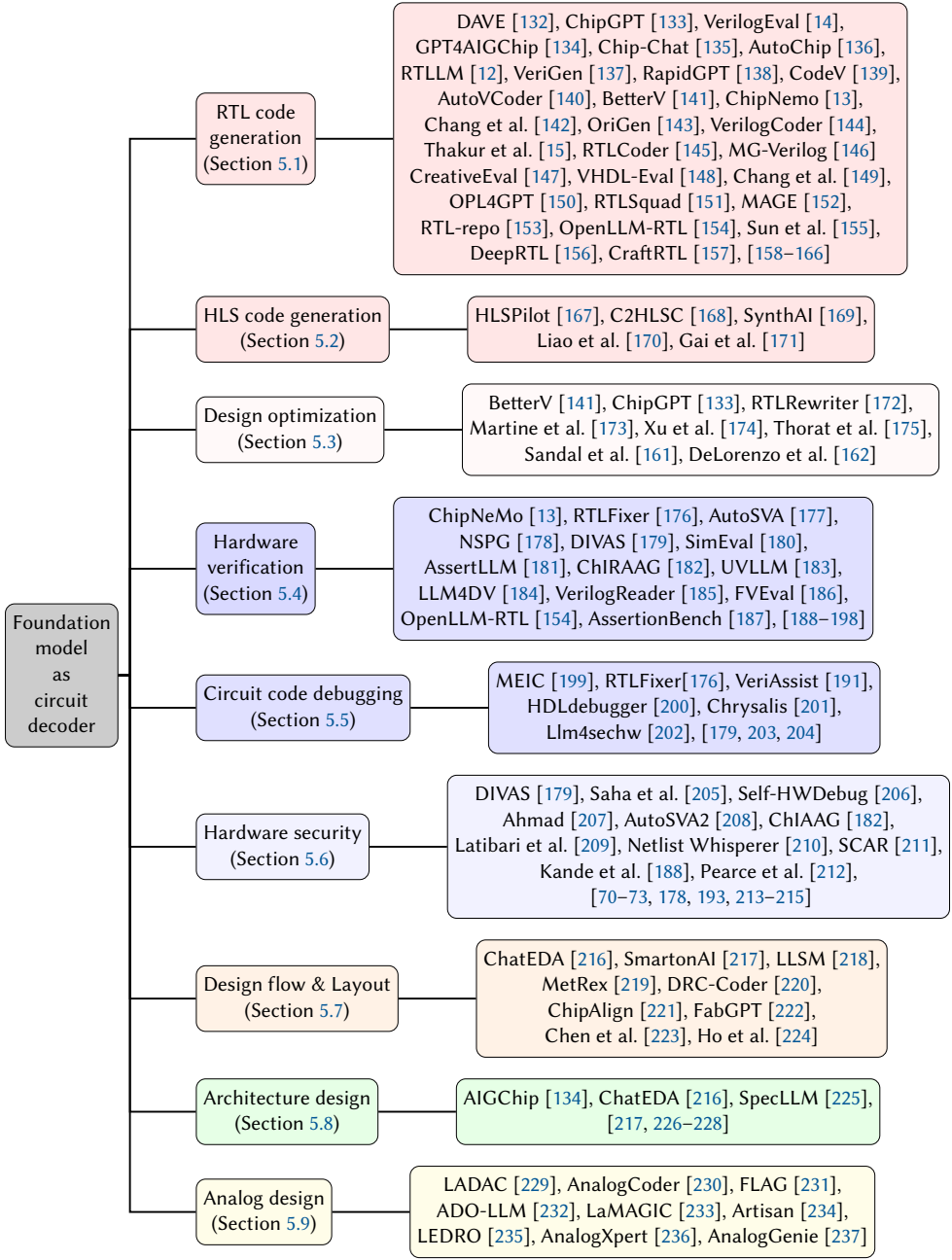


Another major paradigm of circuit foundation models is circuit decoders, which leverage LLMs for the automated generation of circuit-related content. These models facilitate the creation of RTL code (e.g., Verilog or VHDL), HLS code (e.g., SystemC or C++), design scripts (e.g., Tcl or Python), design descriptions, etc. As summarized in Figure~\ref{fig:tree}, this section provides a comprehensive overview of LLM-assisted circuit design techniques, categorizing them into 7 main directions according to their applications in generative EDA tasks:

\begin{enumerate}
    \item \textbf{LLM-assisted hardware code generation and optimization.} This category explores the application of LLMs in generating hardware code across different abstraction levels, such as RTL and HLS code. We will discuss the use of LLMs for RTL code generation in Section~\ref{sec:llm-rtl} and HLS code generation in Section~\ref{sec:llm-hls-gen}. Additionally, we will examine efforts in producing \emph{optimized} hardware in Section~\ref{sec:llm-opt}. Some works in this \emph{optimization} category overlap with the hardware code \emph{generation}. 
    \vspace{.03in}
    \item \textbf{LLM-assisted hardware code verification and debugging.} Beyond code generation, LLMs are employed to verify the correctness of hardware code (HLS or RTL) and to fix potential bugs. Section~\ref{sec:llm-verification} will cover the role of LLMs in hardware code \emph{verification}, while Section~\ref{sec:llm-debug} will focus on hardware \emph{debugging} techniques.
        \vspace{.03in}
    \item \textbf{LLM for hardware security.} 
    Works on \emph{hardware security} 
    focus on security-oriented design, debugging, and verification. We will delve into these topics in Section~\ref{sec:llm-hardware-security}, highlighting the unique challenges in security as distinct from general verification and debugging.        
        \vspace{.03in}
    \item \textbf{LLM for design flow automation and layouts.} Section~\ref{sec:llm-flow} explores the application of LLMs in automating the \emph{design flow} based on natural-language instructions, as well as enhancing circuit \emph{layout} processing for improved manufacturability.
        \vspace{.03in}
        
    \item \textbf{LLM for hardware architecture design.} Section~\ref{sec:llm-hardware-architecture} addresses the application of LLMs at a higher level of abstraction, focusing on design \emph{architecture} and \emph{specifications}. This includes applications for memory design and AI accelerators. 
    
    \vspace{.03in} 
    
    \item \textbf{LLM for analog circuit design.} Beyond the scope of digital VLSI design, LLM-assisted \emph{analog circuit} design is another important direction. 
    In Section~\ref{sec:llm-analog}, we will explore how LLMs can benefit analog circuit design, highlighting the significance of this area in the broader context of hardware development.
\end{enumerate}

\subsection{LLM for RTL Code Generation}
\label{sec:llm-rtl}

RTL design is a crucial step in the whole VLSI design process. This process defines the expected behavior of circuits with hardware description languages (HDLs) like Verilog and VHDL. However, RTL design remains a manual, time-consuming, tedious, and error-prone task. Recently, leveraging LLMs for RTL generation offers a promising automated solution. Specifically, LLM solutions can directly generate expected design RTL in HDL code, typically based on design descriptions in natural language as LLM input. 
Such \emph{RTL code generation} is the most extensively explored application of LLM-assisted EDA techniques. The existing works contribute primarily in two ways: 1) new \emph{benchmarks} evaluating LLM performance, covered in Section \ref{subsub:becnhmark} and listed in Table~\ref{tab:benchmark}; and 2) new \emph{LLM solutions} on RTL code generation, covered in Section \ref{subsub:code-gen} and listed in Table~\ref{tab:llm-rtl} and Figure~\ref{fig:decoder}.


\subsubsection{RTL code generation benchmarks}
\label{subsub:becnhmark}
\

\begin{figure}[!b]
    \centering
    \includegraphics[width=0.9\linewidth]{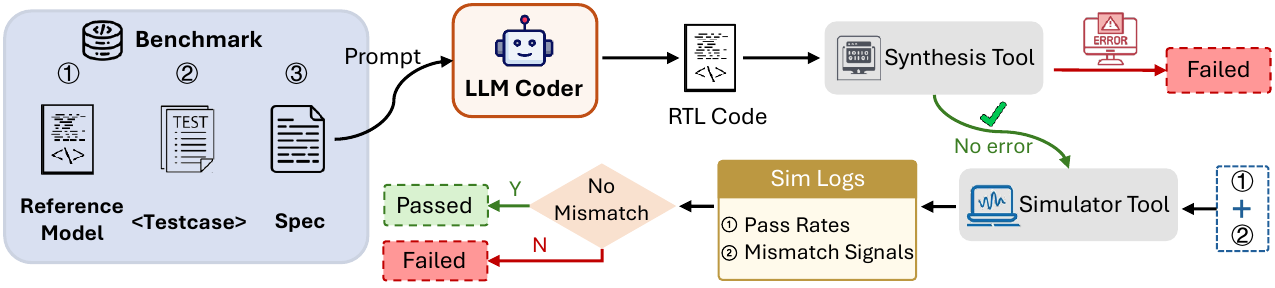}
    \caption{Illustration of the benchmark on LLMs for RTL generation, recent works are covered in Section~\ref{sec:llm-rtl}. Benchmark workflow comprises three steps: 1) LLMs generate RTL code from specifications, 2) The code is input into a synthesis tool to identify syntax errors, and 3) A simulation process checks for mismatches against predefined reference models or test case golden results.}
    \vspace{-.1in}
    \label{fig:rtl-gen-benchmark}
\end{figure}

As LLMs become popular for RTL design generation, \emph{benchmarks} become crucial for assessing the accuracy, efficiency, and reliability of LLM-based solutions for circuits. 
We summarize all benchmarks on RTL generation in Table~\ref{tab:benchmark}, among which 
RTLLM ~\cite{lu2024rtllm} and VerilogEval~\cite{liu2023verilogeval} are two \emph{pioneering and most widely-adopted} benchmarks for evaluating RTL code generation. \Cref{fig:rtl-gen-benchmark} illustrates the evaluation process of the RTL code generation benchmarks. 
A typical benchmark~\cite{lu2024rtllm, liu2023verilogeval} will provide dozens of design cases, each corresponding to one small circuit design or component. For each case, the benchmark will provide three types of files: 1) design descriptions as the LLM input, 2) test benches to verify the correctness of LLM-generated HDL code, and 3) the correct HDL code (i.e., reference model), typically handcrafted by designers, as a reference. 


\begin{table}[!t]
    \centering
    \resizebox{0.9 \textwidth}{!}
    {\begin{tabular}{c|c|c|c}
    \toprule
    \multicolumn{4}{c}{\textbf{Benchmarks for RTL Code Generation}}  \\
        \hline
        Benchmarks & Open-sourced & link & Date\\
        \hline
        \hline
        RTLLM~\cite{lu2024rtllm, liu2024openllm} & \checkmark &  \url{https://github.com/hkust-zhiyao/rtllm} & 2023-10 \\
        \hline
        VerilogEval~\cite{liu2023verilogeval} & \multirow{2}{*}{\checkmark} & \multirow{2}{*}{\url{https://github.com/NVlabs/verilog-eval}} & 2023-12 \\
        VerilogEval v2\cite{pinckney2024revisiting} &  & & 2024-08 \\
        \hline
        CreativeEval~\cite{delorenzo2024creativeval} & \checkmark & \url{https://github.com/matthewdelorenzo/creativeval} & 2024-04\\
        RTL-repo~\cite{allam2024rtl} & \checkmark & \url{https://github.com/AUCOHL/RTL-Repo} & 2024-05 \\
        VHDL-Eval~\cite{vijayaraghavan2024vhdl} &  & & 2024-06 \\
        ChatGPTV~\cite{chang2024natural} & \checkmark & \url{https://github.com/aichipdesign/chipgptv} & 2024-11\\
        \hline

        \new{MMCircuitEval}~\cite{zhao2025mmcircuiteval} &\new{\checkmark} & \url{https://xywen97.github.io/MMCircuitEval/} & \new{2025-05}\\
        \new{CVDP~\cite{Pinckney2025ComprehensiveVD}} & \new{\checkmark} & \url{https://github.com/NVlabs/cvdp_benchmark} & \new{2025-06}\\
        \bottomrule
    \end{tabular}}
    
    \caption{\new{Collection of benchmarks on LLMs for RTL generation in Section~\ref{sec:llm-rtl}. VerilogEval~\cite{liu2023verilogeval} and its second version~\cite{pinckney2024revisiting} share the same open-source link.}}
    \vspace{-.2in}
    \label{tab:benchmark}
\end{table}

RTLLM~\cite{lu2024rtllm} is one of the first benchmarks on design RTL generation based on natural language descriptions. It introduces a comprehensive evaluation framework with three progressive goals: syntax correctness, functionality correctness, and design quality (i.e., PPA metrics). The benchmark includes 30 diverse designs, ranging from simple arithmetic circuits to complex systems like a RISC CPU, and provides automated evaluation pipelines with natural language descriptions, testbenches, and human-crafted reference designs. RTLLM also adopts a self-planning prompt engineering technique, which significantly improves the performance of GPT-3.5~\cite{chatgpt} by decomposing the RTL generation task into planning and code generation steps. The latest version (i.e., RTLLM 2.0) is available in OpenLLM-RTL~\cite{liu2024openllm} and expands the benchmark to 50 designs.

VerilogEval~\cite{liu2023verilogeval} is the other pioneering benchmark on RTL generation based on natural language descriptions. It comprises 156 problems sourced from HDLBits, covering a wide range of topics from combinational circuits to finite state machines. VerilogEval offers two types of problem descriptions: machine-generated (using LLMs) and human-curated, ensuring clarity and reducing ambiguity. The benchmark provides an automated testing environment using the ICARUS Verilog simulator and employs the pass@k metric to evaluate functional correctness. Additionally, VerilogEval explores supervised fine-tuning (SFT) with a synthetic dataset of 8,502 problem-code pairs, demonstrating that fine-tuning can enhance LLM performance, especially for models not originally trained on Verilog. 
Based on VerilogEval~\cite{liu2023verilogeval}, VerilogEval v2~\cite{pinckney2024revisiting} evaluates the performance of new models and enhances the infrastructure and further discusses the importance of prompt engineering for RTL generation task.

In addition to the widely adopted RTLLM and VerilogEval benchmarks, there are several other benchmarks that assess LLMs in RTL code generation. CreativeEval~\cite{delorenzo2024creativeval} evaluates LLM creativity in Verilog generation based on fluency, flexibility, originality, and elaboration, finding GPT-3.5 to be the most creative among tested models. RTL-Repo~\cite{allam2024rtl} collects 4,000 GitHub samples to evaluate LLMs on \emph{`long-range dependency handling'} ability through several metrics. However, the dataset lacks corresponding functional specifications, preventing it from being considered a standard benchmark. VHDL-Eval~\cite{vijayaraghavan2024vhdl} addresses the lack of VHDL-specific benchmarks, offering 202 problems with self-verifying testbenches to evaluate functional correctness through zero-shot generation and fine-tuning. ChatGPTV~\cite{chang2024natural} introduces a multi-modal benchmark for Verilog synthesis, incorporating visual inputs to improve LLM performance in handling spatial circuit complexity, showing significant accuracy gains over text-only approaches.

\begin{table}[!t]
    \centering
    \resizebox{1.0 \textwidth}{!}
    {\begin{tabular}{c|c|c|c|c}
    \toprule
    \multicolumn{5}{c}{\textbf{LLM for RTL Code Generation}}  \\
        \hline
        Works & Open & Link  & Date & Method\\
        \hline
        \hline
        Nair et a.~\cite{ 
            nair2023generating}* & & & 2023-02 &\\
            
          Enrique et al.~\cite{enrique2023adeeplearning} & \checkmark  &\url{https://github.com/99EnriqueD/verilog_autocompletion} & 2023-04 &  \\ChipGPT~\cite{chang2023chipgpt} (Opt) & & & 2023-06 &\\
         
         RTLLM~\cite{lu2024rtllm} & \checkmark &  \url{https://github.com/hkust-zhiyao/rtllm} & 2023-08 & \\
         AutoChip~\cite{thakur2023autochip} & \checkmark  &  \url{https://github.com/shailja-thakur/AutoChip} & 2023-11 & Prompt \\
        
         Chip-Chat~\cite{blocklove2023chip} & & & 2023-11 & engineering\\ 

         Sandal et al.~\cite{zero-shot-rtl-code-generation-attention-sink-aug-llms} (Opt) & & & 2024-01 & \\
          Goh et al.~\cite{goh2024english} & & & 2024-03 & \\ 

          VerilogCoder~\cite{ho2024verilogcoder} & & & 2024-08 & \\

          AIVRIL2~\cite{RapidGPT-eda-aware} & & & 2024-09  &  \\

         Nakkab et al.~\cite{nakkab2024rome} & & & 2024-09 &\\
         
          Vijayaraghavan et al.~\cite{chain-of-descriptions} & & & 2024-09 & \\
         
         MAGE~\cite{zhao2024mage} & & & 2024-12 & \\

         RTLSquad~\cite{wang2025rtlsquad} & \checkmark & \url{https://github.com/observerw/RTLSquad} & 2025-01 & \\

         VRank~\cite{zhao2025vrank} & & & 2025-01 & \\

        \cline{1-1}
          \cline{4-5}
          \multirow{2}{*}{DeLorenzo et al.~\cite{make-every-move-count-llms-mcts} (Opt)} & & & \multirow{2}{*}{2024-02} & Monte Carlo \\
          & & & & tree search \\

        \hline
        \hline
        DAVE~\cite{pearce2020dave} & \checkmark &  & 2020-11 & \\
        VerilogEval~\cite{liu2023verilogeval} & \checkmark & \url{https://github.com/NVlabs/verilog-eval} & 2023-09 &
         SFT with\\
        ChipNemo~\cite{liu2023chipnemo} & & & 2023-10 & private \\
        BetterV~\cite{pei2024betterv} (Opt) & & & 2024-02 & data \\
        Chang et al.~\cite{chang2024data} & & & 2024-05 
         
         & \
        
        \\
        \cline{4-5}
        VeriSeek\cite{wang2024large} &\checkmark & \url{https://huggingface.co/LLM-EDA/VeriSeek} & 2024-08 & SFT with RL \\
        \cline{4-5}

        DeepRTL~\cite{liu2025deeprtl} & & & 2025-02 & Representation \\
        & & & & Learning \\
        \cline{4-5}

        \hline
        \hline
        VeriGen~\cite{thakur2024verigen} & \checkmark & \url{https://github.com/shailja-thakur/vgen} & 2023-07 & UFT \\

        \hline
        
        RTLCoder~\cite{liu2024rtlcoder} & \checkmark & \url{https://github.com/hkust-zhiyao/RTL-Coder}   & 2023-12 &  \\
        CodeV~\cite{zhao2024codev} & \checkmark & \url{https://github.com/IPRC-DIP/CodeV} & 2024-07 &  \\
        MG-Verilog~\cite{zhang2024mg} & \checkmark & \url{https://github.com/GATECH-EIC/mg-verilog} & 2024-07 & SFT with \\
        AutoVCoder~\cite{auto-v-coder} & \checkmark & \url{https://github.com/sjtu-zhao-lab/AutoVCoder} & 2024-07 &  open-sourced \\
        Origen~\cite{cui2024origen} & \checkmark & \url{https://github.com/pku-liang/OriGen} & 2024-09 &  data \\
        
        CraftRTL~\cite{liu2024craftrtl} & \checkmark & \url{https://github.com/NVlabs/CraftRTL} & 2024-09 & \\
        \hline
        \new{VeriRL}~\cite{teng2025verirl} & \checkmark & \url{https://github.com/omniAI-Lab/VeriRL} & \new{2024-09} & \new{\multirow{2}{*}{RL}} \\
        \new{VeriReason}~\cite{wang2025verireason} & \new{\checkmark} & \url{https://nellyw8.github.io/VeriReason/} & \new{2025-05} & \\
        
        \bottomrule
    \end{tabular}}
    \caption{Works on LLMs for RTL generation. In the `Works' column, denotation `*' refers to works on security, while `(Opt)' means the work focuses on design optimization. Though VerilogEval~\cite{liu2023verilogeval, pinckney2024revisiting} is a benchmark (in Table~\ref{tab:benchmark}), it also proposes SFT with problem-pair pairs, while the training dataset is not open-sourced.  DeLorenzo et al.~\cite{make-every-move-count-llms-mcts} introduce an RTL generation framework that integrates the MCTS sampling process, which we classify as a form of prompt engineering since it solely alters the inference process, without necessitating fine-tuning. 
    Some works provide code through GitHub but don't provide open-source models or training datasets. We classify them as SFTs with private data.} 
    \label{tab:llm-rtl}
    \vspace{-.25in}
\end{table}

\begin{table}[!t]
    \centering
    \resizebox{1.0 \textwidth}{!}
    {\begin{tabular}{c|c|c|c|c|c}
    \toprule
    \multicolumn{6}{c}{\textbf{Datasets for Training of RTL Generation}}  \\
        \hline
        Dataset & Open & Link  & Date & Origin \& Feature & Amount \\
        \hline
        \hline
        VeriGen~\cite{thakur2024verigen} & \checkmark & \url{https://github.com/shailja-thakur/vgen}  & 2023-07 & Github \& Unsupervised & [Not listed]]\\
        \hline

        RTLCoder~\cite{liu2024rtlcoder} & \checkmark & \url{https://github.com/hkust-zhiyao/RTL-Coder} & 2023-11 & Synthesized \& SFT & 27K \\

        ChipGPTSeriers~\cite{chang2024data} & \checkmark & \url{https://modelscope.cn/datasets/changkaiyan/chipgptseries} & 2023-12 & synthesized \& SFT & 124K \\

        BetterV~\cite{pei2024betterv} & & & 2024-02 & - & [Not listed]\\

        OriGen~\cite{cui2024origen} & \checkmark & \url{https://github.com/pku-liang/OriGen} & 2024-09 & Synthesized \& SFT  & 222K \\

        AutoVCoder~\cite{auto-v-coder} & & \url{https://github.com/sjtu-zhao-lab/AutoVCoder} & 2024-07 & Github & [Notlisted] \\

        CodeV~\cite{zhao2024codev} && \url{https://github.com/IPRC-DIP/CodeV} & 2024-07 & Synthesized & 165K \\

        CraftRTL~\cite{liu2024craftrtl} &  & \url{https://github.com/NVlabs/CraftRTL} & 2024-09 & Synthesized \& SFT & 80.1K \\

        RTL++~\cite{akyash2025rtl++} & \checkmark & \url{https://huggingface.co/datasets/makyash/RTL-PP} & 2025-05 & Synthesized \& SFT & 200K \\
        
        \hline

        VeriRL~\cite{teng2025verirl} & \checkmark & \url{https://github.com/omniAI-Lab/VeriRL} & 2025-07 & Synthesized \& RL & 53K \\
        
        \bottomrule
    \end{tabular}}
    \caption{\new{Datasets for finetuning on RTL generation.}} 
    \label{tab:llm-rtl-dataset}
    \vspace{-.3in}
\end{table}


\subsection{\new{Training datasets for RTL generatin}}
\label{sec:train-dataset}

\new{The recent year (2025) has witnessed a tremendous boom in new training datasets for RTL generation. We list the most represented datasets in Table~\ref{tab:llm-rtl-dataset}. The earliest dataset (VeriGen~\cite{thakur2024verigen}) uses an unlabeled dataset with an unsupervised fine-tuning strategy. Such an unsupervised dataset is easy to collect but has limited effectiveness on fine-tuning. RTLCoder~\cite{liu2024rtlcoder} and GPTSeriers~\cite{chang2024data} are pioneering works on generating a supervised dataset, featuring \{instruction, reference code\} pairs. The supervised dataset demonstrates strong effectiveness in improving LLM performance. However, generating such a pair-wise dataset is challenging, requiring a significant amount of API calls to commercial LLMs. Additionally, the correctness of the reference code is hard to verify. Targeting the challenge, the following works~\cite{cui2024origen, auto-v-coder, akyash2025rtl++} propose a set of strategies to improve the quality of the generated dataset. For example, OriGen~\cite{cui2024origen} proposes the \emph{code-to-code} dataset augmentation strategy, and RTL++~\cite{akyash2025rtl++} proposes using graph information from the RTL structure to assist the generation of design descriptions. Recently, VeriRL~\cite{teng2025verirl} uses a smaller amount of high-quality data with test sets for RL training. This work uses feedback from the testbench as a reward to fine-tune the model and achieves state-of-the-art performance.}

\subsubsection{RTL code generation techniques}
\label{subsub:code-gen}
\

\begin{figure}
    \centering
    \includegraphics[width=1.0\linewidth]{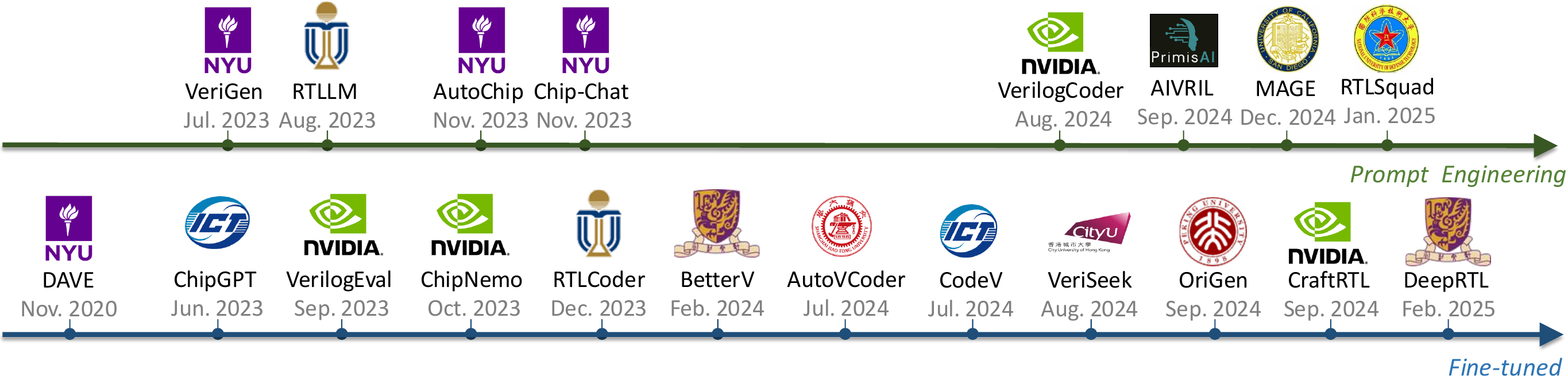}
    \vspace{-.2in}
    \caption{Timeline of works on RTL generation, covered in Section~\ref{sec:llm-rtl}. The timeline also includes the works of generation for design optimization (e.g., BetterV~\cite{pei2024betterv}). Recent works (e.g., Origen~\cite{cui2024origen} and VerilogCoder~\cite{ho2024verilogcoder}) show a trend of utilizing feedback from EDA tools to improve generation quality.}
    \label{fig:decoder}
    \vspace{-.1in}
\end{figure}

\emph{RTL code generation} is the most extensively explored application of LLM-assisted EDA techniques. Given the growing body of work in this area, we categorize existing approaches into four distinct strategies. 
Table~\ref{tab:llm-rtl} and~\Cref{fig:decoder} list the comparison and timeline of all related works, respectively. 

\begin{enumerate}

\item \textbf{Prompt engineering.} This approach designs specific prompts to guide LLMs in generating RTL outputs. The effectiveness of the generated code largely hinges on the quality of these prompts, which often requires iterative refinements and experimental trial-and-error. Well-crafted prompts can lead to correct and even high-quality RTL code. 

\item \new{\textbf{Advanced agentic systems for RTL generation.} With the emergence of LLMs' application on agentic systems for versatile tasks, there is a boost in research on agentic RTL generation.}

\item \textbf{LLMs trained on \emph{private} datasets with \emph{instruction-code pairs}.} This approach fine-tunes LLMs (based on already pre-trained LLMs) using proprietary data, such as industrial in-house circuit designs. It tailors models to an organization's needs, enhancing performance, but requires significant resources and access to high-quality private data. 

\item \textbf{LLMs trained on \emph{open} datasets with \emph{code only}.} This approach uses open-source codebases to fine-tune LLMs, eliminating the need for labor-intensive, high-quality datasets that require manual annotation. Such an unsupervised fine-tuning process can help LLMs capture the inherent structures of RTL code, but is less effective for instruction-following tasks.

\item \textbf{LLMs trained on \emph{open} datasets with \emph{instruction-code pairs}.} Fine-tunes models on pairs of design specifications and RTL implementations, helping them translate specifications into code. This requires a large number of high-quality pairs, which can be challenging to obtain but is quite effective in boosting LLMs' ability on hardware code generation tasks. 

\item \new{\textbf{LLMs finetuned by reinforcement learning (RL) techniques}. Reinforcement learning has demonstrated effectiveness in LLMs' training, especially in code generation. There are also new explorations on RL for RTL generation.}

\end{enumerate}

The first strategy, prompt engineering, primarily leverages commercial LLMs via API calls. In contrast, the other three strategies focus on customizing local LLMs by fine-tuning pre-trained models, mostly from open-source communities. Each strategy has its own advantages and drawbacks. Commercial models (e.g., GPT) reduce the substantial costs of training and deploying LLMs but may raise security and intellectual property concerns. On the other hand, fine-tuning local open-sourced LLMs (e.g., Llama, DeepSeek) can address these security and IP issues but requires significant resources and limits the model size. Smaller-scale customized LLMs tend to be less general compared with large commercial solutions. \new{As the advancement of the commercial models and finetuning technologies, there are two booming strategies for RTL generation. \textbf{Agentic systems} for RTL generation are considered a promising direction for automated RTL generation targeting real industrial applications. The representative work targeting this direction is Spec2RTL-Agent~\cite{yu2025spec2rtl} by Nvidia. Another promising direction is \textbf{RL for RTL generation}~\cite{teng2025verirl, cui2024origen, zhao2024codev}, where the specific RL strategy is applied to finetune LLMs on the RTL generation task and demonstrates significant effectiveness.}

The four distinct strategies for RTL generation using LLMs present unique advantages and challenges. \textbf{Prompt engineering} focuses on crafting precise prompts to guide LLMs in generating RTL code, which usually involves iterative refinement with EDA tools. In contrast, \textbf{LLMs trained on private datasets} (`\emph{Supervised Fine-Tuning}', denoted as `SFT' in Table~\ref{tab:llm-rtl}) enhance model performance by fine-tuning them with proprietary data tailored to specific organizational needs. However, this method demands computational resources for fine-tuning, and the private circuit dataset is not open-sourced to facilitate the advancement of the community. Alternatively, \textbf{LLMs trained on open datasets with code only} (`\emph{Unsupervised Fine-Tuning}', denoted as `UFT' in Table~\ref{tab:llm-rtl}) leverage open-source codebases for unsupervised fine-tuning, reducing the need for labor-intensive dataset preparation. However, this approach is less effective for RTL generation, which requires strict adherence to specific instructions including design descriptions. The UFT trains LLMs to predict the next token based on the preceding context, which aids LLMs in grasping code syntax, but is limited in understanding required functional specifications. Consequently, this UFT approach is typically less effective and thus less adopted. 
Finally, \textbf{LLMs trained on open datasets with instruction-code pairs} (these methods also involve \emph{Supervised Fine-Tuning}, denoted as `SFT') utilize pairs of design specifications and RTL code to train the LLMs, and the dataset is open-sourced. This last strategy is both effective and benefits the community with open-source datasets.



\begin{figure}[!t]
    \centering
    \includegraphics[width=0.95\textwidth]{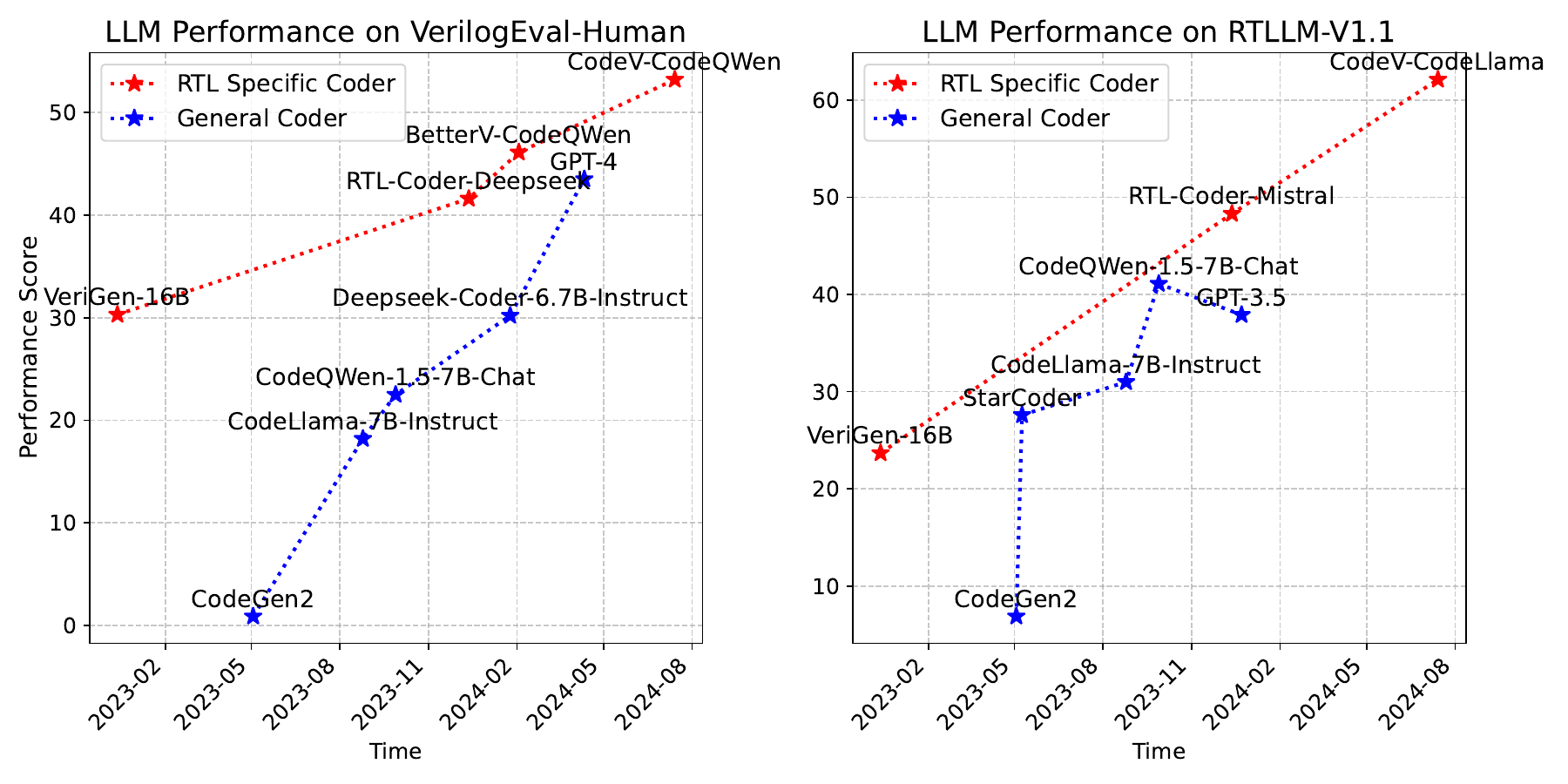}
    \vspace{-.2in}
    \caption{Performance comparison of various approaches to RTL generation tasks. Involved works are covered in Section~\ref{sec:llm-rtl}. The first figure illustrates the performance of different methods over time as evaluated by VerilogEval~\cite{liu2023verilogeval}, while the second figure presents the results obtained from RTLLM~\cite{lu2024rtllm}.}
    \label{fig:performance-comparison}
    \vspace{-.1in}
\end{figure}

Figure~\ref{fig:performance-comparison} compares the performance of various models on VerilogEval-Human~\cite{liu2023verilogeval} and RTLLM~\cite{lu2024rtllm} over time, including both general-purpose coding LLMs (e.g., DeepSeek-Coder, CodeLlama) and RTL-specific coding LLMs (e.g., RTL-Coder~\cite{liu2024rtlcoder}, BetterV~\cite{pei2024betterv}). This comparison highlights the advancements in Verilog code generation, illustrating how domain-specific fine-tuning and architectural modifications enhance the effectiveness of LLMs in hardware design automation.

\begin{figure}
    \centering
    \includegraphics[width=0.88\linewidth]{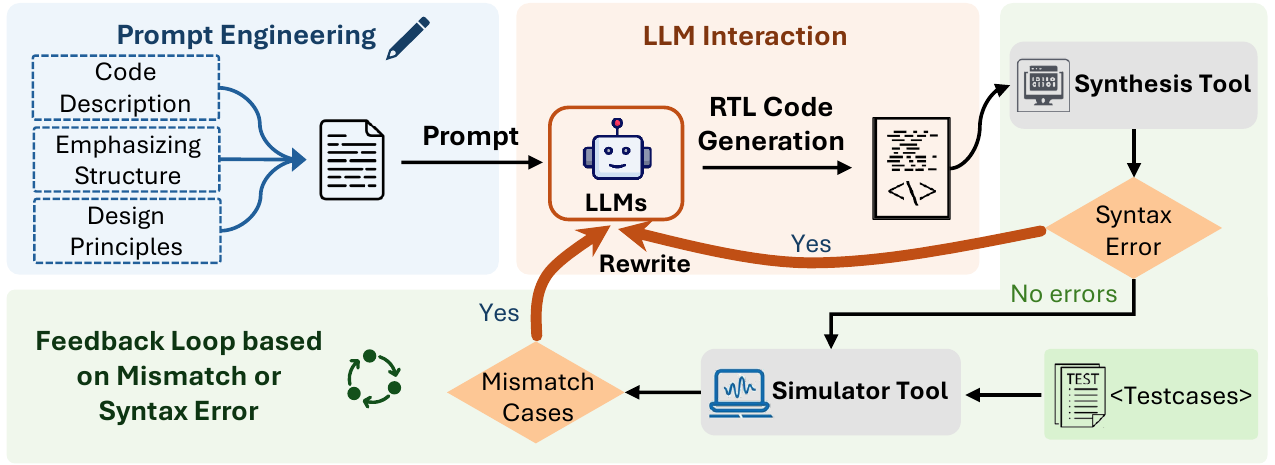}
    \caption{Illustration of the basic flow of RTL design using prompt engineering. Related works are covered in Section~\ref{sec:llm-rtl}. The design specification will be combined with manually designed structure analysis and design principles (in VerilogCoder~\cite{ho2024verilogcoder}), and then the LLMs will take the prompt to generate corresponding RTL code. Most early works (RTLLM~\cite{lu2024rtllm}, AutoVCoder~\cite{auto-v-coder}) on RTL generation stop at this step. The more recent works incorporate the feedback from EDA tools into the design flow, for example, utilizing the error information and mismatch log to prompt LLMs for rewriting.}
    \label{fig:prompt-engineering-flow}
    \vspace{-.1in}
\end{figure}

\textbf{Strategy 1: Prompt engineering.} 
Prompt engineering is one of the earliest strategies used for RTL generation due to its simplicity and effectiveness in applying LLMs to circuit design, including commercial LLMs. While recent efforts have focused on developing new fine-tuned LLMs for RTL generation, research about prompt engineering continues to evolve.
Prompt engineering~\cite{survey-on-prompt-engineering} mainly focuses on designing and optimizing input prompts to effectively communicate with LLMs and elicit desired design generations. The advantage of this methodology is the elimination of the requirement for fine-tuning and adaptability across different LLMs. Tons of works~\cite{chang2023chipgpt, blocklove2023chip, lu2024rtllm, thakur2023autochip, nair2023generating, nakkab2024rome, RapidGPT-eda-aware, enrique2023adeeplearning} have explored applying and customizing advanced prompt engineering techniques for RTL code generation by LLMs, as demonstrated in~\Cref{fig:prompt-engineering-flow}.

Chip-Chat~\cite{blocklove2023chip}, as a pioneering work, investigates the use of conversational LLMs, such as OpenAI's ChatGPT, in translating natural language specifications into HDLs for circuit design. Through a case study, the authors explore the collaborative design of an 8-bit accumulator-based microprocessor with GPT-4. The methodology involves breaking down the design into subtasks managed through conversation threads, where GPT-4 generates Verilog code guided by a human engineer who verifies and refines the output. The study finds that while LLMs can produce high-quality code and act as effective design assistants, they require human oversight for specification corrections and struggle with verification tasks. This research highlights the potential of LLMs to enhance productivity in circuit design when used as a complement to human expertise.

Recently, VerilogCoder~\cite{ho2024verilogcoder} presents a novel framework utilizing multiple AI agents to automate Verilog code generation and correction. It introduces a task and circuit relation graph for structured task decomposition, ensuring the inclusion of essential signal and state transition details. The system incorporates an AST-based waveform tracing tool for debugging, allowing agents to identify and correct functional errors. Using the ReAct~\cite{yao2023react} technique, agents iteratively interact with Verilog tools, including syntax checkers and simulators, to refine the code. This methodology significantly enhances the automation of circuit design. 

\new{\textbf{Strategy 2: Advanced agentic systems for RTL generation.} Different from the prompt engineering techniques that rely only on a single LLM to implement the RTL code, agentic frameworks~\cite{zhao2024mage, ho2024verilogcoder, wang2025rtlsquad, yu2025spec2rtl} orchestrate more than one LLM (agent) to implement a single design cooperatively. One of the most representative works is MAGE~\cite{zhao2024mage}, which orchestrates four agents: (1) testbench agent, (2) RTL agent, (3) judge agent, and (4) debug agent. Different agents focus on a specific task. In this way, each agent performs better than a single-agent system. Additionally, the agents utilize feedback from tools, for e.g., the results from the testbench, and invoke the (4) debug agent to conduct self-correction. Such a multi-agent-based self-correction paradigm for RTL generation demonstrates significant effectiveness and encourages extensive exploration in this direction. Since such a paradigm relies solely on commercial LLMs and requires no additional fine-tuning. The multi-agent system at this early stage is still a static workflow, where each step clearly specifies the agent and the task. The recent agentic system enables more flexible usage of the LLM agents, where each agent has more choices of the next step. For e.g., Spec2RTL-agent~\cite{yu2025spec2rtl} implements a more complex agentic system, involving more agents. Each agent has more choices, for e.g., involving human intervention or self-reflection.}

\textbf{Strategy 3: LLMs trained on private dataset with instruction-code pairs.} 
Aside from prompt engineering for circuit design generation, another powerful technique is fine-tuning. Fine-tuning can be categorized into unsupervised fine-tuning and supervised fine-tuning. Here we focus on models that utilize supervised fine-tuning on private datasets. Many solutions in this domain are developed by industrial companies or in collaboration with industrial organizations. As a result, the models or datasets are often not open-sourced.
DAVE~\cite{pearce2020dave} presents a pioneering custom dataset generation process that employs a template-based approach to create instruction-code pairs. They frame the Verilog generation task as a machine translation problem, fine-tuning a GPT-2 model to produce Verilog code from English descriptions. The training dataset generation process utilizes ``Task/Result metastructure'' that outlines the type of digital design task and relevant details, together with templates representing various scenarios such as combinational assignments and registers. The generated dataset is not open-sourced but includes diverse task instances to aid in fine-tuning the model.
AutoVCoder~\cite{auto-v-coder} is a framework designed to improve the accuracy of LLMs in generating Verilog code. It addresses the challenges of low syntactic and functional correctness in LLM-generated RTL code by employing three key techniques: a high-quality hardware dataset generation method, a two-round LLM fine-tuning process, and a domain-specific RAG mechanism. The framework uses a code scorer to filter a large dataset of Verilog code from GitHub and generates a synthetic dataset using GPT-3.5. The two-round fine-tuning leverages these datasets, and the RAG module is designed to enhance the process by providing relevant context during code generation.
OriGen~\cite{cui2024origen} is a new open-source framework for generating RTL Verilog code. It addresses the limitations of existing open-source LLMs by incorporating a novel code-to-code augmentation technique and a self-reflection mechanism. The augmentation method uses a commercial LLM (Claude3-Haiku) as a ``teacher'' to improve the quality of open-source RTL datasets. To address the scarcity of high-quality Verilog data, CodeV~\cite{zhao2024codev} leverages the observation that LLMs excel at summarizing Verilog code, rather than generating it from scratch. The system operates by first collecting and filtering a large corpus of high-quality Verilog modules from open-source repositories. These modules are then fed into GPT-3.5, which generates multi-level summaries—detailed functional descriptions and higher-level problem statements—for each module. These description-code pairs form a high-quality dataset used to fine-tune base LLMs (CodeLlama, DeepSeekCoder, and CodeQwen), resulting in the CodeV series of models.\looseness=-1

DeepRTL~\cite{liu2025deeprtl} introduces a unified representation model to enhance both understanding and generation of Verilog code. The model addresses limitations in previous approaches, which focus primarily on Verilog code generation, neglecting the critical task of understanding. DeepRTL~\cite{liu2025deeprtl} is fine-tuned on a comprehensive dataset that aligns Verilog code with multi-level natural language descriptions, covering line, block, and module levels with both detailed and high-level functional descriptions. The dataset includes both open-source and proprietary Verilog code, annotated using a CoT approach with GPT-4 and verified by human experts. The authors introduce a novel benchmark for Verilog understanding and propose using semantic evaluation metrics like embedding similarity and GPT score, which capture semantic coherence more effectively than traditional methods like BLEU and ROUGE. Additionally, the paper employs curriculum learning, allowing the model to incrementally build knowledge from simpler to more complex tasks, enhancing its performance in both understanding and generation of Verilog code.

\textbf{Strategy 4: LLMs trained on open datasets with code only.}
In the initial phase of fine-tuning LLMs for RTL design generation, early efforts primarily relied on \emph{unsupervised} data sourced from platforms like GitHub and other open-source code repositories. The key benefit of using unsupervised datasets is their ease of training and the large volume of existing unlabeled data, eliminating the need for labor-intensive labeling tasks. However, the limited label alignment of these datasets limits their effectiveness in training LLMs for RTL generation.
These techniques mainly enable LLMs to understand language patterns, structures, and semantics by processing vast amounts of text data. LLMs are tasked to predict the next token given the previous context~\cite{chatgpt}. 

For example, VGen~\cite{thakur2023benchmarking}, as the pioneering work, first evaluates the ability of unsupervised LLMs to generate Verilog code, a critical aspect of circuit design. The authors fine-tune several pre-trained LLMs on a large dataset of Verilog code collected from GitHub and textbooks, creating the largest training corpus for this purpose. They develop an evaluation framework that includes test benches for assessing both the syntactic and functional correctness of the generated code across various problem scenarios.
Wang et al.~\cite{wang2024large} explores the use of LLMs for automatically generating Verilog code from natural language specifications. It introduces a novel approach that employs reinforcement learning with golden code feedback, specifically using proximal policy optimization and a reward function based on the similarity of abstract syntax trees between generated and reference code. This method enhances the semantic evaluation of the generated code and addresses the limitations of existing open-source models, which often lack performance compared to commercial alternatives.
The authors present their model, VeriSeek, which has 6.7 billion parameters and achieves state-of-the-art results.

\begin{figure}
    \centering
    \includegraphics[width=1.0\linewidth]{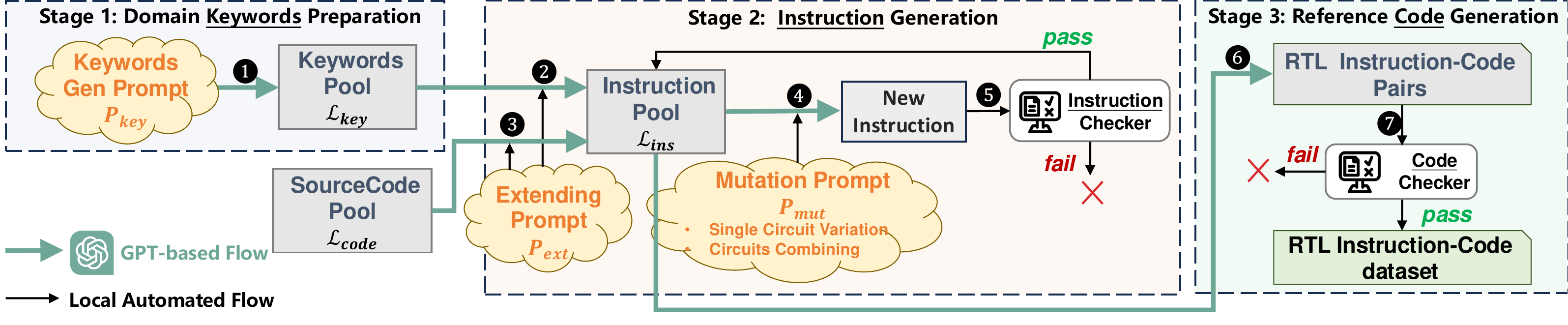}
    \vspace{-.2in}
    \caption{Data generation flow of RTLCoder~\cite{liu2024rtlcoder}. Following works (like AutoVCoder~\cite{auto-v-coder}, Origen~\cite{cui2024origen}) also adopts similar methodologies for dataset generation. The dataset generation consists of three steps. The first two steps are designed to generate diverse instructions (design specifications) and the final step is to generate high-quality reference code for the next fine-tuning step.}
    \label{fig:rtlcoder-data-gen}
    \vspace{-.2in}
\end{figure}

\textbf{Strategy 5: LLMs trained on open datasets with instruction-code pairs.} 
This process, known as `\emph{instruction fine-tuning}' or `\emph{supervised fine-tuning'~\cite{instruction-finetuning}}, adjusts the model to follow specific instructions or prompts more effectively. 
Different from adopting in-house private datasets in strategy 2, this strategy tries to provide open-source datasets to benefit the community. 
To address the scarcity of high-quality training data, RTLCoder~\cite{liu2024rtlcoder} introduces an automated dataset generation workflow. As illustrated in Figure\ref{fig:rtlcoder-data-gen}, the dataset generation workflow follows a three-step process, establishing a foundational paradigm for dataset creation. Its generated dataset enables a fine-tuned LLM coder that outperforms GPT-3.5~\cite{chatgpt} and achieves performance comparable to GPT-4~\cite{chatgpt}. This pioneering work provides the first open-source RTL LLM coder with instruction fine-tuning.  
Similar to the paradigm introduced by RTLCoder, CraftRTL~\cite{liu2024craftrtl} further analyzes existing LLMs' performance on Verilog, identifying two key weaknesses: poor handling of non-textual representations (like Karnaugh maps and waveforms) and inconsistent performance due to minor coding errors.
Targeting these weaknesses, CraftRTL creates a "correct-by-construction" synthetic dataset that includes Karnaugh maps, finite state machines, and waveform representations.  They develop an automated framework for generating detailed error reports that identify common minor mistakes in code completions, which are then used to create a targeted code repair dataset by injecting errors into correct open-source code.

\new{\textbf{Strategy 6: LLMs finetuned by reinforcement learning (RL) techniques.} The recent works~\cite{teng2025verirl, wang2025verireason} explore the integration of RL strategies for finetuning LLMs on RTL generation. The standard method is to use the feedback from the testbench and curate the reward based on different situations (e.g., syntax correctness, functional correctness, etc). The synthetic dataset is important for the RL training, especially the high-quality testbenches. For example, VeriRL~\cite{teng2025verirl} curates a high-quality training dataset with testbenches and designs a specialized \emph{4-stage} training process. Specifically, VeriRL introduces a sample-balanced weighting strat
egy that adaptively balances learning dynamics based on reward
probability distributions.}

        

\subsection{LLMs for HLS Code Generation}\label{sec:llm-hls-gen}

Similar to \emph{RTL code} generation, several works have explored LLM-based \emph{HLS code} generation to improve automation and efficiency when designing hardware with high-level programming languages. As summarized in~\Cref{tbl:llm-hls-gen}, existing solutions primarily rely on prompt engineering without fine-tuning the models. Additionally, many works are open-sourced. Notably, HLSPilot~\cite{xiong2024hlspilot} and Liao et al.~\cite{are-llms-any-good-for-hls} introduce new benchmarks for evaluating HLS code generation performance.

For example, SynthAI~\cite{sheikholeslam2024synthai} introduces a multi-agent generative AI framework for modular HLS design, integrating ReAct agents, CoT prompting, RAG, and web search capabilities to enhance decision-making. By systematically planning and executing modular designs, SynthAI improves design quality and scalability. 
HLSPilot~\cite{xiong2024hlspilot} focuses on hybrid CPU-FPGA architectures, proposing a three-stage approach: C/C++ to HLS translation, design space exploration, and LLM-based profiling. It integrates C-to-HLS optimization strategies to generate complex circuit designs, employs a DSE tool for pragma parameter tuning, and leverages LLMs for performance profiling to identify bottlenecks and optimize HLS designs. 
Liao et al.~\cite{are-llms-any-good-for-hls} investigate the translation of natural language specifications or C code into RTL, evaluating the capability of LLMs to automate hardware design. C2HLSC~\cite{collini2024c2hlsc} explores fully automated C-to-HLS transformation, refactoring generic C code into an HLS-compatible format while supporting hierarchical designs and pragma generation for optimizing area and throughput. These works collectively highlight the potential of LLMs in improving HLS design automation, enabling more efficient translation from high-level code to synthesizable hardware descriptions. \new{LHS~\cite{LHS} proposes a framework that LLMs to automate the HLS process for deep learning tasks, particularly those involving complex convolutional neural networks. Traditional HLS methods require significant manual intervention to resolve synthesizability issues and optimize source code, which can be time-consuming and error-prone. LHS addresses these challenges by using LLMs to automatically identify and rectify synthesis issues in C code, perform necessary code modifications, and insert optimization pragmas, thereby streamlining the design of efficient hardware accelerators.}

\begin{table}[!t]
\small
      \centering
     \hspace{-.1in}
     \setlength{\tabcolsep}{0.5em}
      \renewcommand{\arraystretch}{1.2}
     \resizebox{1.0 \textwidth}{!}{
        \begin{tabular}{ c||c|c|c|c|c|c|c } 
        \toprule
        \multicolumn{7}{c|}{\textbf{LLM for HLS Generation}} \\
        \hline
        \multirow{2}{*}{Method} & \multirow{1}{*}{New} & \multirow{1}{*}{New}  &  \multirow{1}{*}{Open} & \multirow{1}{*}{New} & Prompt & \multirow{2}{*}{Link} & \multirow{2}{*}{Date}\\
        & Model & Dataset  & Method & Benchmark & Engineering & & \\
        \hline
        \hline
        SynthAI~\cite{sheikholeslam2024synthai} &  & & \checkmark & & \checkmark & \url{https://github.com/sarashs/FPGA_AGI} & 2024-05\\
         HLSPilot~\cite{xiong2024hlspilot} & &  & \checkmark &  \checkmark & \checkmark & \url{https://github.com/xcw-1010/HLSPilot} & 2024-08\\
         Liao et al.~\cite{are-llms-any-good-for-hls} & & & & \checkmark & \checkmark & & 2024-08\\
         C2HLSC~\cite{collini2024c2hlsc} &  & & \checkmark & & \checkmark & \url{https://github.com/Lucaz97/c2hlsc} & 2024-11\\
        \bottomrule
        \end{tabular}
      }
        \caption{Existing explorations in LLM-aided HLS code generation, covered in Section~\ref{sec:llm-hls-gen}.}
        \label{tbl:llm-hls-gen}
        \vspace{-.2in}
\end{table}

\subsection{LLMs for Design Optimizations} 
\label{sec:llm-opt}

During circuit code generation using LLMs, besides functional correctness focused by Section~\ref{sec:llm-rtl} and \ref{sec:llm-hls-gen}, design quality metrics such as power, performance, and area (PPA) are also critical for ensuring efficiency and practicality. Recent advancements have explored optimizing LLM-generated circuits, focusing on both \textbf{RTL code optimization} and \textbf{HLS code optimization} to enhance hardware performance. A detailed comparison of existing works in this domain is provided in~\Cref{tab:llm-opt}.

%
\textbf{RTL code optimization.}
RTL code optimization leverages LLMs to refine hardware designs for better efficiency, focusing on improving PPA metrics. For example, ChipGPT~\cite{chang2023chipgpt} employs an enumerative search strategy, generating multiple design variations and selecting the one with the best PPA. BetterV~\cite{pei2024betterv} fine-tunes LLMs on domain-specific Verilog datasets, applying instruct-tuning and generative discriminators to improve Verilog code quality and optimize synthesis outcomes. However, current evaluations in BetterV primarily assess design quality based on AIG node reduction during synthesis, without directly considering final PPA metrics.
RTLRewriter~\cite{yao2024rtlrewriter} introduces a framework for RTL code rewriting, breaking down large circuits into smaller segments to enhance synthesis efficiency and leveraging multi-modal program analysis to incorporate visual and textual information. Its benchmark demonstrates superior performance compared to traditional RTL compilers such as Yosys and E-graph. Additionally, the work by Martínez et al.~\cite{martinez2024code} focuses on identifying key computational patterns like GEMM, convolution, and FFT within hardware code using LLM-based prompting techniques. Their method reduces false positives by employing a two-phase prompting approach, first interpreting the code and then verifying algorithm presence, highlighting the importance of prompt engineering for optimizing LLM-driven hardware design.

\textbf{HLS code optimization.}
Besides RTL code optimization, optimizing HLS code, particularly pragma optimization, is a crucial task in high-level synthesis. Xu et al.~\cite{xu2024optimizing} propose RALAD (Retrieve Augmented Large Language Model Aided Design), a framework leveraging LLMs and RAG to optimize HLS programs without requiring computationally expensive fine-tuning. HLS allows circuit design using high-level languages like C/C++, but manual optimization remains highly expertise-driven. RALAD mitigates this challenge by embedding user code and a knowledge base (e.g., FPGA textbooks), retrieving relevant code snippets via a top-k search, generating prompts that incorporate user instructions and retrieved snippets, and using an LLM like CodeLlama to produce optimized code. The study also explores the impact of manual annotations to further refine optimization quality, demonstrating the framework’s effectiveness in automating HLS code improvements.
\new{HLSRewritter~\cite{HLSRewriter} introduces an LLM-aided framework designed to automate the refactoring and optimization of standard C/C++ code for HLS, addressing common incompatibility issues. It employs a step-wise reasoning process to systematically analyze and detect HLS-incompatible errors, enhancing the accuracy of refactoring. The framework integrates a repair library created from HLS tool manuals, utilizing an RAG approach to guide the LLM in generating correct HLS-compatible code. Additionally, it incorporates a pipeline-aware decomposition strategy to break down complex loop structures for efficient pipelining and parallel execution, along with a bit width adjuster to optimize the precision of floating-point variables.}
\looseness=-1


\begin{table}[]
    \centering
    \resizebox{0.85 \textwidth}{!}
    {\begin{tabular}{c|c|c|c}
   \toprule
   \multicolumn{4}{c}{\textbf{LLM for Design  Optimization}}  \\
        \hline
        Works & Open & Link & Date  \\
         \hline
        \hline
        
        Martine et al.~\cite{martinez2024code}& & &2023-07  \\
        Sandal et al.~\cite{zero-shot-rtl-code-generation-attention-sink-aug-llms} & & & 2024-01  \\
        BetterV~\cite{pei2024betterv} & & & 2024-02 \\
        DeLorenzo et al.~\cite{make-every-move-count-llms-mcts} & & &2024-02 \\
        RTLRewriter~\cite{yao2024rtlrewriter} & \checkmark & \url{https://github.com/yaoxufeng/RTLRewriter-Bench} & 2024-09  \\ 
        Xu et al.~\cite{xu2024optimizing} & & & 2024-10 \\
        \bottomrule
    \end{tabular}}
    \caption{Collection of works on design optimization, covered in Section~\ref{sec:llm-opt}.}
    \label{tab:llm-opt}
    \vspace{-.25in}
\end{table}

\subsection{LLM for Hardware Code Verification} 
\label{sec:llm-verification}

\begin{figure}
    \centering
    \includegraphics[width=0.95\linewidth]{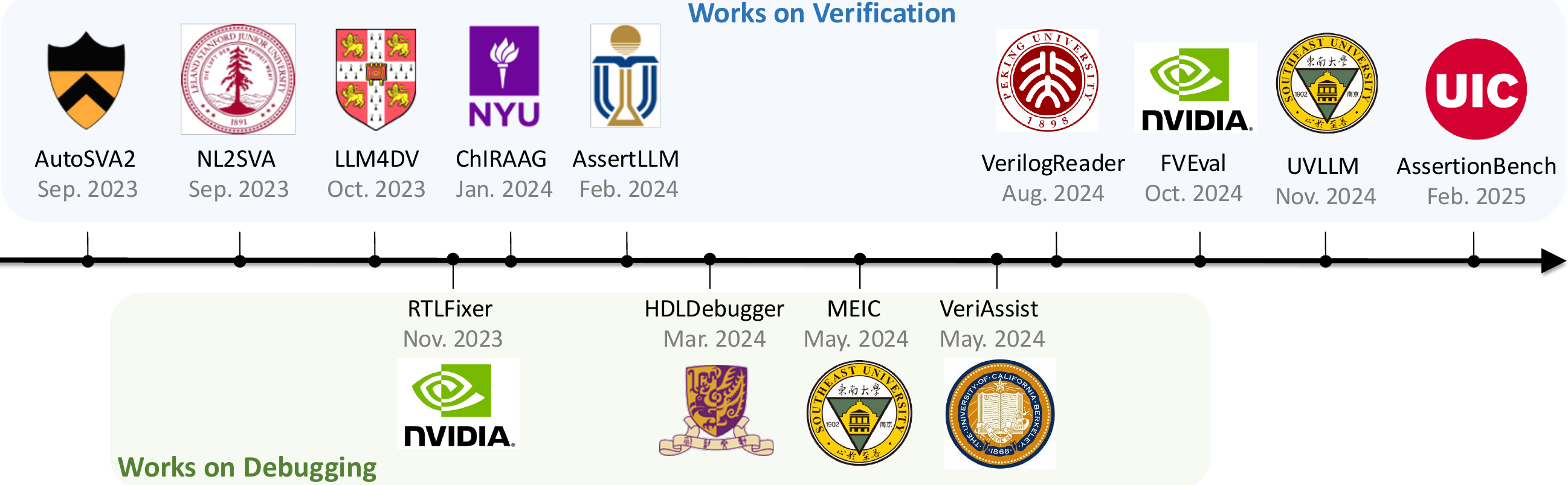}
    \vspace{-.15in}
    \caption{Timeline of RTL verification (\Cref{sec:llm-verification}) and debugging works (\Cref{sec:llm-debug}).}
    \label{fig:time-line-rtl-verify}
    \vspace{-.2in}
\end{figure}

In addition to circuit code generation, verifying the functional correctness of circuit designs is a critical yet highly labor-intensive task that heavily relies on human engineers. To address this challenge, LLM-based solutions have been explored to automate hardware verification. Table~\ref{tbl:llm-rtl-veri} summarizes existing works in this direction.
Current LLM-based verification approaches focus on two primary directions: 1) \textbf{Assertion generation with LLMs.} 
These approaches leverage LLMs to generate assertions based on design specifications or RTL code~\cite{fang2024assertllm, kande2024security, orenes2023using, sun2023towards, liu2024domain, huang2024towards, mali2024chiraag, meng2023unlocking}. The generated assertions are then used to validate whether the design under test (DUT) complies with its specifications, with either formal verification tools (e.g., Cadence JasperGold) for static formal property verification or simulation tools (e.g., Synopsys VCS) for dynamic verification on test benches.
2) \textbf{Test bench generation with LLMs.} LLMs are also employed to generate test stimuli, enhancing the simulation-based verification process~\cite{huang2024towards, UVLLM, ma2024verilogreader, zhang2023llm4dv, bhandari2024llm}. 
The comparison of these explorations is listed in~\Cref{tbl:llm-rtl-veri}, with the timeline demonstrated in~\Cref{fig:time-line-rtl-verify}, almost all existing explorations directly employ prompt engineering due to the lack of high-quality verification data for fine-tuning.
Some of these verification efforts focus specifically on security verification, which will be further discussed in Section~\ref{sec:llm-hardware-security}.\looseness=-1

\subsubsection{Assertion generation with LLMs} 

\

We categorize existing works on assertion generation into two main types: assertion generation benchmarks and assertion generation techniques. The former focuses on evaluating the effectiveness of LLMs in generating functionally correct assertions, while the latter explores various methodologies to improve the accuracy and reliability of LLM-generated assertions. We detail these two categories below.\looseness=-1

\textbf{Benchmarking LLM-aided assertion generation.}
Similar to RTL code generation, benchmarking is crucial for evaluating the quality of LLM-generated assertions. The evaluation process involves three key aspects: syntax correctness, functional correctness, and overall assertion quality. Syntax correctness can be verified using RTL code compilers, while functional correctness can be validated through simulation-based verification or formal property checking based on the golden RTL implementations. However, assessing assertion quality remains an open challenge, as it depends on multiple factors, such as completeness and relevance to the design specification.

Currently, key assertion generation benchmarks include AssertionBench~\cite{pulavarthi2024assertionbench}, AssertEval~\cite{liu2024openllm}, and FVEval~\cite{kang2024fveval}, all of which use Cadence JasperGold for formal property verification of generated assertions against golden RTL implementations.
Specifically, AssertionBench~\cite{pulavarthi2024assertionbench} consists of 100 Verilog hardware designs from OpenCores~\cite{URL:opencore}, with formally verified assertions derived from GOLDMINE~\cite{vasudevan2010goldmine} and HARM~\cite{germiniani2022harm} tools. The evaluation metrics include syntax correctness and functional correctness. AssertEval~\cite{liu2024openllm} from OpenLLM-RTL~\cite{liu2024openllm} includes 17 OpenCores~\cite{URL:opencore} designs, each accompanied by a natural language specification and golden RTL implementation. It evaluates assertions based on syntax correctness, functional correctness, and COI (cone-of-influence) coverage. FVEval~\cite{kang2024fveval} assesses assertions in three scenarios: (1) NL2SVA-Human, generating assertions from human-written specifications and real-world testbenches; (2) NL2SVA-Machine, translating formal logic from synthetic natural language descriptions to SystemVerilog assertions; and (3) Design2SVA, directly generating assertions from RTL designs. It evaluates various LLMs (e.g., GPT-4o, Gemini, LLaMA3) based on syntax correctness, full functional correctness, and partial correctness (assertions that are logically related but not fully equivalent to the reference).

\textbf{Generate design assertions with LLMs.}
LLMs automate hardware verification by leveraging natural language specifications and RTL code to produce SystemVerilog Assertions (SVA). These techniques can be categorized based on their input types: (1) Natural language specifications alone (e.g., ChIRAAG~\cite{mali2024chiraag}, AssertLLM~\cite{fang2024assertllm}). (2) RTL code alone (e.g., AutoSVA2~\cite{orenes2023using}). (3) Both specification and RTL code (e.g., NL2SVA~\cite{sun2023towards}).
Due to the scarcity of high-quality assertion datasets, most works employ prompt engineering rather than fine-tuning LLMs.
Evaluation methods typically rely on formal property verification (FPV) using tools like Cadence JasperGold, ensuring that generated assertions maintain logical correctness. Some works, such as ChIRAAG~\cite{mali2024chiraag}, also incorporate simulation-based validation using Synopsys VCS with test benches.

For example, AutoSVA2~\cite{orenes2023using} prompts GPT-4 with RTL code and a refined rule-based system to generate valid SVAs, validated through FPV.
NL2SVA~\cite{sun2023towards} employs few-shot prompting with both RTL and natural language descriptions to guide assertion generation, also evaluated via FPV. 
ChIRAAG~\cite{mali2024chiraag} relies solely on natural language specifications, using prompt engineering for assertion synthesis, with validation conducted through simulation.
AssertLLM~\cite{fang2024assertllm} processes entire specification documents, utilizing a three-phase approach where different LLMs handle specification extraction, waveform analysis, and assertion generation, with verification performed through FPV.
\new{AssertionForge~\cite{bai2025assertionforge} proposes a novel approach for generating SystemVerilog Assertions (SV As) by integrating information from natural language specifications and RTL code through a structured knowledge graph. This method addresses the challenges of ambiguity and incompleteness in specifications by constructing a unified KG that captures both high-level design intent and low-level implementation details. The approach employs a multi-resolution context synthesis process, which includes global summarization, signal-specific retrieval, and a Guided Random Walk with Adaptive Sampling to explore the KG, enabling the generation of precise and contextually rich prompts for LLMs. This structured representation enhances the quality of generated assertions by effectively bridging the gap between design specifications and implementation behavior.}
\looseness=-1

\begin{table}[!t]
\small
      \centering
     \hspace{-.1in}
     \setlength{\tabcolsep}{0.5em}
      \renewcommand{\arraystretch}{1.2}
     \resizebox{1.0 \textwidth}{!}{
        \begin{tabular}{ c||c|c|c|c|c|c|c } 
        \toprule
        \multicolumn{8}{c}{\textbf{LLM for RTL Verification}}  \\
        \hline
        \multirow{2}{*}{Method} & New & New & Open  & Open & Prompt & \multirow{2}{*}{Link} & \multirow{2}{*}{Date}\\
        & Model & Dataset & Model  & Benchmark & Engineering &  & \\
        \hline
        \hline
         AutoSVA2~\cite{orenes2023using} & & & & & \checkmark & & 2023-09\\
        NL2SVA~\cite{sun2023towards} & & & & & \checkmark & & 2023-09 \\
        LLM4DV~\cite{zhang2023llm4dv} & & & &  \checkmark & \checkmark & \url{https://github.com/ZixiBenZhang/ml4dv} & 2023-10 \\
        ChIRAAG~\cite{mali2024chiraag} & & & & &  \checkmark & & 2024-01 \\
         AssertLLM~\cite{fang2024assertllm} & & & \checkmark  & \checkmark & \checkmark & \url{https://github.com/hkust-zhiyao/AssertLLM} & 2024-02 \\
         Xiao et al.~\cite{xiao2024llm} & & & & & \checkmark & & 2024-03 \\
         Blocklove et al.~\cite{evaluating-llms-for-hardware-design-test} & & & & & \checkmark & & 2024-04 \\
         Liu et al.~\cite{liu2024domain} & \checkmark & \checkmark & & &  & & 2024-04 \\
         Huang et al.~\cite{huang2024towards} & & & & & \checkmark & & 2024-05 \\

         Bhandari et al.~\cite{bhandari2024llm} & & & & 
        \checkmark & \checkmark & & 2024-06 \\
        
         VerilogReader~\cite{ma2024verilogreader} & & & & & \checkmark & \url{github.com/magicYang1573/llm-hardware-test-generation} & 2024-06\\
         FVEval~\cite{kang2024fveval} & & \checkmark & & \checkmark & \checkmark  &\url{https://github.com/NVlabs/FVEval} & 2024-10\\
         
        UVLLM~\cite{UVLLM} & & & \checkmark & & \checkmark & \url{https://github.com/amyuch/UVLLM} & 2024-11 \\

         AssertionBench~\cite{pulavarthi2024assertionbench} & & & &  \checkmark & \checkmark & & 2025-02 \\
        \bottomrule 
        \end{tabular}
      }
        \caption{Explorations in LLM-aided RTL code verification (\Cref{sec:llm-verification}).} 
        \label{tbl:llm-rtl-veri}
        \vspace{-.35in}
\end{table}

\subsubsection{Test bench generation with LLMs} 

\

In addition to assertion generation, recent advancements in LLM-based verification have introduced automated test bench generation~\cite{huang2024towards, UVLLM, ma2024verilogreader, zhang2023llm4dv, bhandari2024llm}, significantly reducing the manual effort involved in verifying RTL designs. These approaches aim to enhance coverage metrics, including code coverage and functional coverage, by generating high-quality test benches for simulation. Existing explorations also fall into these main categories: (1) Test bench generation for \textit{code coverage}. This type focuses on measuring how thoroughly the RTL code is exercised during simulation. This includes metrics such as statement coverage, branch coverage, toggle coverage, and FSM state coverage. Achieving high code coverage ensures that most structural elements of the design have been tested but does not guarantee full functional correctness. (2) Test bench generation for \textit{functional coverage}. This type ensures that all intended design behaviors are tested according to the specification. Functional coverage is often defined using assertions and covergroups, verifying that different functional scenarios, corner cases, and expected behaviors are exercised. Unlike code coverage, functional coverage validates the correctness of the design beyond just its structural execution. We detail the two categories below.

\textbf{Test bench generation for code coverage.}
VerilogReader~\cite{ma2024verilogreader} integrates LLMs into coverage-directed test generation, focusing on achieving code coverage closure by generating test stimuli that target uncovered RTL lines and branches. It takes as input a Verilog design under test (DUT), natural language descriptions, and code coverage reports from simulations. The output consists of automatically generated test stimuli designed to improve code coverage in RTL verification. To generate test inputs effectively, VerilogReader employs prompt engineering with a Prompt Generator that structures LLM interactions in two stages: first, understanding the DUT and its current coverage status, and second, generating test inputs in a structured JSON format. Additionally, it includes a Coverage Explainer, which transforms raw simulator coverage reports into an LLM-readable format, and a DUT Explainer, which enhances LLM comprehension of Verilog code by providing natural language descriptions and test guidance.

\textbf{Test bench generation for functional coverage.}
Most existing explorations~\cite{huang2024towards, UVLLM, zhang2023llm4dv, bhandari2024llm} focus on functional coverage, as LLMs more excel in understanding RTL functionality and specifications rather than analyzing RTL code structure, which is required for code coverage.
For example, VeriAssist~\cite{huang2024towards} takes the design specification as input and generates initial RTL code along with corresponding test cases. It employs a self-verification process, where the generated RTL is simulated with test cases while considering timing constraints. This is followed by a self-correction mechanism, where the LLM refines the RTL design based on simulation feedback, addressing compilation and functional errors. By mimicking a human-in-the-loop design approach, VeriAssist~\cite{huang2024towards} improves the accuracy and correctness of both RTL code and test benches.
Another work UVLLM~\cite{UVLLM} integrates LLMs with Universal Verification Methodology (UVM) to automate test case generation and RTL code repair. The framework consists of four steps: pre-processing, where linters and LLMs eliminate syntax errors; UVM processing, which generates and runs test cases within a UVM testbench; post-processing, which analyzes simulation logs to identify errors; and repair, where LLMs generate RTL patches based on detected issues. While UVLLM is open-sourced and showcases LLMs’ potential in verification automation, challenges remain, including the need for extensive training data and the high computational cost of large-scale LLM inference.\looseness=-1

\subsection{LLM for Hardware Code Debugging}  
\label{sec:llm-debug}

Debugging in hardware design involves identifying and fixing both syntax and functional errors in circuit implementations. Traditionally, engineers conduct this process of fixing bugs manually, making it a tedious and labor-intensive task. Recent advancements in LLMs automate hardware debugging, reducing human intervention and improving efficiency. Existing research explores LLM-assisted debugging for both RTL and HLS code, offering new methodologies for error detection, root cause analysis, and automated patch generation, as detailed below.

\begin{figure}
    \centering
    \includegraphics[width=0.92\linewidth]{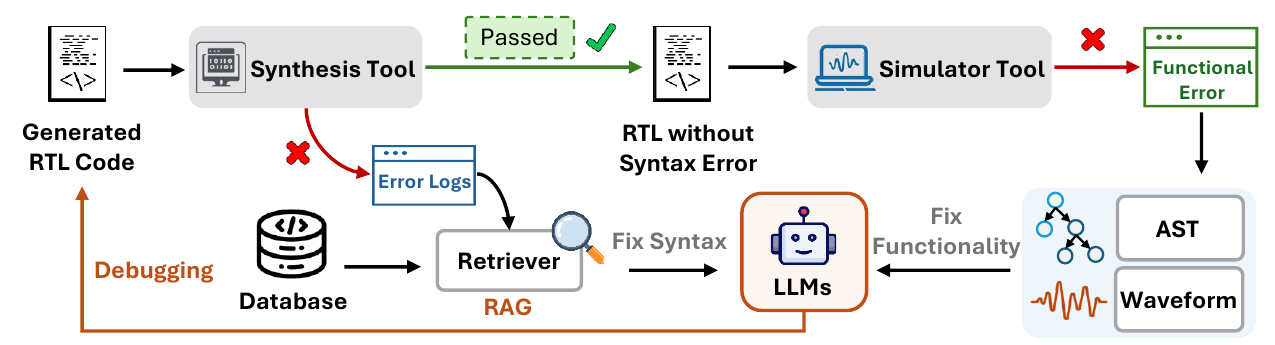}
    \caption{Overview of the LLM-based flow for RTL debugging, recent works are covered in Section~\ref{sec:llm-debug}. This approach includes two input methods for LLM solutions: the first method assigns debugging tasks directly to the LLM without supplementary information, while the second method enhances the debugging process by incorporating error information from EDA tools. This error log can be input directly into the LLM or used to query a pre-defined debugging database (RAG). Additionally, some approaches utilize AST or waveform tracing tools to more effectively identify problematic code segments.}
    \label{fig:flow-llm-rtl-debugging}
    \vspace{-.1in}
\end{figure}

\subsubsection{LLM for RTL code debugging}
\label{sec:rtl-debug}
\  

RTL debugging focuses on resolving errors identified during the verification process. Unlike verification, which primarily detects inconsistencies, debugging involves both locating and correcting these issues to ensure functional correctness. For example, representative debugging works~\cite{xu2024meic, tsai2023rtlfixer, huang2024towards} leverage LLMs for both bug detection and bug fixing, emphasizing the automated correction of RTL errors. The debugging process typically consists of two key steps: (1) identifying the bug by pinpointing the exact error location within the RTL code and (2) fixing the bug by generating corrected RTL logic. While verification highlights potential failures, debugging requires deeper reasoning to determine the root cause of errors and propose appropriate fixes. \Cref{tbl:llm-rtl-debug} and~\Cref{fig:time-line-rtl-verify} demonstrate the comparison and timeline of these explorations, respectively.

RTL bugs are broadly categorized into syntax bugs and functional bugs~\cite{xu2024meic}. Syntax bugs, such as missing semicolons or incorrect module instantiations, can be directly flagged by compilers (i.e., synthesis tools). Functional bugs, on the other hand, require executing test cases or formal verification to identify behavioral mismatches. Based on this classification, we categorize existing works in LLM-assisted RTL debugging into \textbf{syntax debugging} and \textbf{functional debugging}. Since functional debugging involves more complex reasoning and deeper analysis of design behavior, tools capable of addressing both syntax and functional errors are classified as functional debuggers.

\begin{table}[!t]
\small
      \centering
     \hspace{-.1in}
     \setlength{\tabcolsep}{0.5em}
      \renewcommand{\arraystretch}{1.2}
     \resizebox{0.9 \textwidth}{!}{
        \begin{tabular}{ c||c|c|c|c|c|c|c } 
        \toprule 
        
        \multicolumn{8}{c}{\textbf{LLM for RTL Debugging}}  \\
        \hline
        \multirow{2}{*}{Method} & New & New & Open  & Open & Prompt & \multirow{2}{*}{Link} & \multirow{2}{*}{Date}\\
        & Model & Dataset & Model  & Benchmark & Engineering &  & \\
        \hline
        \hline
        RTLFixer~\cite{tsai2023rtlfixer} & & & \checkmark & \checkmark & \checkmark & \url{https://github.com/NVlabs/RTLFixer} & 2023-11 \\
        HDLDebugger~\cite{yao2024hdldebugger} & & & & & \checkmark & & 2024-03 \\
        VeriAssist~\cite{huang2024towards} & & & & & & & 2024-05 \\
        MEIC~\cite{xu2024meic} & & & & & & &2024-05 \\
        Qayyum et al.~\cite{qayyum2024bugs} & & & & & & & 2024-06 \\
        VerilogCoder~\cite{ho2024verilogcoder} & & &  & & \checkmark & & 2024-08 \\
        UVLLM~\cite{UVLLM} & & & \checkmark & & \checkmark & \url{https://github.com/amyuch/UVLLM} & 2024-11 \\
        
        \bottomrule
        \end{tabular}
      }
        \caption{Explorations in LLM-aided RTL code debugging (\Cref{sec:llm-debug}).} 
        \label{tbl:llm-rtl-debug}
        \vspace{-.25in}
\end{table}


\textbf{Syntax debugging.}
RTLFixer~\cite{tsai2023rtlfixer} and HDLDebugger~\cite{yao2024hdldebugger} are among the pioneering works to explore LLM-assisted RTL syntax debugging. Both works leverage the RAG technique to improve debugging accuracy by transforming syntax-buggy RTL code into syntax-correct RTL designs.
RTLFixer~\cite{tsai2023rtlfixer} integrates RAG and ReAct prompting, creating an autonomous debugging agent that retrieves expert guidance and applies iterative reasoning to correct syntax errors effectively. It also introduces VerilogEval-Syntax, a debugging dataset consisting of 212 erroneous Verilog implementations to benchmark LLM performance in syntax correction.
HDLDebugger~\cite{yao2024hdldebugger}, developed around the same time, similarly employs RAG to retrieve relevant debugging information from documentation and code databases. Additionally, it incorporates a self-guided fine-tuning process to improve LLM-based debugging accuracy. The framework also includes a novel data generation module that synthetically creates pairs of buggy and corrected HDL code using a reverse engineering approach. Both methods significantly enhance LLM capabilities in syntax debugging by integrating retrieval-based contextual learning and structured reasoning techniques.



\textbf{Functional debugging.} 
Compared with syntax debugging, functional debugging is significantly more challenging as it requires deep reasoning about circuit functionality and identifying the root cause of errors. Recent works~\cite{huang2024towards, xu2024meic, qayyum2024bugs, UVLLM, ho2024verilogcoder} have explored LLM-based approaches to address functional RTL debugging.
VeriAssist~\cite{huang2024towards} enhances pre-trained LLMs with self-verification and self-correction techniques. The framework generates test cases alongside RTL code and simulates the generated design to detect functional errors. If discrepancies are identified, self-correction mechanisms iteratively refine the RTL code based on simulation feedback, improving debugging accuracy. MEIC~\cite{xu2024meic} proposes an LLM-based iterative debugging framework and introduces a new debugging benchmark based on RTLLM-v1.0~\cite{lu2024rtllm}, an RTL generation dataset containing 15 source designs. By introducing 178 buggy variations of these designs, MEIC categorizes errors into syntax and functional bugs, providing a structured evaluation dataset for LLM-based debugging research. Qayyum et al.~\cite{qayyum2024bugs} integrate RAG into functional debugging by retrieving relevant RTL specifications and comparing them with the RTL implementation. This enables LLMs to detect inconsistencies and suggest fixes based on the intended circuit behavior, significantly improving debugging accuracy through formal specification guidance.\looseness=-1



Beyond direct RTL code analysis, some works incorporate auxiliary sources such as abstract syntax tree (AST) and waveform analysis to enhance functional debugging. UVLLM~\cite{UVLLM} introduces an LLM-based unified verification methodology, leveraging AST representations to improve error localization and code corrections. Similarly, VerilogCoder~\cite{ho2024verilogcoder} employs a rewriting mechanism that enhances debugging accuracy. This process integrates information from EDA tools, such as ASTs and waveform tracing tools, to refine LLM-driven RTL debugging. By combining LLM reasoning with structured program analysis, these methods offer improved robustness in identifying and correcting functional design errors.

\subsubsection{LLM for HLS debugging} 
\label{sec:llm-hls-debug}
\  


Compared to RTL and HLS generation tasks, significantly fewer works focus on debugging at the HLS level. One of the pioneering efforts in this direction is Chrysalis~\cite{deming-chen-hls-debug}, a benchmark designed for training and evaluating LLMs’ capability to identify functional bugs in HLS code. Unlike syntax errors that can be easily detected by compilers, many functional bugs at the HLS level require deeper semantic analysis and program reasoning. Chrysalis provides a structured dataset that allows LLMs to learn patterns of common HLS-specific issues and evaluate their debugging performance in terms of both syntax correctness and functional accuracy. This benchmark sets the foundation for future research in LLM-assisted HLS debugging by offering a standardized dataset for evaluating model capabilities in detecting and resolving high-level synthesis errors.
\new{HLSDebugger~\cite{wang2025hlsdebugger} proposes a novel solution for identifying and correcting logic bugs in HLS code using LLMs. It addresses three primary challenges: the scarcity of high-quality circuit data, the complexity of debugging logic bugs compared to syntax errors, and the need for a multi-tasking approach that combines bug identification and correction. HLSDebugger generates a large labeled dataset of approximately 300,000 samples to train the model. It employs a customized encoder-decoder architecture, where the encoder identifies bug locations and types, while the decoder generates corrected code snippets. This integrated approach enhances the model's performance by leveraging contextual information during the debugging process, ultimately aiming to automate and improve the efficiency of HLS code debugging.}

\subsection{LLMs for Hardware Security}\label{sec:llm-hardware-security}

Besides functional verification and debugging, a growing number of research explore the use of LLMs for hardware security verification and threat detection. Recent works~\cite{paria2023divas, ahmad2023fixing, kokolakis2024harnessing, saha2024llm, wang2024llms, ahmad2024hardware, kande2024security, paria2024navigating, tarek2024socurellm, kande2024llms, saha2024empowering, akyash2024self} integrate LLMs into automated security analysis, detection of vulnerabilities, and protection of hardware designs. As summarized in \Cref{tbl:llm-security}, most of these approaches rely on prompt engineering to enhance security verification and threat detection. \looseness=-1

LLM-based research in hardware security can be divided into two primary directions. \textbf{Protective hardware security} focuses on using LLMs to detect vulnerabilities, generate security patches, and implement secure-by-design methodologies at the RTL and gate levels. These approaches aim to proactively mitigate security risks through automated analysis and verification techniques. \textbf{Offensive hardware security}, in contrast, explores how LLMs can facilitate attack strategies and identify potential hardware exploits. 
Together, these two research directions contribute to the development of more resilient defense mechanisms by providing insights into adversarial techniques and enabling the design of effective countermeasures. Below, we introduce representative works about both directions in detail.


\begin{table}[!t]
\small
      \centering
     \hspace{-.1in}
     \setlength{\tabcolsep}{0.5em}
      \renewcommand{\arraystretch}{1.2}
     \resizebox{0.8 \textwidth}{!}{
        \begin{tabular}{ c||c|c|c|c|c|c|c } 
        \toprule
        \multicolumn{8}{c}{\textbf{LLM for Hardware Security}}  \\
        \hline
        \multirow{2}{*}{Method} & New & New & Open & Open & Open & Prompt & Date\\
        & Model & Dataset & Model & Dataset & Benchmark & Engineering & \\
        \hline
        \hline

         Pearce et al.~\cite{pearce2023examining} & & & & \checkmark & &  \checkmark & 2021-12 \\
         Baleegh et al.~\cite{on-hardware-security-bug-code-fixes-by-prompting-llms} & & & & & &  \checkmark & 2023-02 \\
         Kande et al.~\cite{kande2024security} & & & & & \checkmark  & \checkmark & 2023-06 \\
         DIVAS~\cite{paria2023divas}& & & & & & \checkmark & 2023-08 \\
         NSPG~\cite{meng2023unlocking} & \checkmark & \checkmark & & & & & 2023-08 \\
          SCAR~\cite{srivastava2024scar} & & & & & & \checkmark & 2023-10 \\
         Netlist Whisperer~\cite{whisperer} & & & & & & \checkmark & 2023-11 \\
        SecRT-LLM~\cite{saha2024empowering} & & \checkmark & & & & \checkmark & 2024-05 \\
        Self-HWDebug~\cite{akyash2024self} & & & & & & \checkmark & 2024-05  \\
         Qayyum et al.~\cite{qayyum2024bugs} & & & & & & & 2024-06 \\
        \bottomrule
        \end{tabular}
      }
       \vspace{.05in}
        \caption{Existing explorations in LLMs for security, covered in Section~\ref{sec:llm-hardware-security}.} 
        \label{tbl:llm-security}
        \vspace{-.3in}
\end{table}

\textbf{LLM-aided protective hardware security.}
Research on LLM-assisted protective hardware security can be categorized into two key areas: \textbf{\textit{security bug detection}} through security assertion generation and \textbf{\textit{security bug fixing}}, which involves identifying and debugging vulnerabilities. These approaches highlight the potential of LLMs in automating security analysis, enhancing verification processes, and mitigating hardware vulnerabilities.


In \textbf{\textit{security bug identification}}, LLMs have been explored to automate the detection of hardware vulnerabilities and generate security assertions. For instance, Kande et al.~\cite{kande2024security} demonstrate the potential of LLMs in generating hardware security assertions, a task that traditionally requires significant expertise. Similar to the functional assertion generation process, their framework employs LLM to generate security assertions based on security specifications and evaluates LLM performance using a benchmark suite of real-world designs and corresponding golden reference security assertions, analyzing the impact of prompt detail on accuracy. 
DIVAS~\cite{paria2023divas} introduces an LLM-powered framework that automates SoC security analysis and policy-based protection. The system maps vulnerabilities to Common Weakness Enumerations (CWEs), generates verifiable SVAs, and implements security policies through security modules or wrappers. Evaluated on open-source benchmarks, DIVAS demonstrates effectiveness in automating SoC security analysis, policy enforcement, and vulnerability detection using the DiSPEL tool. Similarly, SecRL-LLM~\cite{saha2024empowering} propose a database containing 10,000 vulnerable finite state machine designs incorporating 16 security weaknesses. They further develop an LLM-based framework, integrating in-context learning and fidelity-check mechanisms to enhance both vulnerability insertion and detection in hardware designs.
\new{HWREx~\cite{Todaes-Security-HWREx} proposes an AI-enabled framework designed to address hardware vulnerabilities and IoT security through an ontology-driven storytelling approach. It combines ML and NLP techniques to analyze and extract meaningful insights from cybersecurity databases, specifically focusing on the relationships between Common Weakness Enumeration (CWE), Common Vulnerability Exposure (CVE), and Common Attack Pattern Enumeration and Classification (CAPEC). HWREx automates the identification of vulnerability patterns, provides mitigation strategies using LLMs, and enhances understanding of the evolving nature of hardware weaknesses, thereby supporting security assessments and improving overall cybersecurity posture.}
\looseness=-1

Beyond vulnerability detection, some works also explore \textbf{\textit{security bug fixing}} by leveraging LLMs for automated debugging and countermeasure implementation.
Pearce et al.~\cite{pearce2023examining} conducted one of the earliest studies in this area, prompting LLMs to automatically repair software security vulnerabilities as early as 2021, prior to the emergence of today’s more powerful models. Their work presents a comprehensive empirical evaluation of multiple commercial and open-source LLMs, including OpenAI Codex, AI21’s Jurassic models, Polycoder, and gpt2-csrc, assessing their ability to generate secure and functional patches for synthetic, handcrafted, and real-world security bugs.
Recently, Self-HWDebug~\cite{akyash2024self} introduces a framework leveraging LLMs to generate debugging instructions for security issues. By defining a set of CWEs and corresponding mitigation strategies, the framework enhances LLM prompt effectiveness, enabling automated security debugging and vulnerability resolution.
In the domain of side-channel attack (SCA) mitigation, Netlist Whisperer~\cite{whisperer} and SCAR~\cite{srivastava2024scar} propose LLM-driven solutions to enhance security at the hardware level. Netlist Whisperer~\cite{whisperer} adopts a two-phase, pre-silicon LLM-based approach: first, a GPT-3 model identifies power leakage-inducing nets in a circuit, and then a second GPT-3 model generates an SCA-resistant netlist, eliminating the need for traditional power trace collection. SCAR~\cite{srivastava2024scar} focuses on cryptographic accelerators, utilizing control-data flow graphs to identify and localize SCA vulnerabilities. It then employs a deep-learning explainer to analyze the vulnerabilities and leverages an LLM to automatically generate and insert security patches into the RTL code.\looseness=-1

\textbf{LLM-aided offensive hardware security.} 
Besides protective solutions, research also explores how LLMs can be leveraged to execute security threats, such as automated hardware trojan insertion.
For example, Kokolakis et al.~\cite{kokolakis2024harnessing} propose an LLM-based framework for automating hardware trojan insertion and evaluating its impact on a modern RISC-V microarchitecture. Their method begins with a filtering process to identify suitable modules for Trojan insertion. The RTL code of selected modules is provided to the LLM, which assists in implanting hardware trojans by modifying the design. While this approach demonstrates the feasibility of hardware trojan insertion using fine-tuned LLMs, further research is needed for more complex attack scenarios.



\begin{table}[]
    \centering
    \resizebox{0.7\textwidth}{!}{
    \begin{tabular}{c|c|c|c}
    \toprule
    \multicolumn{4}{c}{\textbf{LLM for Design Flow Automation}}  \\
        \hline
        Method & Open & Link & Date \\
        \hline
        \hline
        ChatEDA~\cite{he2023chateda} & \checkmark &\url{https://github.com/wuhy68/ChatEDAv1} & 2023-08 \\
        RAG-EDA~\cite{pu2024customized} &  \checkmark &\url{https://github.com/lesliepy99/RAG-EDA} & 2024-07 \\
        ChipAlign~\cite{deng2024chipalign} & & & 2024-12 \\
          
          \bottomrule
          \multicolumn{4}{c}{} \\
          \toprule
          \multicolumn{4}{c}{\textbf{LLM for Layout Design}} \\
          \hline
        Method & Open & Link & Date \\ \hline \hline
        Ho et al.~\cite{ho2024large} & & & 2024-05 \\
        FabGPT~\cite{jiang2024fabgpt} & & & 2024-07 \\
        Chen et al.~\cite{chen2024intelligent} & & & 2024-08 \\ 
        DRC-Coder~\cite{chang2024drc} & & & 2024-11 \\
        \bottomrule
    \end{tabular}
    }
    \caption{Explorations on LLMs for design flow automation and layout design, covered in~\Cref{sec:llm-flow}.}
    \label{tab:llm-flow}
    \vspace{-.4in}
\end{table}

\subsection{LLM for Design Flow Automation and Layout Design}
\label{sec:llm-flow}  

\subsubsection{LLM for design flow automation}

\

LLMs have also been explored for automating design flow processes, primarily in two key areas: \textbf{\textit{design flow script synthesis}}~\cite{he2023chateda} and \textbf{\textit{chip-related question-answering}} (QA)~\cite{pu2024customized, deng2024chipalign}. These applications aim to reduce human effort in configuring and optimizing EDA toolchains while improving accessibility to chip design knowledge. We compare existing works in~\Cref{tab:llm-flow}.

For \textbf{\textit{design flow script synthesis}}, ChatEDA~\cite{he2023chateda} is a pioneering work that leverages LLMs for automating EDA design flow execution. The framework decomposes user requests into structured sub-tasks, generates EDA scripts, and autonomously executes them using EDA tools. To enhance the model’s understanding of EDA workflows, instruction tuning techniques are applied. Additionally, ChatEDA introduces a benchmark suite comprising 50 tasks that include simple flow calls, complex multi-step flow executions, and parameter-tuning scenarios. The evaluation process assesses both the correctness and executability of the generated scripts using real EDA tools, followed by manual scoring to ensure logical coherence and practical usability.

For \textbf{\textit{chip-related question-answering}}, RAG-EDA~\cite{pu2024customized} presents a customized RAG framework for EDA tool documentation QA. The framework processes EDA tool documentation and user queries, employing hybrid information retrieval methods that combine lexical search (i.e., TF-IDF, BM25) and semantic retrieval (i.e., vector embeddings). A contrastive learning-based reranker is trained to filter relevant documents, improving retrieval accuracy. The LLM is then fine-tuned through a two-stage approach: (1) domain knowledge pretraining using EDA textbooks and (2) instruction tuning with QA datasets. The evaluation metrics include retrieval recall (recall@k) for the retriever and reranker, and BLEU, ROUGE-L, and UniEval scores to assess the accuracy and factual consistency of generated answers.
Additionally, ChipAlign~\cite{deng2024chipalign} extends LLM capabilities for chip-related QA tasks by addressing the challenge of aligning domain-adapted large language models for chip design with strong instruction-following abilities. It proposes a training-free model merging approach that combines a domain-specific chip LLM with a general instruction-aligned LLM. Instead of retraining on instruction-following data, ChipAlign employs a novel geodesic interpolation technique in the weight space to produce a merged model that maintains chip design expertise while significantly improving instruction alignment.

\subsubsection{LLM for netlist generation}

\new{There are also pioneering works on generating the netlist directly. For example, Circuit Transformer~\cite{li2024circuit-transformer} introduces a generative neural model that produces logic circuits that strictly preserve logical equivalence with given Boolean functions. Traditional generative models struggle with this task because they often make incorrect predictions that violate equivalence constraints. The Circuit Transformer addresses this challenge through a novel decoding mechanism that incrementally builds circuits by generating tokens while incorporating beneficial "cutoff properties" to block invalid candidates. The study also formulates the optimization of equivalent circuit generation as a Markov decision process, demonstrating the potential of this model for circuit design and optimization.}

\subsubsection{LLM for layout design.}

\

Some recent works employ foundation models for circuit layouts to enhance the \textbf{physical design process and manufacturability}, as demonstrated in~\Cref{tab:llm-flow}. Since circuit layouts are typically represented in the format like images, these works typically integrate vision models with LLMs to better understand and process circuit layouts.\looseness=-1

For instance, FabGPT~\cite{jiang2024fabgpt} introduces a large multimodal model designed for wafer defect knowledge querying in semiconductor fabrication. It processes Scanning Electron Microscope (SEM) images of wafers alongside textual metadata extracted using Optical Character Recognition (OCR) and predefined label sets. By fusing visual and textual information, the model enhances defect detection and knowledge retrieval. A pre-trained multimodal encoder captures critical wafer defect features, while a prediction module identifies defect types. Additionally, the model incorporates a Q\&A system with a modulation module that aligns visual and textual representations to improve interpretability in querying fabrication processes.
Ho et al.~\cite{ho2024large} propose an LLM-based optimization framework for standard cell layout design, incrementally generating clustering constraints to enhance PPA and routability. Their study assesses existing LLMs’ understanding of SPICE netlists, clustering constraints, and physical layout descriptions. Leveraging ReAct prompting, the model iteratively refines clustering decisions, improving standard cell layout optimization.
Chen et al.~\cite{chen2024intelligent} integrate reinforcement learning (RL) for OPC recipe optimization with a multi-modal LLM-backed agent system for recipe summarization. The RL component fine-tunes OPC parameters such as edge placement error (EPE) measurement points and polygon fragmentation to improve lithography accuracy. Meanwhile, the LLM-based agent extracts features, summarizes results, and generates structured OPC recipes, enhancing automation in semiconductor manufacturing.
DRC-Coder~\cite{chang2024drc} presents a multi-agent framework for automating design rule checking (DRC) code generation using LLMs and vision-language models. It mimics human DRC coding by decomposing tasks into interpretation and coding, assigning two specialized LLM agents to reduce hallucinations and enhance reasoning accuracy. The framework also integrates domain-specific functions, including foundry rule analysis, layout design rule violation (DRV) analysis, and automated debugging loops to refine DRC rule generation. By incorporating vision models, DRC-Coder can interpret design rule illustrations and layout structures, ensuring accurate and executable DRC scripts.

\begin{table}[!t]
\small
      \centering
     \hspace{-.1in}
     \setlength{\tabcolsep}{0.5em}
      \renewcommand{\arraystretch}{1.2}
     \resizebox{0.7\textwidth}{!}{
        \begin{tabular}{ c||c|c|c } 
        \toprule
        \multicolumn{4}{c}{\textbf{LLM for Architecture Design}}  \\
        \hline
        Method & Open & Link & Date \\
        
        \hline
        \hline
          
          Yan et al.~\cite{yan2023viability} & & & 2023-06 \\
          Liang et al.~\cite{liang2023unleashing} & & & 2023-07\\
          GPTAIGChip~\cite{fu2023gpt4aigchip} & &  & 2023-09 \\
          SpecLLM~\cite{li2024specllm} & \checkmark & \url{https://github.com/hkust-zhiyao/SpecLLM} & 2024-01\\
          
        \bottomrule
        
        \end{tabular}
      }
       \vspace{.05in}
        \caption{Explorations in LLM-aided architecture design, covered in~\Cref{sec:llm-hardware-architecture}}
        \label{tbl:llm-arch}
        \vspace{-.25in}
\end{table}

\subsection{LLMs for Hardware Architecture Design}
\label{sec:llm-hardware-architecture}

For hardware architecture design, LLMs have been explored in two primary areas: \textbf{\textit{circuit architecture design}}~\cite{yan2023viability, fu2023gpt4aigchip, liang2023unleashing} and \textbf{\textit{specification document processing}}~\cite{li2024specllm}. The comparison and timeline of these works are shown in~\Cref{tbl:llm-arch}.
These works aim to leverage LLMs to enhance automation, reduce design complexity, and improve efficiency in architectural decision-making.

In \textbf{\textit{circuit architecture design}}, GPT4AIGChip~\cite{fu2023gpt4aigchip} proposes an automated prompt-generation pipeline using in-context learning to guide LLMs in generating high-quality AI accelerator designs. This approach enables the structured decomposition of hardware design tasks, improving the consistency and efficiency of generated architectures.
LCDA~\cite{yan2023viability} applies LLMs to accelerate the software-hardware co-design process, particularly for compute-in-memory architectures in AI accelerators. It addresses the cold-start problem in traditional co-design approaches by leveraging LLMs to guide design space exploration, significantly reducing the search time. 
QGAS~\cite{liang2023unleashing} extends the application of LLMs to quantum computing, using GPT-4 to iteratively design variational quantum algorithm ansatz architectures and translate the architecture into quantum assembly language code.
\new{ChatArch~\cite{TODAES-ChatArch} is a knowledge-driven Graph-of-Thought (GoT) multi-LLM-agent framework developed for efficient processor architecture optimization, aiming to accelerate design iterations and meet stringent PPA objectives. Its core methodology involves decomposing the complex processor architecture design space into modular subspaces, which are then iteratively optimized by specialized LLM agents within a GoT framework that captures microarchitecture dependencies. The framework is underpinned by a comprehensive RISC-V design knowledge repository, consolidating expertise from textbooks, empirical data, and expert insights. LLM agents leverage this knowledge through structured prompts, few-shot learning, and RAG to propose, implement, and evaluate microarchitectural designs, generating corresponding behavioral models.}
\looseness=-1

For \textbf{\textit{specification document processing}}, 
SpecLLM~\cite{li2024specllm} tackles the inefficiencies and error-prone nature of developing architecture specifications in architecture design. It explores the use of LLMs to automate both the generation of specifications and the review of existing documentation. To structure the problem, the authors categorize architecture specifications into three levels, covering different degrees of design abstraction. They also introduce a dataset of 46 documents to evaluate the effectiveness of their approach. By leveraging LLMs, SpecLLM enhances both efficiency and accuracy in specification drafting and validation, demonstrating the potential for further automation in this critical aspect of hardware design.

\begin{table}[!t]
    \centering
    \resizebox{0.8\textwidth}{!}{
    \begin{tabular}{c|c|c|c}
    \toprule
    \multicolumn{4}{c}{\textbf{LLM for Analog Circuit Design}}  \\
        \hline
        Method & Open & Link & Date  \\
        \hline \hline
        \multicolumn{4}{c}{\new{Topology generation}}\\
        \hline
        LADAC~\cite{Liu2024ladac} & & & 2023-12\\
        AnalogCoder~\cite{lai2024analogcoder} & \checkmark & \url{https://github.com/anonyanalog/AnalogCoder}& 2024-05 \\
        FLAG~\cite{mao2024flag} &  & & 2024-05 \\
       
        LaMAGIC~\cite{chang2024lamagic} & & & 2024-07 \\
        Artisan~\cite{chen2024artisan} & & & 2024-11 \\
        
        AnalogXpert~\cite{zhang2024analogxpert} & & & 2024-12\\
        AnalogGenie~\cite{topologiesanaloggenie} & \checkmark & \url{https://github.com/xz-group/AnalogGenie} & 2025-01 \\
       
        \new{LEDRO~\cite{kochar2024ledro}} & \new{\checkmark} & \url{https://github.com/dimplekochar/LEDRO} & \new{2025-05} \\
        \new{MenTeR~\cite{chen2025menter}} &  &  & \new{2025-05} \\
        \new{LATENT~\cite{chaudhuri2025latent}} & & & \new{2025-05} \\

        \hline

        \multicolumn{4}{c}{\new{Topology optimization}}\\

        \hline
         ADO-LLM~\cite{yin2024ado} & & & 2024-06 \\
         LEDRO~\cite{kochar2024ledro} & & & 2024-11 \\
        
        \bottomrule
    \end{tabular}
    }
    \caption{Works on LLMs for analog circuit design, covered in~\Cref{sec:llm-analog}.}
    \label{tab:llm-analog}
    \vspace{-.25in}
\end{table}

\subsection{LLMs for Analog Circuit Design.}
\label{sec:llm-analog}

While most research on LLMs for hardware design has focused on \emph{digital} VLSI circuits, recent studies have started exploring LLM's potential in \emph{analog circuit design}. Unlike digital design, which follows well-defined logic rules, analog circuits typically require tuning and optimization based on human expertise, making LLM-assisted automation more challenging. A summary of existing works is provided in~\Cref{tab:llm-analog}. These studies introduced LLM-powered frameworks targeting different analog circuit types, such as power converters and amplifiers, focusing on knowledge-based reasoning, topology synthesis, and circuit optimization. These approaches aim to enhance the efficiency of analog design, addressing its inherent complexities. \new{LLM-based solutions for analog circuit design mainly focus on two important aspects: \emph{(1) topology generation} and \emph{(2) optimization}.}

\new{\textbf{Topology generation}~\cite{zhang2024analogxpert, chang2024lamagic, lai2024analogcoder, topologiesanaloggenie, Liu2024ladac} is the primary focus of the analog circuit, which mainly involves SPICE code generation. Works in this domain typically employ training-free methodologies and employ advanced prompt engineering strategies.} For example, LADAC~\cite{Liu2024ladac} introduces an LLM-driven decision-making agent for analog design, incorporating a knowledge library and interactive tools to assist in transistor sizing and simulation. AnalogCoder~\cite{lai2024analogcoder} employs a training-free LLM approach for Python-based circuit generation, integrating prompt engineering and feedback mechanisms to refine designs. LaMAGIC~\cite{chang2024lamagic} fine-tunes LLMs for analog topology generation, particularly for power converters, developing structured input-output representations that enhance circuit synthesis accuracy.
Artisan~\cite{chen2024artisan} focuses on operational amplifier (opamp) design, integrating topology selection and parameter tuning while employing Tree-of-Thought (ToT) and Chain-of-Thought (CoT) reasoning to improve structured decision-making. AnalogXpert~\cite{zhang2024analogxpert} streamlines analog circuit topology synthesis by incorporating a subcircuit library for design space reduction and using CoT prompting and iterative proofreading for better design accuracy. AnalogGenie~\cite{topologiesanaloggenie} introduces a generative AI framework that utilizes a large dataset of over 3,000 analog circuit topologies. It employs a GPT-based model for sequential pin connection prediction, offering a scalable and flexible approach to analog circuit generation. These works collectively demonstrate the growing potential of LLMs in automating complex aspects of analog design, paving the way for more efficient and scalable circuit synthesis methodologies.\looseness=-1

\new{\textbf{Optimization} of analog circuits is another important aspect, which usually involves multi-objective design spaces.} ADO-LLM~\cite{yin2024ado} combines Bayesian Optimization with LLMs to improve circuit design efficiency, leveraging Gaussian Process models for systematic design space exploration and in-context learning for guided optimization.
\new{LEDRO~\cite{kochar2024ledro} enhances analog circuit sizing by using LLMs to refine design search regions, improving the efficiency of existing optimization methods.}\looseness=-1





\section{Challenges, Discussion, and Potential Directions}
\label{sec:challenges}

Despite the significant advancements in circuit foundation models, several challenges remain in terms of model performance, scalability, data availability, and the integration of predictive circuit encoders and generative circuit decoders. Moreover, these challenges are closely interrelated, often affecting and amplifying one another. We believe addressing these challenges is crucial to further enhancing the effectiveness and applicability of foundation AI models in EDA.
In this section, we discuss our observed challenges and potential research directions to further improve the effectiveness and applicability of foundation AI models in EDA.

\subsection{Challenge 1: Circuit Foundation Model Generalization and Scalability}\label{sec:c1}
The development of circuit foundation models presents several challenges regarding generalization, performance, and scalability. If we can address these challenges, circuit foundation models can effectively support a wider range of design tasks while maintaining computational efficiency.

\textbf{Towards more generalized circuit embeddings from circuit encoders.}
One of the key challenges is designing circuit encoders that generate generalized embeddings capable of capturing both the semantic and structural intrinsic properties of circuits. These embeddings should effectively support largely different downstream tasks. For instance, design quality evaluation relies heavily on structural characteristics, while functional reasoning and verification require a deep understanding of circuit semantics.
This requires innovations in ML model architectures, self-supervised learning techniques tailored for circuits, multimodal fusion strategies, and cross-design-stage alignment. The integration of graph-based encoders with text-based LLMs has shown promise in capturing both structural and semantic information, as seen in works like CircuitFusion~\cite{fang2025circuitfusion}, NetTAG~\cite{fang2025nettag}, and ProgSG~\cite{qin2024cross}. However, further advancements are needed to enhance representation learning across different abstraction levels while ensuring alignment between circuit modalities.

\textbf{Towards reducing hallucinations of circuit decoders.}
Decoder-based models, particularly those leveraging LLMs, are prone to generating hallucinated or syntactically incorrect HDL code. Unlike natural language, circuit descriptions have strict syntax and correctness constraints, requiring verification and refinement mechanisms to ensure reliability. Approaches such as reinforcement learning with human feedback (RLHF), constraint-aware decoding, and post-generation validation can help mitigate hallucinations and improve the practicality of decoder-based circuit models.

\textbf{Towards more scalable circuit foundation models.}
Scalability remains a critical challenge, particularly for large-scale circuit designs. Current models struggle with handling industrial-scale designs due to the complexity and size of modern VLSI circuits. Divide-and-conquer strategies, such as hierarchical modeling, circuit partitioning~\cite{yao2024rtlrewriter}, and subgraph-based processing~\cite{fang2025circuitfusion}, offer potential solutions to improve scalability. By segmenting circuits into smaller, more manageable sub-circuits and processing them independently, models can maintain computational efficiency without sacrificing accuracy. Techniques such as progressive training, adaptive resolution encoding, and distributed processing can further enhance the scalability of circuit foundation models, enabling their deployment in large-scale EDA workflows.

\new{
\textbf{Towards standardized benchmarks and evaluation protocols.}
A further challenge is the lack of widely adopted benchmarks to systematically evaluate CFMs. At present, common benchmarks are mainly available for LLM-based RTL code generation (e.g., VerilogEval, RTLLM, and CVDP), where multiple methods can be compared under a shared evaluation protocol. However, for most encoder-based CFMs and many decoder-based models beyond RTL generation, evaluation is typically performed on different datasets with heterogeneous metrics. This makes it difficult to compare models fairly or to track progress across tasks, modalities, and design stages. There is a pressing need for more general benchmark suites and standardized evaluation protocols that cover a broader spectrum of EDA tasks (e.g., PPA prediction, verification, functional reasoning), include both predictive and generative objectives, and reflect realistic industrial design constraints. Establishing such benchmarks is a key step toward objectively assessing CFM generalization and guiding future model and data design.
}

\subsection{Challenge 2: Circuit Data Avaliability}
The effectiveness of circuit foundation models heavily depends on access to large and diverse datasets for pre-training and fine-tuning. While efforts such as OpenABC-D~\cite{chowdhury2021openabc}, CircuitNet~\cite{chai2023circuitnet}, and DeepCircuitX~\cite{li2025deepcircuitx} have contributed by collecting open-source circuit designs, obtaining a sufficiently large and labeled dataset remains a challenge. Privacy concerns, proprietary design restrictions, and the high cost of generating high-quality annotated data further limit dataset availability. Overcoming these barriers may require advancements in synthetic dataset generation or novel circuit data augmentation techniques.

\textbf{Towards generating synthetic circuit datasets.}
One emerging approach to overcoming data scarcity is the generation of unlimited synthetic circuit datasets, which can be created using graph-based or text-based methods, as explored in~\cite{fang2024transferable, liu2025towards, liu2025syncircuit}. Graph-based approaches can generate large-scale circuit graphs with diverse topologies but often lack meaningful semantics, making it difficult to ensure functional correctness. On the other hand, text-based synthesis methods, such as automated HDL code generation, can produce realistic functional modules but typically lack scalability and diversity in structural variations. Bridging the gap between these two approaches by incorporating both functional correctness and large-scale diversity remains an open research challenge.\looseness=-1

\begin{figure}[!t]
    \centering
    \includegraphics[width=0.7\linewidth]{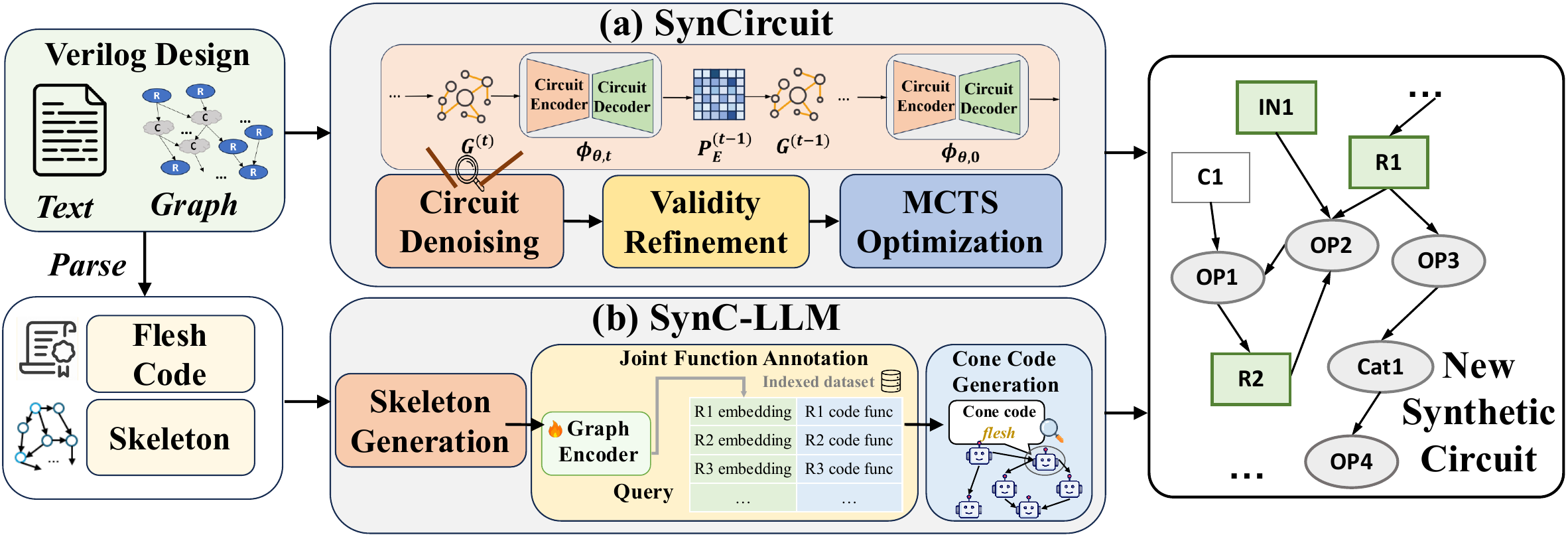}
    \vspace{-.1in}
    \caption{Comparison of synthetic circuit dataset generation paradigms. (a) Graph-based framework exemplified by SynCircuit~\cite{liu2025syncircuit}. (b) Hierarchical graph+LLM framework exemplified by SynC-LLM~\cite{liu2025sync}.}
    \label{fig:sync}
    \vspace{-.1in}
\end{figure}

\new{In our recent attempts, SynCircuit~\cite{liu2025syncircuit} and SynC-LLM~\cite{liu2025sync} show that carefully designed synthetic RTL corpora can substantially improve the downstream performance of circuit learning and foundation models (e.g., for PPA prediction tasks). SynCircuit follows a graph-centric paradigm: it learns to sample new circuit graphs that resemble real designs in terms of structural statistics and synthesizability, and then converts them back to RTL. In contrast, SynC-LLM adopts a hierarchical generation strategy that first produces high-level circuit skeletons and then uses language models to generate local RTL blocks, making the overall pipeline more scalable to very large designs while injecting richer local semantic cues.}

\new{At the same time, both approaches highlight important open challenges. Although they can generate structurally realistic and compilable circuits at scale, the global functional behavior of the synthesized designs is only weakly controlled, since no explicit end-to-end specifications are enforced. Understanding how to better steer the functionality and distribution of synthetic designs, and how to mix them with limited real industrial circuits during pre-training, remains an open problem for circuit foundation models.}

\textbf{Towards more advanced circuit data augmentation.}
Data augmentation techniques have been widely explored in machine learning to improve model generalization. In the circuit domain, functionally equivalent transformations, such as logic optimization from the logic synthesis tools, have been used to create diverse training samples~\cite{wang2022functionality, fang2025circuitfusion, fang2025nettag}. However, existing augmentation strategies primarily focus on structural transformations while maintaining functional equivalence. Future advancements could explore more sophisticated augmentation techniques, such as e-graph rewriting for RTL designs~\cite{coward2024combining} and netlists~\cite{chen2024syn} for broader design space exploration. These techniques can further enhance the robustness of circuit foundation models, ensuring they learn richer representations while preserving key design constraints such as timing, power, and area.

\subsection{Challenge 3: Bridging the Gap Between Circuit Encoder and Decoder}\label{sec:c3}
While circuit encoders and decoders have been developed separately to support predictive and generative tasks, unifying these two paradigms presents an opportunity to create a more powerful circuit foundation model. By integrating learned embeddings from circuit encoders into decoder-based generative models, and leveraging synthetic circuit generation from decoders to enhance circuit foundation model pre-training, the capabilities of both sides can be significantly improved.

\textbf{Towards leveraging encoder embeddings for decoder generation.}
Current circuit decoders, often based on pre-trained LLMs, generate circuit contexts without explicitly considering circuit embeddings learned by encoders. By leveraging circuit encoders to generate structured, functionally meaningful embeddings, decoders can refine their generation process to ensure greater correctness and design feasibility. One potential approach is integrating decoder-based circuit text generation with graph embeddings learned from circuit encoders, allowing decoders to generate RTL, netlist, or layout designs that align with realistic circuit representations. 

\begin{figure}[!t]
    \centering
    \includegraphics[width=0.7\linewidth]{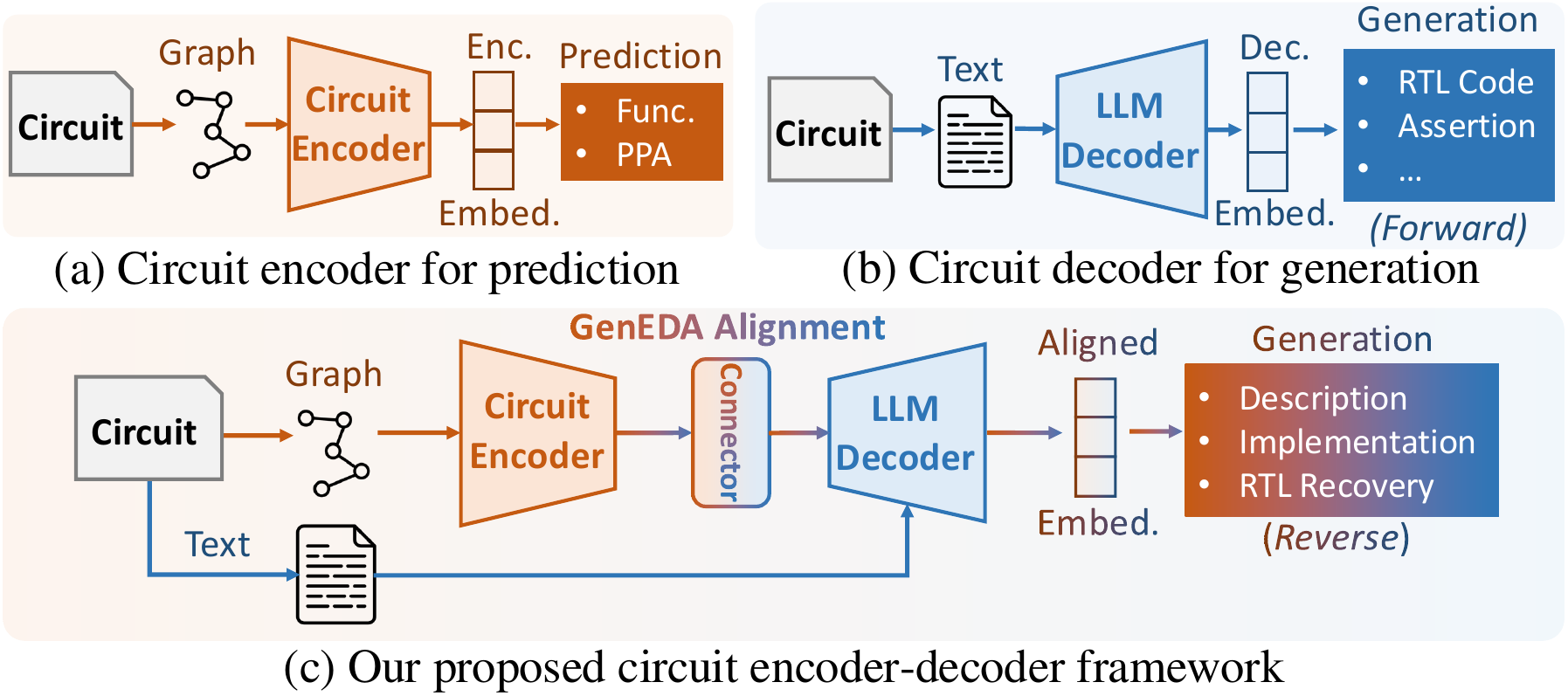}
    \vspace{-.1in}
    \caption{For bridging the gap between circuit encoders and decoders, our GenEDA framework~\cite{fang2025geneda} shows that integrating encoder representations into LLM decoders can significantly enhance model capability.}
    \label{fig:GenEDA}
    \vspace{-.35in}
\end{figure}

\new{
As shown in~\Cref{fig:GenEDA}, in our recent work GenEDA~\cite{fang2025geneda}, we integrate our latest netlist encoder NetTAG~\cite{fang2025nettag} into advanced LLMs via a dedicated circuit connector.
Specifically, GenEDA aligns the graph latent space of NetTAG with the text latent space of LLM decoders through two paradigms: (i) an embedding-based alignment that injects NetTAG embeddings into trainable LLMs during instruction tuning, and (ii) a prediction-based alignment that feeds encoder-generated functional predictions as textual prompts to frozen commercial LLMs.%
This combined encoder–decoder framework enables challenging reverse netlist functional reasoning tasks, such as generating high-level specifications, natural-language functional descriptions, and even RTL code directly from low-level gate-level netlists, and consistently improves LLM performance over using either the encoder or the LLM independently.
}

\textbf{Towards leveraging decoder generated circuits for enhancing circuit foundation models.}
Generating synthetic circuits at different abstraction levels (e.g., RTL, netlist, layout) not only improves data availability but also provides a valuable resource for pre-training both encoders and decoders. By training foundation models on synthetically generated yet functionally diverse circuits, models can capture richer design patterns and structural relationships. Additionally, synthetic circuits can be used to fine-tune models for specific design tasks, enhancing their generalization across unseen circuit designs. Future research could explore hybrid approaches that combine rule-based generation, reinforcement learning, and generative models to create high-quality synthetic datasets that support both encoder and decoder training.

\new{\subsection{Potential Future Directions for Circuit Foundation Models}}

\new{
As circuit foundation models are still in their early stages, there remain many open opportunities to deepen their impact on practical chip design. Below, we highlight two promising directions in the authors' view: using CFMs as core engines for circuit optimization, and extending them into agentic systems that can actively plan and manage chip design workflows.
}

\new{
\textbf{Leveraging circuit foundation models for circuit optimization.}
Most existing CFMs are currently used as \emph{predictors} (e.g., for PPA estimation) or \emph{generators} (e.g., for RTL code snippets), but they are rarely integrated into closed-loop \emph{optimization} frameworks that directly improve circuit implementations.
A promising direction is to couple CFMs with reinforcement learning (RL) or other search-based techniques so that the model not only predicts or generates, but also actively proposes and evaluates optimization actions~\cite{hsiao2025buffalo} (e.g., buffering, sizing, placement, or routing decisions).
In such a setup, CFMs provide rich, reusable circuit representations and fast surrogates for PPA feedback, while RL explores the design space guided by these priors, ultimately targeting end-to-end PPA improvement under real design constraints.
}

\new{
\textbf{Agentic CFMs for chip design.}
Another exciting direction is to develop \emph{agentic} CFMs that can autonomously plan and coordinate multi-step chip design workflows.
One line of work is \emph{flow-level agentic CFMs}, which act as high-level agents that orchestrate EDA tools, maintain long-term memory of design states, and adaptively select optimization passes or flow configurations based on CFM predictions and historical results.
Another line is \emph{algorithm-level agentic CFMs}, where the model is used to analyze, refine, and even self-evolve EDA algorithms themselves~\cite{yu2025autonomous} (e.g., proposing new heuristics, tuning cost functions, or discovering novel transformations) in a closed feedback loop.
Both directions aim to move from ``CFMs as passive models'' to ``CFMs as active decision-making agents'' that continuously interact with tools, data, and designs to improve the overall chip design process.
}

\section{Conclusion}\label{sec:concl}

In this survey, we provide a systematic review of the latest progress in circuit foundation models, categorizing them into encoder-based and decoder-based approaches. Encoders aim to learn generalized circuit embeddings through self-supervised pre-training techniques, supporting predictive tasks such as design quality estimation and functional verification. Decoders, primarily built upon pre-trained LLMs, focus on generative tasks such as HDL code generation and verification automation.
As AI techniques continue to transform the EDA landscape, circuit foundation models hold the potential to significantly reduce design effort, accelerate the chip design process, and improve design quality. Future potential research may target enhancing scalability, generalization, and efficiency, ultimately driving AI-powered innovation in modern VLSI design.

\section{Acknowledgments}

This work is supported by Guangdong Science and Technology Department (No. 2025A0505000035), HKUST–SEU Joint Research Seed Fund, and Hong Kong Research Grants Council (RGC) CRF-YCRG C6003-24Y. It was partially conducted by ACCESS – AI Chip Center for Emerging Smart Systems, supported by the InnoHK initiative of the Innovation and Technology Commission of the Hong Kong Special Administrative Region Government.



\bibliographystyle{unsrt}

\bibliography{ref, ref-llm-security, ref-llm-arch, ref-encoder, ref-ai}  


\end{document}